\documentclass{aa}
\def \eg {e.g.}

\def \ie {i.e.}
\def \cf {cf.}

\def \omegam {{\hbox{$\Omega_{\rm m}$}}}
\def \omegal {{\hbox{$\Omega_\Lambda$}}}
\def \hzero {{\hbox{$H_0$}}}

\def \arcsec {\hbox{$^{\prime\prime}$}}

\def \msun {\hbox{${\rm M_\odot}$}}

\def \mfive {\hbox{$M_{500}$}}

\newcommand{\ergs }{\mbox{erg s$^{-1}$}}

\newcommand{\kmsmpc }{\mbox{km s$^{-1}$ Mpc$^{-1}$}}

\newcommand{\mujyb }{\mbox{$\mu$Jy beam$^{-1}$}}

\newcommand{\acis }{ACIS}

\newcommand{\obsid }{ObsID}

\newcommand{\uv }{\textit{uv}}

\newcommand{\pybdsf }{\textsc{pybdsf}}
\newcommand{\pybdsfE }{PYthon Blob Detector and Source Finder}

\newcommand{\ciao }{\textsc{ciao}}

\newcommand{\caldb }{\textsc{caldb}}

\newcommand{\pyproffit }{\textsc{pyproffit}}

\newcommand{\halofdca }{\textsc{Halo-FDCA}}
\newcommand{\halofdcaE }{Halo-Flux Density CAlculator}

\newcommand{\xmm }{{\em XMM-Newton}}
\newcommand{\chandra }{{\em Chandra}}

\newcommand{\einstein }{{\em Einstein}}

\newcommand{\asca }{{\em ASCA}}

\newcommand{\rosat }{{\em ROSAT}}

\newcommand{\lofar }{LOFAR}

\newcommand{\ska }{SKA}
\newcommand{\skaE }{Square Kilometer Array}

\newcommand{\meerkat }{MeerKAT}

\newcommand{\mgcls }{MGCLS}
\newcommand{\mgclsE }{MeerKAT Galaxy Cluster Legacy Survey}

\usepackage{graphicx}
\usepackage{placeins}
\usepackage{comment}
\usepackage{amssymb}
\usepackage{multirow,bigdelim}
\usepackage{txfonts}
\usepackage[export]{adjustbox}
\usepackage{hyperref}
\hypersetup{colorlinks,citecolor=blue,filecolor=black,linkcolor=red,urlcolor=black}
\usepackage{color}

\begin{document} 

\title{Surface brightness discontinuities in radio halos}
\subtitle{Insights from the MeerKAT Galaxy Cluster Legacy Survey}

\authorrunning{A. Botteon et al.} 
\titlerunning{Surface brightness discontinuities in radio halos}

\author{Andrea Botteon\inst{\ref{ira}}, Maxim Markevitch\inst{\ref{goddard}}, Reinout J. van Weeren\inst{\ref{leiden}}, Gianfranco Brunetti\inst{\ref{ira}}, \and Timothy W. Shimwell\inst{\ref{astron},\ref{leiden}}}

\institute{
INAF - IRA, via P.~Gobetti 101, I-40129 Bologna, Italy \label{ira} \\
\email{andrea.botteon@inaf.it} 
\and
NASA/Goddard Space Flight Center, Greenbelt, MD 20771, USA \label{goddard} 
\and
Leiden Observatory, Leiden University, PO Box 9513, NL-2300 RA Leiden, The Netherlands \label{leiden} 
\and
ASTRON, the Netherlands Institute for Radio Astronomy, Postbus 2, NL-7990 AA Dwingeloo, The Netherlands \label{astron}
}

\date{Received XXX; accepted YYY}

\abstract
{Dynamical motions in the intra-cluster medium (ICM) can imprint distinctive features on X-ray images that map the thermal bremsstrahlung emission from galaxy clusters, such as sharp surface brightness discontinuities due to shocks and cold fronts. The gas dynamics during cluster mergers may also drive large-scale turbulence in the ICM, which in turn generates extended (megaparsec-scale) synchrontron sources known as radio halos.}
{Surface brightness edges have been found numerous times  in the thermal gas of clusters based on X-ray observations. In contrast, edges in radio halos have only been observed in a handful of cases. Our goal is to search for new radio surface brightness discontinuities in the ICM.}
{We inspected the images of the Bullet Cluster and the other 25 radio halos reported in the \mgclsE. To aid the identification of surface brightness discontinuities, we applied a gradient-filtering edge-detection method to the radio images.}
{We find that the adopted filtering technique is helpful in identifying surface brightness edges in radio images, allowing us to identify at least one gradient in half of the radio halos studied. For the Bullet Cluster, we find excellent agreement between the locations of the four radio discontinuities detected and the X-ray edges. This similarity informs us that there is substantial interplay between thermal and nonthermal components in galaxy clusters. This interplay is likely due to the frozen-in ICM magnetic field, which mediates the advection of cosmic rays while being dragged by thermal gas flows.}
{We conclude that radio halos are shaped by dynamical motions in the ICM and that they often display surface brightness discontinuities, which appear to be co-located with edges in the thermal gas emission. Our results demonstrate that new and future generations of radio telescopes will provide an approach to efficiently detecting shocks and cold fronts in the ICM that is  complementary to X-rays.}


\maketitle

\section{Introduction}

\begin{table*}
 \centering
 \caption{List of the galaxy clusters in the \mgcls\ that host a radio halo. Elongated halos and halos with embedded bright AGN sources are indicated with a dagger symbol. All values are from \citet{knowles22}. Abell 85 is included in the list because it hosts a type of diffuse emission bridging both halo and mini-halo classifications \citep[see][]{knowles22}.}
 \label{tab:sample}
 \begin{tabular}{lrrrrrrrr}
  \hline
  \hline
  \multicolumn{1}{l}{Cluster name} &
  \multicolumn{1}{r}{RA} &
  \multicolumn{1}{r}{Dec} &
  \multicolumn{1}{r}{$z$} &
  \multicolumn{1}{r}{LAS} &
  \multicolumn{1}{r}{LLS} &
  \multicolumn{1}{r}{rms} &
  \multicolumn{1}{r}{Beam} 
  \\
  &
  \multicolumn{1}{r}{[deg]} &
  \multicolumn{1}{r}{[deg]} &
  &
  \multicolumn{1}{r}{[arcmin]} &
  \multicolumn{1}{r}{[Mpc]} &
  \multicolumn{1}{r}{[\mujyb]} &
  \multicolumn{1}{r}{[$\arcsec \times \arcsec$]} 
  \\
  \hline
  Abell 85$^\dagger$ & 10.453 & -9.318 & 0.056 & 4.0 & 0.26 & 3.3 & $7.7\times7.6$ \\
  Abell 209 & 22.990 & -13.576 & 0.209 & 7.9 & 1.62 & 3.6 & $7.8\times7.7$ \\
  Abell 521 & 74.536 & -10.244 & 0.248 & 5.3 & 1.23 & 3.4 & $8.5\times7.9$ \\
  Abell 545b & 83.102 & -11.543 & 0.154  &4.7 & 0.75 & 3.1 & $8.3\times8.0$ \\
  Abell 2667$^\dagger$ & 357.920 & -26.084 & 0.232  &4.6 & 1.02 & 2.7 & $7.6\times7.5$ \\
  Abell 2744 & 3.567 & -30.383 & 0.307 & 7.5 & 2.04 & 2.9 & $7.1\times7.1$ \\
  Abell 2811 & 10.537 & -28.536 & 0.108 & 5.2 & 0.62 & 2.6 & $9.2\times8.3$ \\
  Abell 3558 & 201.978 & -31.492 & 0.048 & 4.5 & 0.25 & 2.9 & $7.4\times7.2$ \\
  Abell 3562 & 202.783 & -31.673 & 0.050 & 11.5 & 0.67 & 3.3 & $7.3\times7.2$ \\
  Abell S295 & 41.400 & -53.038 & 0.300 & 4.0 & 1.07 & 2.3 & $8.1\times7.9$ \\
  Abell S1063 & 342.181 & -44.529 & 0.348 & 5.5 & 1.62 & 2.6 & $7.1\times7.0$ \\
  Bullet Cluster & 104.658 & -55.950 & 0.297 & 8.5 & 2.26 & 2.8 & $7.9\times7.6$ \\
  El Gordo & 15.719 & -49.250 & 0.870 & 3.3 & 1.53 & 1.5 & $7.2\times6.6$ \\
  MACS J0417.5-1154$^\dagger$ & 64.394 & -11.909 & 0.443 & 4.9 & 1.68 & 2.9 & $7.9\times7.8$ \\
  RXC J1314.4-2525 & 198.599 & -25.256 & 0.244 & 4.7 & 1.08 & 4.2 & $7.5\times7.4$ \\
  J0145.0-5300 (Abell 2941) & 26.260 & -53.014 & 0.117 & 4.9 & 0.62 & 2.6 & $7.5\times7.4$ \\
  J0225.9-4154 (Abell 3017) & 36.478 & -41.910 & 0.220 & 2.4 & 0.51 & 2.7 & $7.6\times7.4$ \\
  J0232.2-4420 (PSZ2 G259.98-63.43) & 38.070 & -44.348 & 0.286 & 5.8 & 1.49 & 2.6 & $7.5\times7.3$ \\
  J0303.7-7752 (PSZ1 G294.68-37.01) & 45.943 & -77.869 & 0.274 & 3.8 & 0.95 & 2.9 & $7.5\times7.5$ \\
  J0352.4-7401 (Abell 3186) & 59.123 & -74.031 & 0.127 & 10.7 & 1.46 & 2.6 & $7.8\times7.6$ \\
  J0516.6-5430 (Abell S520) & 79.158 & -54.514 & 0.295 & 5.7 & 1.51 & 3.1 & $7.7\times7.4$ \\
  J0528.9-3927 (Abell 3343) & 82.235 & -39.463 & 0.284 & 4.0 & 1.06 & 2.6 & $7.3\times7.3$ \\
  J0638.7-5358 (Abell S592)$^\dagger$ & 99.694 & -53.972 & 0.227 &6.4 & 1.40 & 3.4 & $7.3\times7.2$ \\
  J0645.4-5413 (Abell 3404) & 101.372 & -54.219 & 0.164 & 7.7 & 1.30 & 3.4 & $7.7\times7.6$ \\
  J1601.7-7544 (PSZ2 G313.88-17.11) & 240.445 & -75.746 & 0.153 & 5.9 & 0.94 & 3.7 & $8.5\times7.8$ \\
  J2023.4-5535 (PSZ1 G342.33-34.92) & 305.852 & -55.592 & 0.232  &4.7 & 1.04 & 2.7 & $7.5\times7.5$ \\
  \hline
 \end{tabular}
 \tablefoot{Col. 1: cluster name; Cols. 2 and 3: coordinates; Col. 4: redshift; Col. 5: largest angular size; Col. 6: largest linear size; Cols. 7 and 8: noise and resolution of the \meerkat\ image.}
\end{table*}

\begin{figure*}
  \centering
  \includegraphics[width=\hsize,trim={0cm 0cm 0cm 0cm},clip,valign=c]{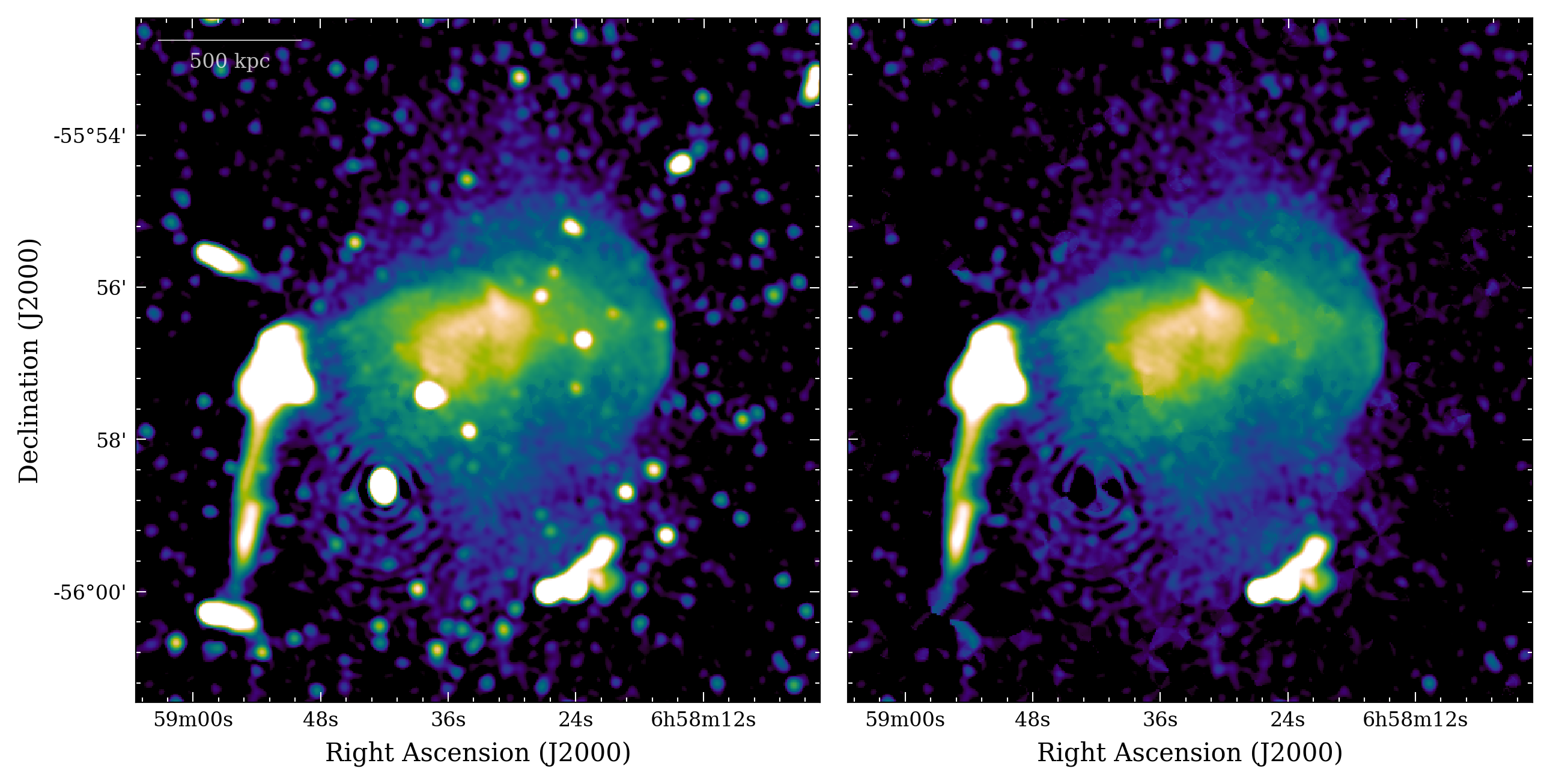}
  \includegraphics[width=\hsize,trim={0cm 0cm 0cm 0cm},clip,valign=c]{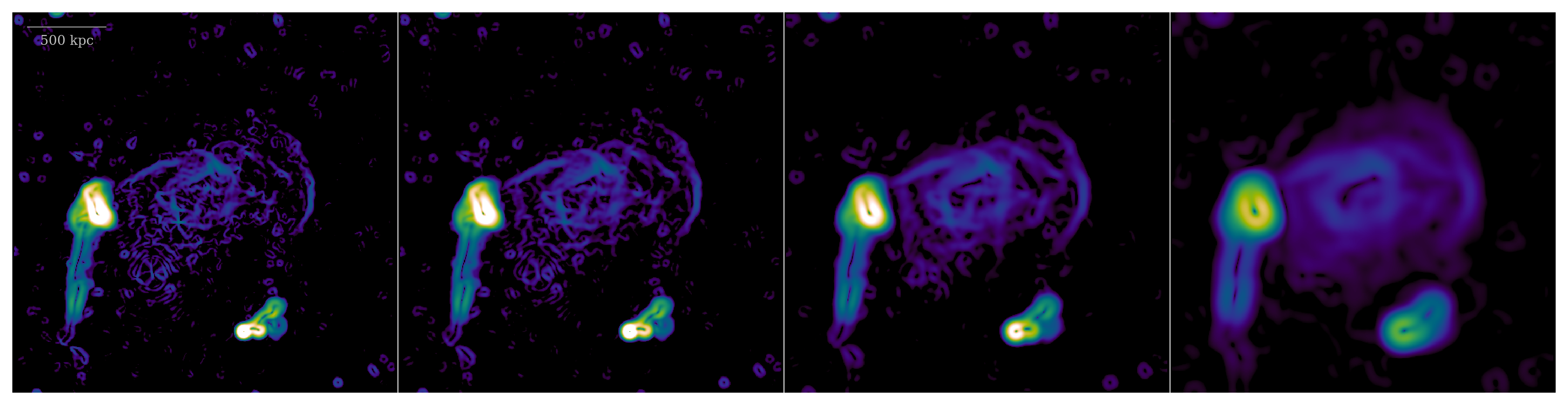}
  \caption{Radio images of the Bullet Cluster. \textit{Top panels}: \meerkat\ image at 1.28 GHz of the radio halo with (\textit{left}) and without (\textit{right}) discrete sources. \textit{Bottom panels}: GGM-filtered images obtained with $\sigma=1,2,4,$ and $8$ pixels (from \textit{left} to \textit{right}).}
  \label{fig:bullet_meerkat}
\end{figure*}

\begin{figure*}
  \centering
  \includegraphics[width=\hsize,trim={0cm 0cm 0cm 0cm},clip,valign=c]{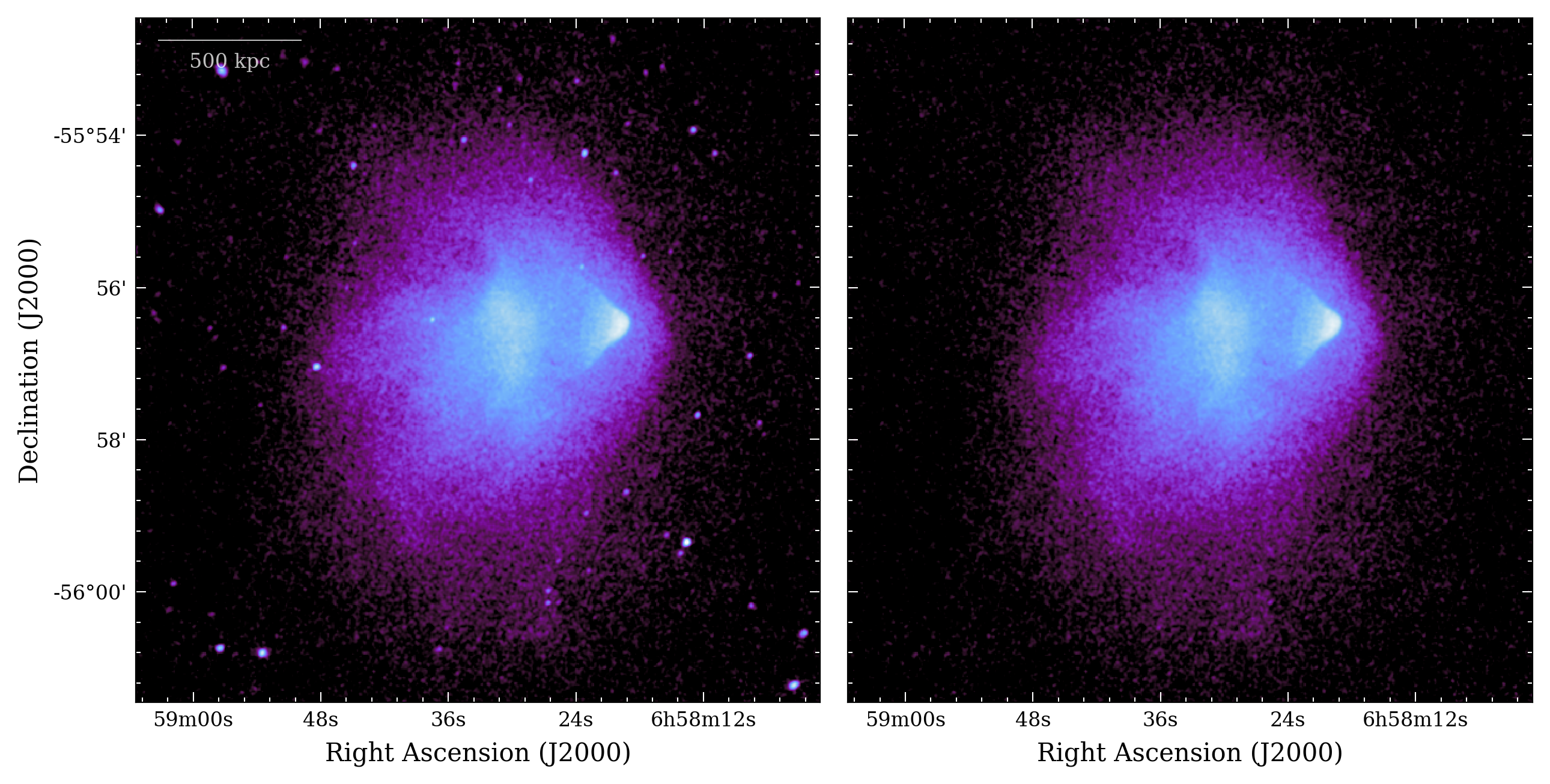}
  \includegraphics[width=\hsize,trim={0cm 0cm 0cm 0cm},clip,valign=c]{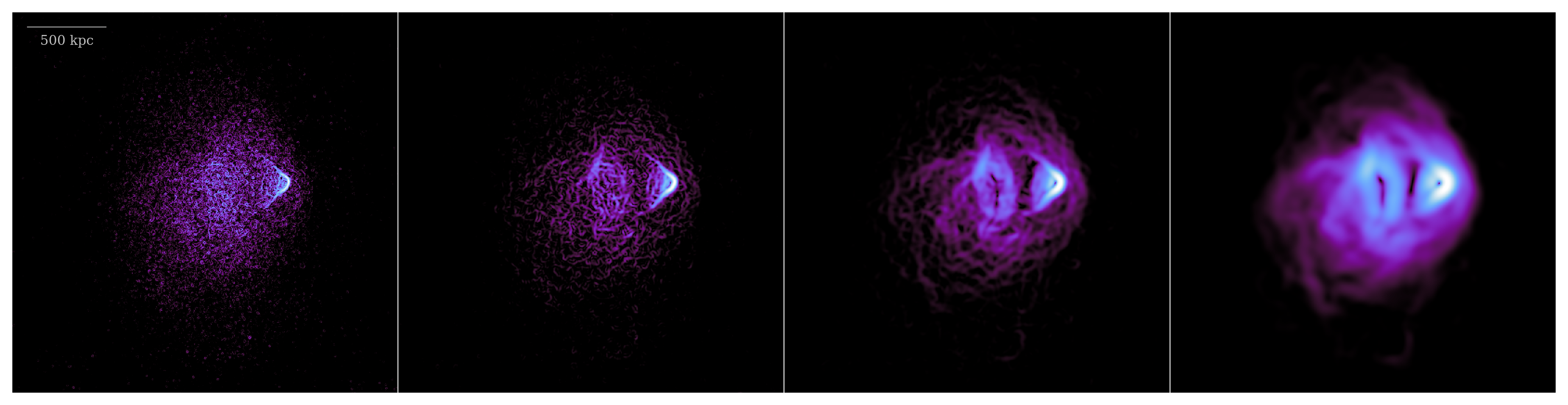}
  \caption{X-ray images of the Bullet Cluster. \textit{Top panels}: \chandra\ image in the 0.5$-$2.0 keV band of the thermal ICM with (\textit{left}) and without (\textit{right}) discrete sources. \textit{Bottom panels}: GGM-filtered images obtained with $\sigma=1,2,4,8$ pixels (from \textit{left} to \textit{right}).}
  \label{fig:bullet_chandra}
\end{figure*}

The intra-cluster medium (ICM) is an X-ray-luminous (${\sim}10^{44}$ \ergs\ in the 0.1$-$2.4 keV band), hot ($T \sim 10^8$ K), and tenuous ($n_e \sim 10^{-3}$ particles cm$^{-3}$) plasma that comprises ${\sim}75\%$ of the baryonic mass of galaxy clusters \citep[\eg,][]{sarazin86rev}. It acquires its high temperature from the energy released during the formation process of clusters, where gravitational energy is dissipated by shocks and large-scale turbulent motions and goes mainly into heating the ICM \citep[\eg,][]{norman99, ricker01}. A small fraction of the kinetic energy dissipated in cluster mergers is eventually channeled into nonthermal components (relativistic particles and magnetic fields), generating cluster-scale synchrotron sources \citep[e.g.,][for reviews]{brunetti14rev, vanweeren19rev}. \\ %
\indent
One of the major findings of galaxy cluster science enabled by the superb angular resolution of \chandra\ is the characterization of sharp X-ray surface brightness discontinuities in the ICM due to shocks and cold fronts \citep[e.g.,][]{markevitch00, vikhlinin01cold, markevitch02bullet}. These features play an important role in our understanding of different aspects of the physics that governs the thermal gas (e.g.,\ the plasma equipartition time, heat conductivity, and viscosity) as well as the processes that appear to be responsible for the generation of the nonthermal components mixed therein. For example, \chandra\ helped to establish the spatial coincidence between: radio relics and shock fronts in cluster outskirts \citep[e.g.,][]{giacintucci08, macario11, shimwell15, botteon16a115}; the boundaries of radio mini-halos and sloshing cold fronts \citep[e.g.,][]{mazzotta08, giacintucci14rxj1720, giacintucci14minihalos}; and the edges in the brightness distribution of some radio halos and X-ray-detected shocks \citep[e.g.,][]{markevitch05, markevitch10arx, shimwell14}. These results demonstrate the interplay between nonthermal components and dynamical motions in the ICM, whose signatures in the thermal gas can only be efficiently unveiled when X-ray images reach sufficiently high resolution and sufficient count statistics are available. \\ %
\indent
Historically, the morphology of diffuse radio sources in the ICM has generally been observed as smooth in appearance. This was likely due to limitations of the data used, which were only able to recover the extended faint emission at low resolution and with low signal-to-noise ratio (S/N). Indeed, deeper and higher resolution observations in recent years (mainly performed with the new generation of radio interferometers) have unveiled an increasing number of substructures in radio halos \citep[e.g.,\ A2255,][]{botteon20a2255, botteon22a2255}, mini-halos \citep[e.g.,\ Perseus Cluster,][]{gendronmarsolais17}, and relics (e.g.,\ Sausage Cluster, \citealt{digennaro18sausage}; Toothbrush Cluster, \citealt{rajpurohit18}; A2256, \citealt{owen14, rajpurohit22a2256}; A3667, \citealt{degasperin22}). Understanding the origin of these features may provide important insights into the generation, distribution, and evolution of cosmic rays and magnetic fields in the ICM. As particle acceleration of thermal electrons is an inefficient process, a preexisting population of mildly relativistic seed electrons in the ICM is required to produce the observed level of synchrotron emission in clusters \citep[e.g.,][]{brunetti14rev}. In the case of radio halos, the seed electrons are believed to be reaccelerated in situ by the turbulence injected during mergers \citep[e.g.,][]{brunetti01coma, petrosian01, fujita03, brunetti07turbulence, brunetti11mfp, brunetti16stochastic, pinzke17, nishiwaki22}. \\
\indent
\meerkat\ \citep{jonas09}, located
in South Africa, is a precursor of the \skaE\ (\ska), and is to be integrated into the mid-frequency component of \ska\ in the future. With its 64 dishes, \meerkat\ offers high sensitivity, dynamic range, and resolution, which, coupled with its densely populated core (48 antennas are concentrated within a diameter of $\sim$1 km), provide exquisite image quality with which to study large-scale diffuse emission from clusters in great detail. While inspecting the recently released image at 1.28 GHz of the Bullet Cluster from the \mgclsE\ \citep[\mgcls;][]{knowles22}, we noticed that in addition to the edge\footnote{This is coincident with the famous shock front detected toward the west of the system \citep{markevitch02bullet}.} of the radio halo emission already pointed out by \citet{shimwell14} but revealed in much more detail in the new \meerkat\ image \citep[see also][]{sikhosana23}, there are other surface brightness discontinuities in the radio halo. This motivated us to characterize them and search for similar edges in the entire sample of 26 radio halos detected in the \mgcls. In this paper, we present the results of this search. The paper is structured as follows. In Section~\ref{sec:data} we provide details of the data used for the analysis. In Section~\ref{sec:techniques}, we describe the edge-detection technique(s) employed to search for discontinuities in the radio images. In Section~\ref{sec:results}, we present the results of our analysis of the Bullet Cluster and of our search for radio edges in the other radio halos in the sample. In Section~\ref{sec:discussion}, we discuss our findings and their implications. Finally, in Section~\ref{sec:conclusions}, we summarize the most important results of our study. We assume a $\Lambda$ cold dark matter cosmology with $\omegal = 0.7$, $\omegam = 0.3$, and $\hzero = 70$ \kmsmpc\ throughout.

\section{Data}\label{sec:data}

\subsection{\meerkat}

The \mgcls\ \citep{knowles22} is a large program ($\sim$1000 h) consisting of observations of  $\sim$6$-$10 h in duration  with \meerkat\ \textit{L}-band (900$-$1670 MHz) of 115 galaxy clusters in the southern sky. We used the full-resolution ($\sim$8\arcsec) images that were made available with the first legacy product data release (DR1). These images have typical root-mean-square (rms) noise of $\sim$2$-$5 \mujyb\ and reference frequency of 1.28 GHz. We retrieved these images for the Bullet Cluster and for the other 25 galaxy clusters reported to host a radio halo in \citet{knowles22}  from the \mgcls\ website\footnote{\url{http://mgcls.sarao.ac.za/}}. The resolution and noise of the images used in this work are summarized in Table~\ref{tab:sample} together with the main properties of the clusters and halos studied. In all images, 1 pixel corresponds to $\sim$1.2\arcsec\ except for El Gordo in which 1 pixel is equal to 1.5\arcsec. \\
\indent
Discrete radio sources interfere with the analysis of the diffuse emission in which they are embedded but this can be mitigated by removing them from the images. Subtracting discrete sources precisely is a difficult task  and one of two different methods is generally adopted 
in the literature to perform this step. The first method is to subtract the clean components of the discrete sources from the \uv\ data and to reimage the subtracted visibilities to obtain source-subtracted images. In the second method, discrete sources are cosmetically removed directly from the images that contain them. The first approach confers the advantage that discrete sources and diffuse emission can be somewhat separated in interferometric data with sensitivity to a wide range of angular scales. However, the calibrated measurement sets are not available in the \mgcls-DR1 and performing the first procedure would require recalibration of all the data from raw visibilities. For this reason, we followed the second approach, the details of which are provided below. \\
\indent
\meerkat\ 1.28 GHz images are extremely sensitive to the emission from active galactic nuclei (AGN), most of which are unresolved \citep[e.g.,][]{heywood22mightee}. Source-detection algorithms such as \pybdsfE\ \citep[\pybdsf;][]{mohan15} and \textsc{Aegean} \citep{hancock12, hancock18} are generally useful to identify and characterize these compact sources. Nonetheless, in the presence of significant diffuse emission in the field-of-view (FoV) and/or AGN with extended and distorted morphology, which are conditions that both often occur in galaxy clusters, source detection is challenging and the automated packages can struggle to accurately characterize the emission from physically distinct but co-located objects. For this reason, we identified contaminating discrete sources by visually inspecting the radio images for local intensity peaks and drawing circular or elliptical region files encompassing the source emission. The visual search was carried out meticulously because of the small number of images to inspect (26, corresponding to the 26 galaxy clusters in our sample) and because we restricted the search to small regions covering the entire radio halo (the FoV typically inspected ranges from 0.03 to 0.22 deg$^2$). Once identified, pixels inside the contaminating discrete source regions were first replaced with NaN values, then they were filled with values interpolated from neighboring data points assuming a 2D Gaussian kernel with the \texttt{interpolate\_replace\_nans} function from the Astropy package. A similar approach is commonly used in X-ray images (see e.g.,\ the following subsection) and produces satisfactory results also in the radio images, as shown in the Fig.~\ref{fig:bullet_meerkat} (top panels). Still, we note in the case of very bright or extended sources it may introduce artifacts and remove any structure of the radio halo that may exist in the excised region. Depending on the specific case, we decided either to not remove difficult sources from the original images or to not consider in our analysis regions affected by subtraction (or calibration) artifacts that would lead to unreliable results.

\subsection{Chandra}

In this work, we also used archival \chandra\ \acis-I observations of the Bullet Cluster (\obsid s: 554, 3184, 4984, 4985, 4986, 5355, 5356, 5357, 5358, 5361). Data were processed using \ciao\ v4.13 \citep{fruscione06} with \caldb\ v4.9.0 \citep{graessle06} following standard data reduction procedures. The filtered exposure time amounts to $\sim$517 ks. We produced exposure-corrected mosaic images of the cluster in the 0.5$-$2.0 keV energy band with and without point sources removed (Fig.~\ref{fig:bullet_chandra}, top panels). Point sources were automatically detected with the \texttt{wavdetect} task, visually validated, and thus cosmetically removed by replacing them with random values extracted from neighboring pixels. The pixel size in our \chandra\ images is $\sim$1\arcsec.

\begin{figure*}
  \centering
  \includegraphics[width=.348\hsize,trim={0cm 0cm 0cm 0cm},clip,valign=c]{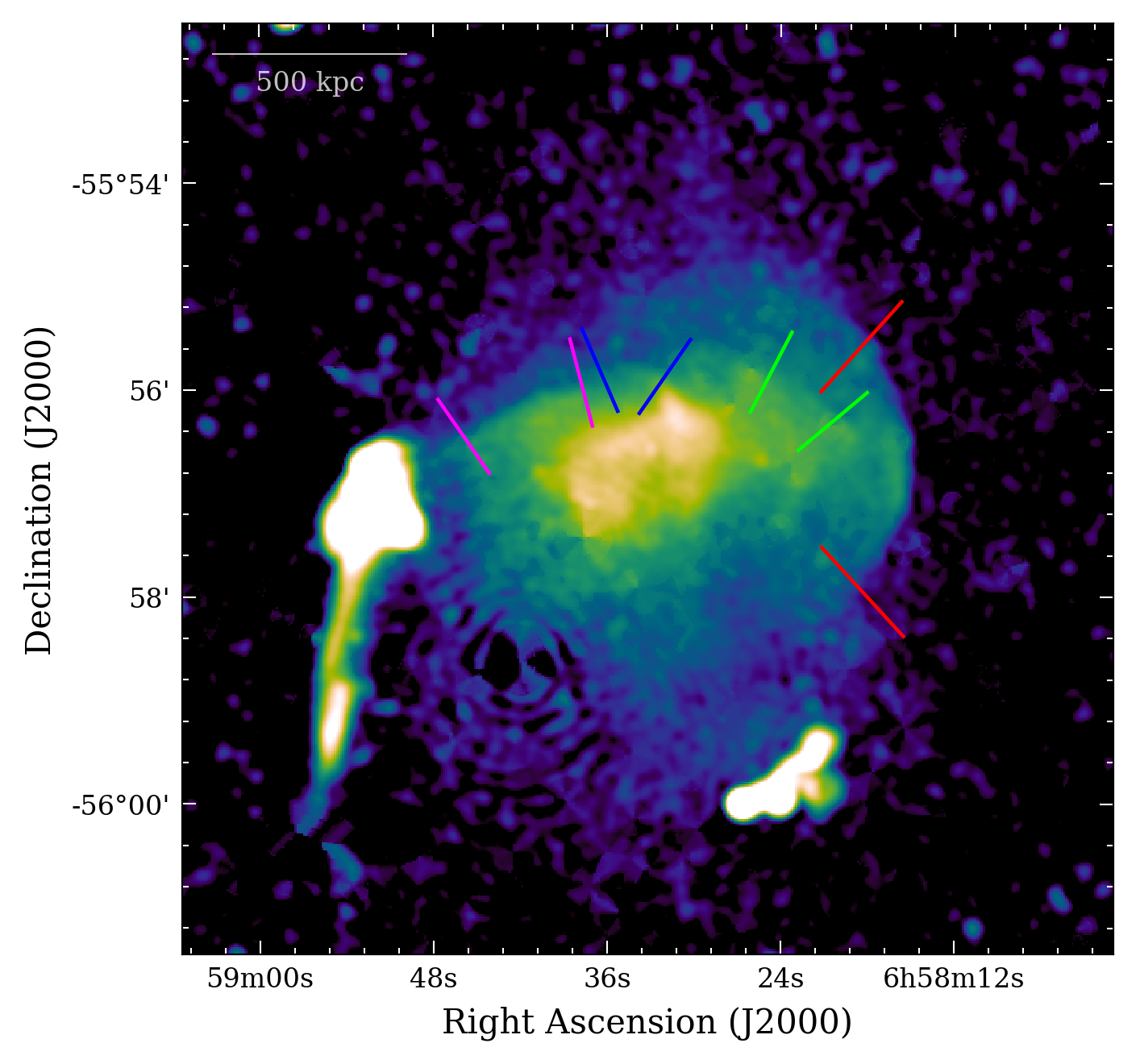}
  \includegraphics[width=.32\hsize,trim={0cm 0cm 0cm 0cm},clip,valign=c]{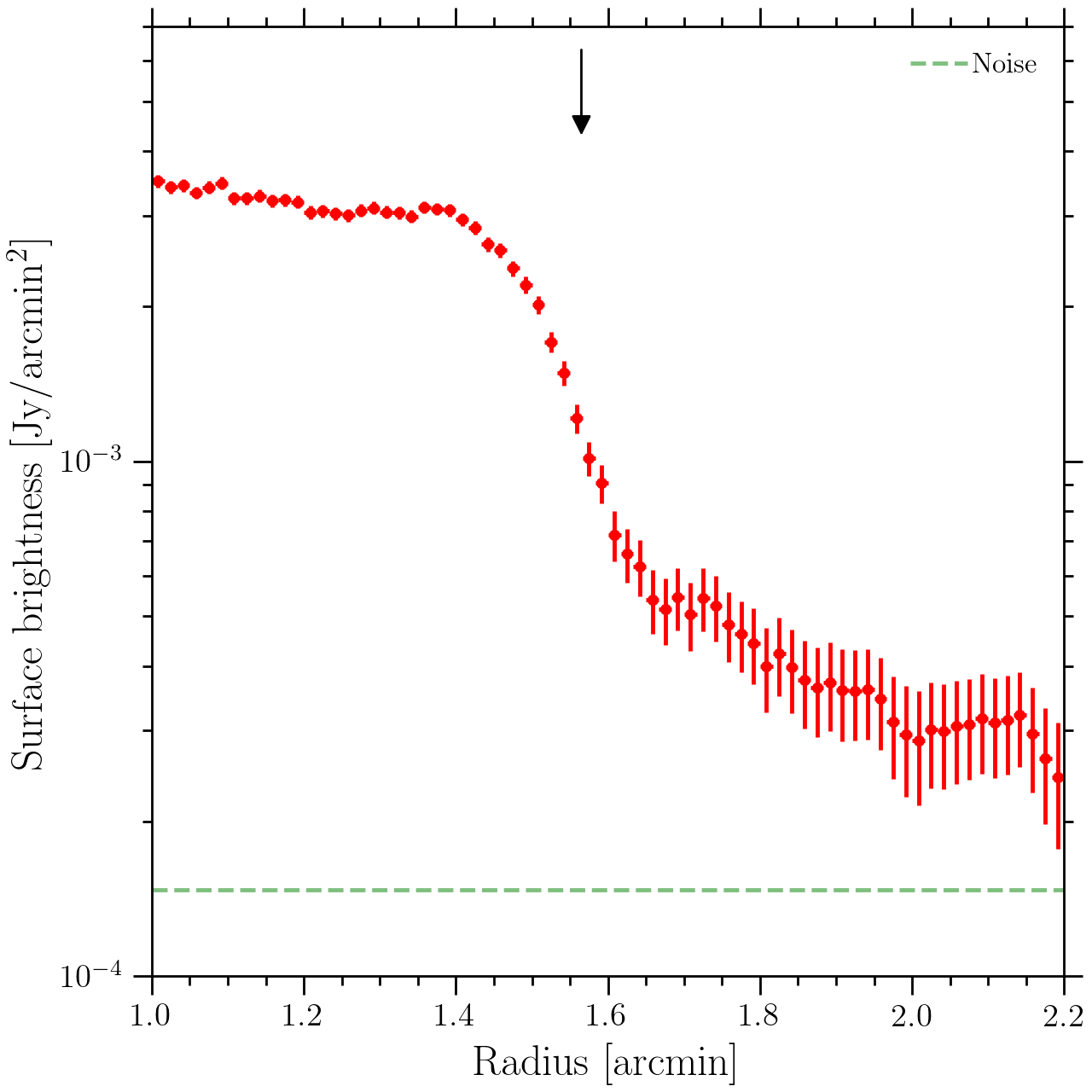}
  \includegraphics[width=.32\hsize,trim={0cm 0cm 0cm 0cm},clip,valign=c]{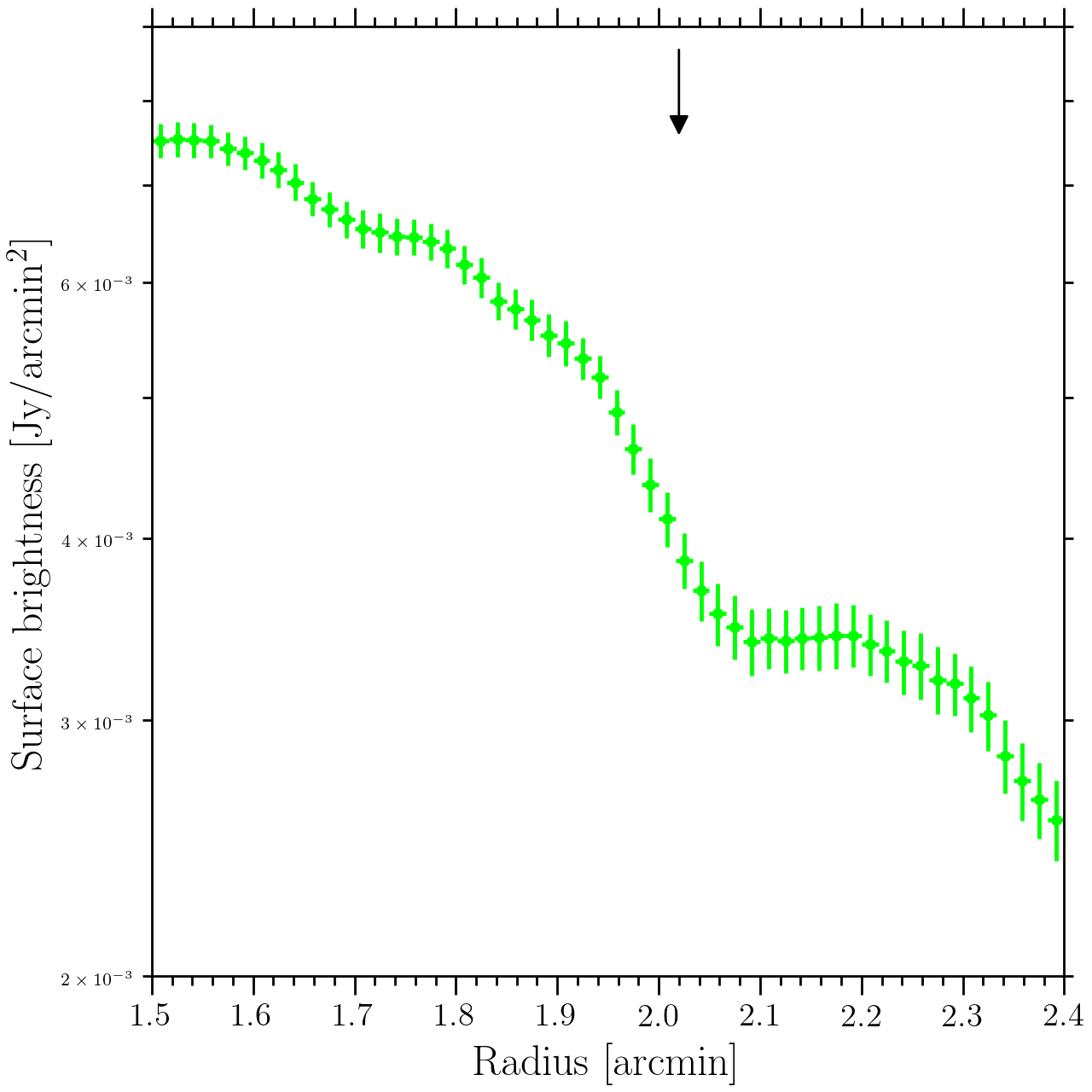}
  \includegraphics[width=.32\hsize,trim={0cm 0cm 0cm 0cm},clip,valign=c]{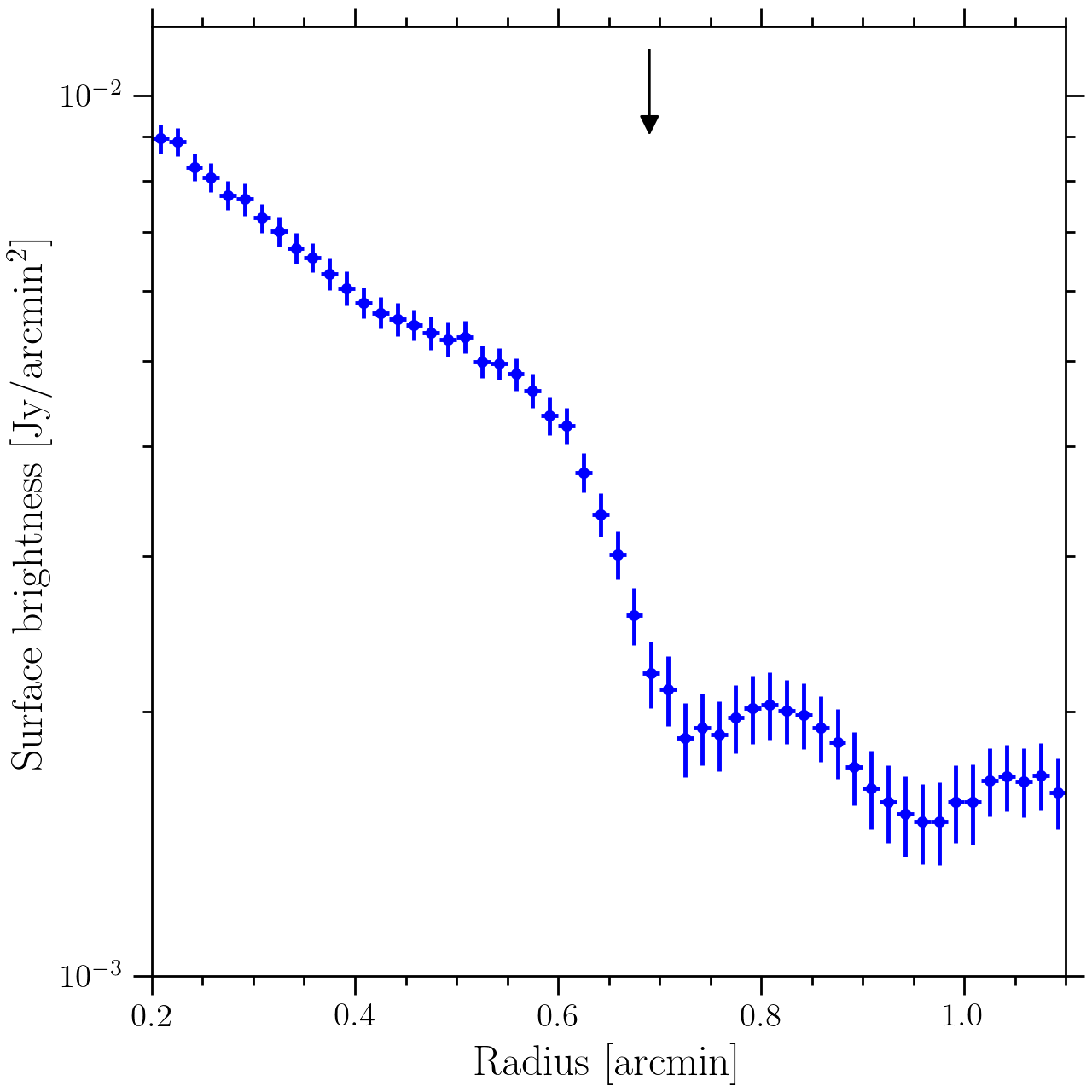}
  \includegraphics[width=.32\hsize,trim={0cm 0cm 0cm 0cm},clip,valign=c]{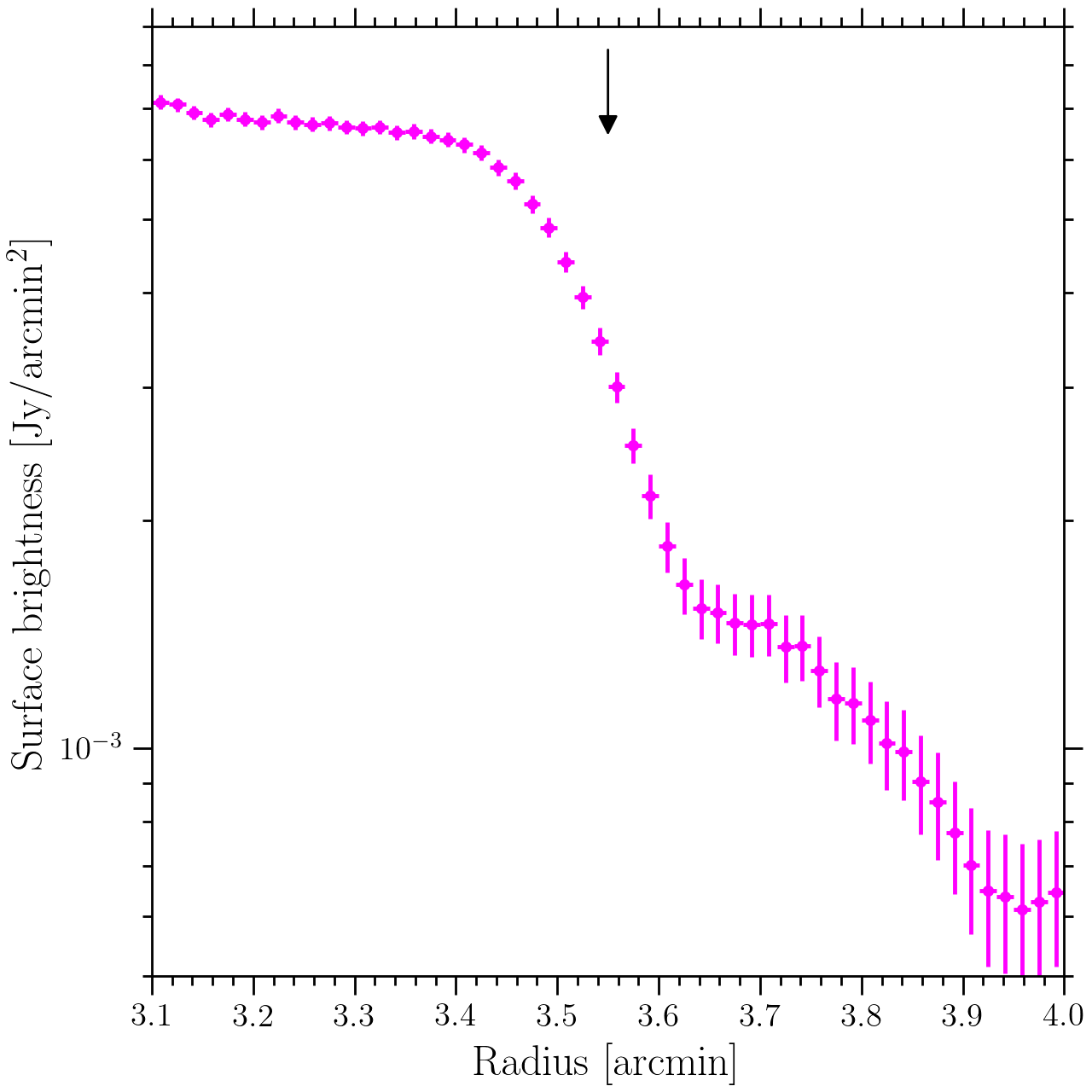}
  \caption{Radio surface brightness profiles extracted from the discrete source-subtracted \meerkat\ image of the Bullet Cluster across the most prominent features highlighted by the GGM-filtered images. Colored segments in the top-left panel denote the sectors used to extract the surface brightness profiles plotted following the same color scheme in the other panels of the figure. Arrows highlight the discontinuities in the profiles.}
  \label{fig:bullet_sb}
\end{figure*}

\begin{figure*}
  \centering
  \includegraphics[width=\hsize,trim={0cm 0cm 0cm 0cm},clip,valign=c]{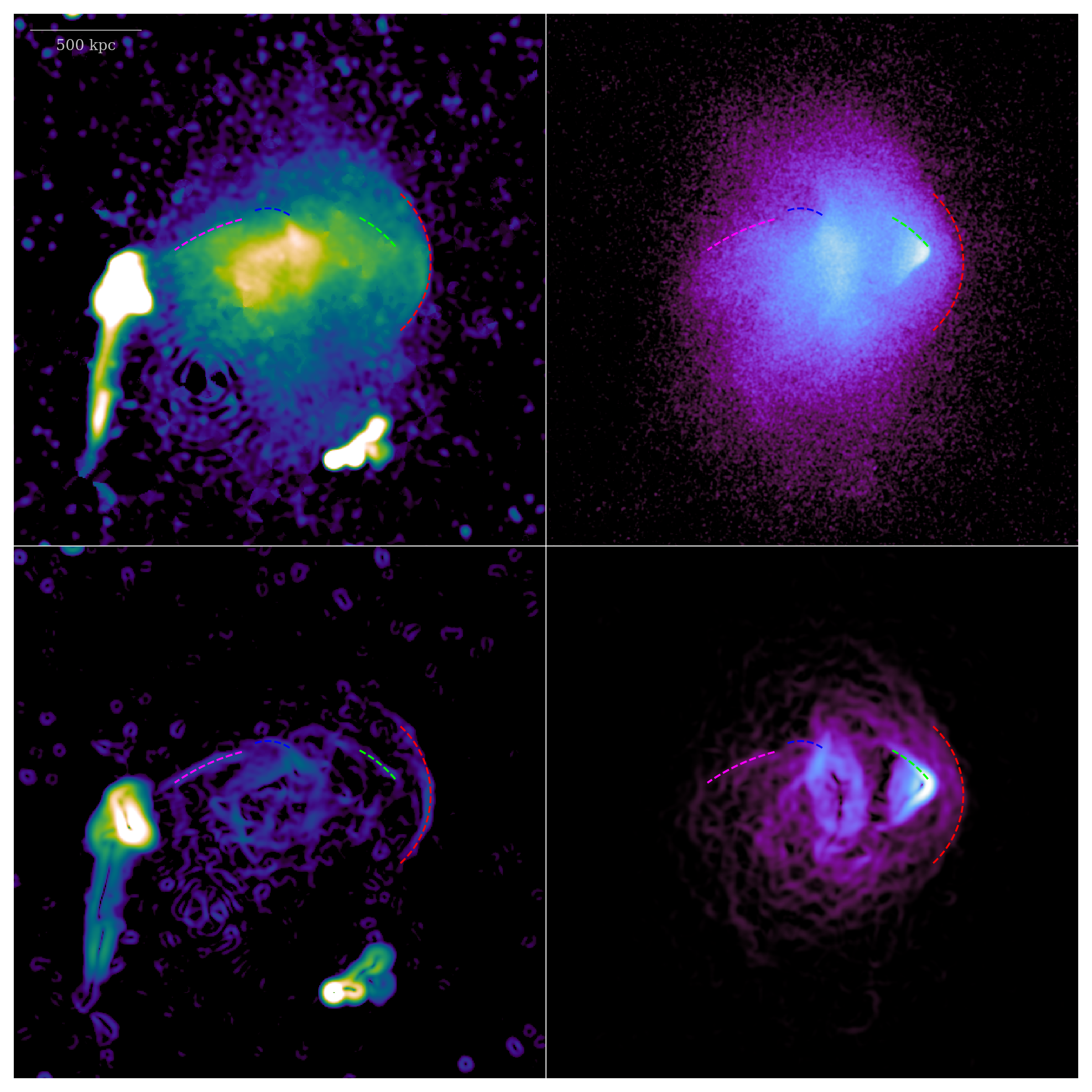}
  \caption{\meerkat\ and \chandra\ images of the Bullet Cluster (\textit{top panels}) and corresponding GGM-filtered images (\textit{bottom panels}). Dashed arcs mark the position and extent of the radio surface brightness discontinuities.}
  \label{fig:bullet_comparison}
\end{figure*}

\begin{figure*}
  \centering
  \includegraphics[width=.49\hsize,trim={0cm 0cm 0cm 0cm},clip,valign=c]{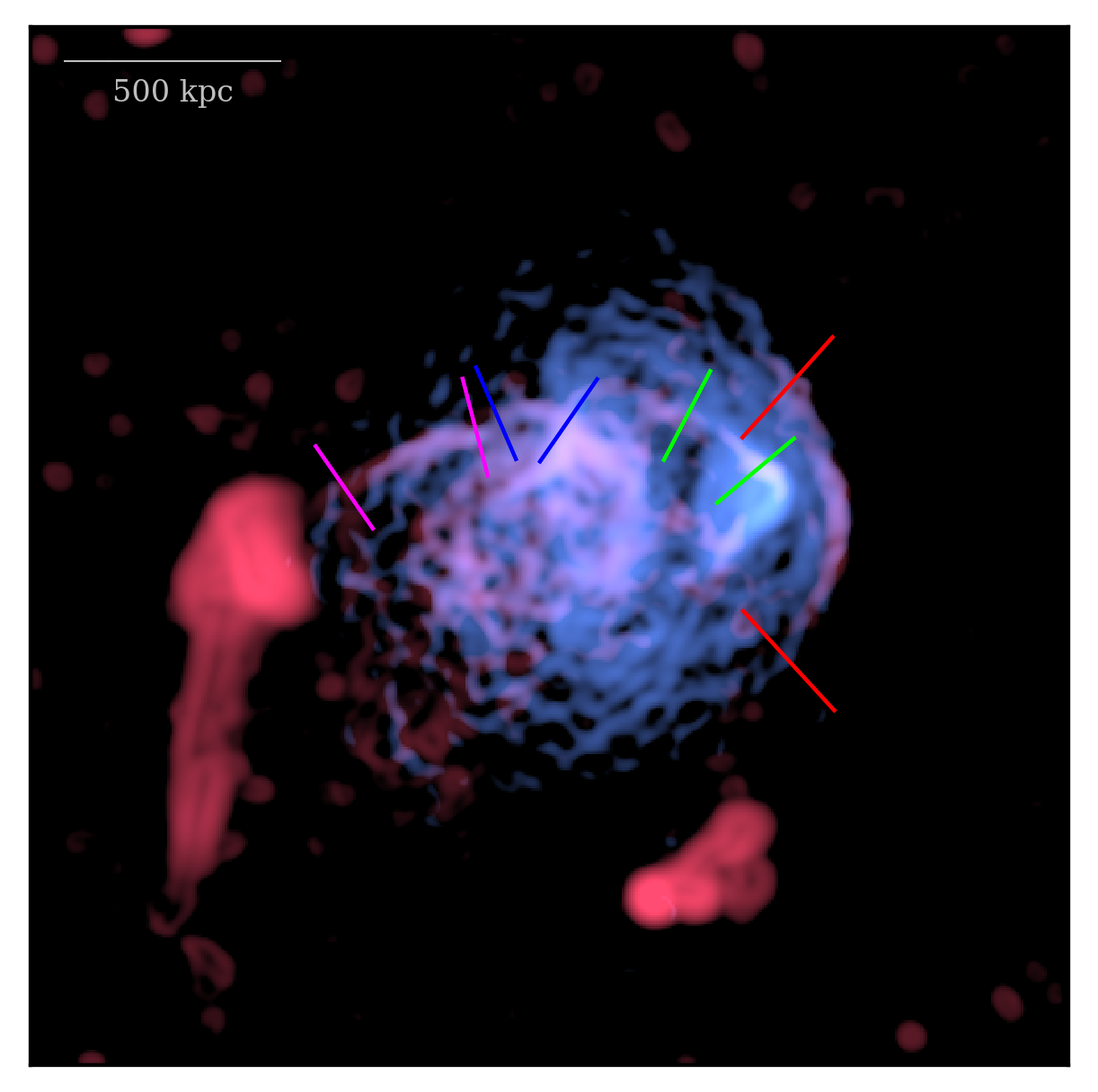}
  \includegraphics[width=.49\hsize,trim={0cm 0cm 0cm 0cm},clip,valign=c]{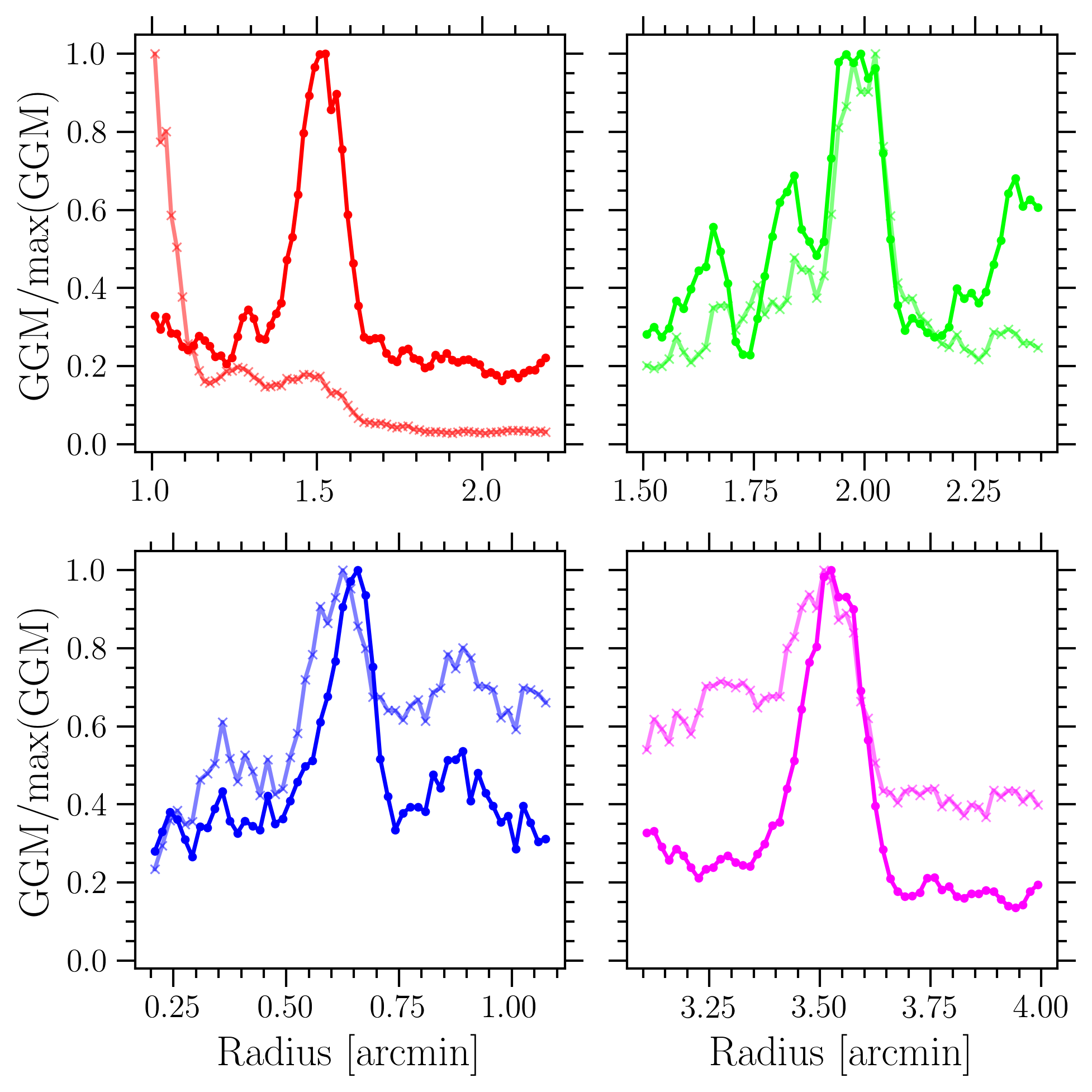}
  \caption{Comparison between radio and X-ray surface brightness gradients. \textit{Left panel}: Composite image of the gradients in the Bullet Cluster. The \meerkat\ GGM-filtered image is reported in \textit{red} while the \chandra\ GGM-filtered image is reported in \textit{blue}. \textit{Right panels}: Profiles extracted from the \meerkat\ (\textit{dots}, darker colors) and \chandra\ (\textit{crosses}, lighter colors) GGM-filtered images in the four analyzed sectors, following the same color scheme. The profiles are normalized to the maximum value in the sector.}
  \label{fig:ggm_profiles}
\end{figure*}

\begin{figure*}
  \centering
  \includegraphics[width=\hsize,trim={0cm 0cm 0cm 0cm},clip,valign=c]{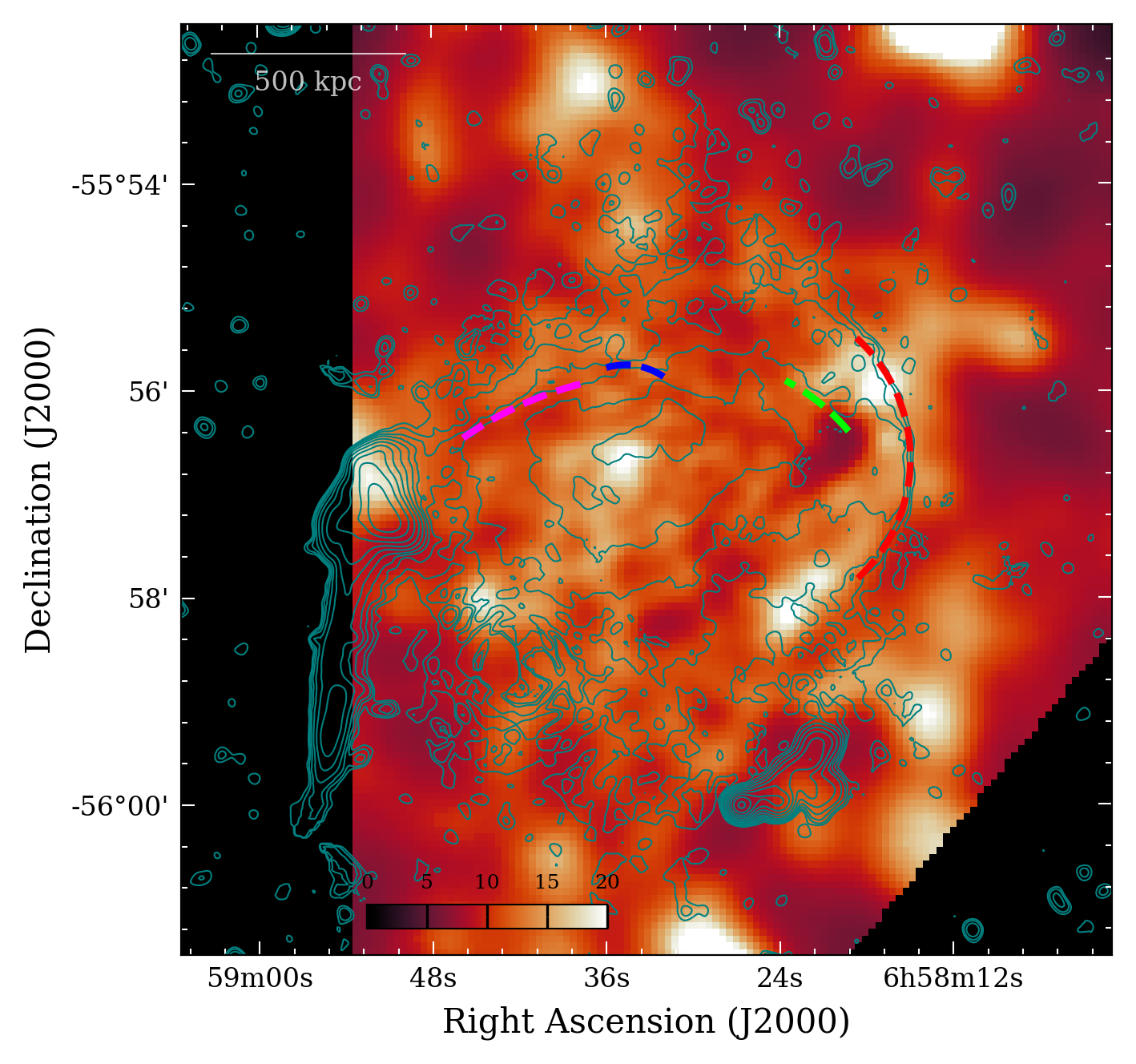}
  \caption{Temperature map in keV units of the Bullet Cluster derived from the \chandra\ data following the method described in \citet{markevitch00} and \citet{wang16a520}. Contours show the radio emission from the \meerkat\ discrete source-subtracted image spaced by a factor of 2 starting from 3 $\times$ rms. Dashed arcs mark the positions of the radio surface brightness discontinuities. The corresponding error map is reported in Appendix.~\ref{app:bullet_filters}.}
  \label{fig:bullet_kt}
\end{figure*}

\section{Edge-detection techniques}\label{sec:techniques}

Surface brightness discontinuities are regions characterized by sharp and significant brightness changes, and therefore gradient-filtering techniques are particularly useful in helping to identify them. Among the numerous edge-detection methods developed in digital image processing \citep[e.g.,][]{gonzalez02}, the most commonly used are summarized below. \\
\indent
The \textit{Prewitt/Roberts/Sobel} operators are based on first-order derivatives \citep{prewitt70, roberts63, sobel68}. They are discrete differentiation operators and differ in the approximation made to compute the gradient. The results of these filters are very similar. \\
\indent
The \textit{Gaussian Gradient Magnitude} (GGM) filter uses Gaussian derivatives to compute the gradient. By changing the width of the Gaussian, it is possible to search for gradients over different scales. \\
\indent
The \textit{Canny edge detection} is multi-step algorithm \citep{canny86}. Firstly, the image is smoothed with a Gaussian to reduce noise. Secondly, a first-order derivative operator (generally Sobel) is used to find gradients. Thirdly, local maxima are searched for in the gradient image in order to thin out the edges (this step is called nonmaximum suppression). Finally, edge points between two thresholds are tracked by hysteresis and are then connected.\\
\indent
The \textit{Laplace} operator is based on second-order derivatives \citep{marr80}. Contrary to the techniques described above, this is a zero-crossing-based method (namely, edges are identified when the second derivative takes values of zero). The input image is often smoothed with a Gaussian before applying this filter to reduce noise. \\
\indent
The \textit{unsharp mask} is obtained by subtracting an image smoothed with a wide Gaussian from the same image smoothed with a narrower Gaussian. This is a band-pass filter that is sensitive to spatial scales between the two adopted Gaussian widths. \\
\indent
The usage of a certain edge-detection method depends on the specific application and the characteristics of the image. Among these methods, the GGM filter is known to be robust to noise, to produce thin edges, and to be computationally efficient. Thus, after a visual comparison with the results provided by the other algorithms (see bottom panels in Fig.~\ref{fig:bullet_meerkat} and Appendix~\ref{app:bullet_filters}), we elected the GGM filter as the reference technique to aid our search for surface brightness discontinuities in radio halos. The advantage of this filter in applications to astronomical images (from \chandra\ and \xmm) has already  been pointed out in the literature \citep{sanders16ggm, walker16}, and consists in its flexibility when measuring gradients over more or fewer pixels depending on the scale of the features of interest and image S/N. Indeed, the GGM filter has been used intensively in X-ray images of galaxy clusters to find surface brightness edges associated with shocks and cold fronts \citep[e.g.,][]{sanders16centaurus, sanders22, walker17, walker18split, botteon18edges, douglass18, schellenberger19}. To the best of our knowledge, it has not been applied yet to radio images of cluster diffuse synchrotron sources\footnote{But the Sobel filter has been already used in the past in the study of radio galaxies \citep[e.g.,][]{murgia01, laing11, horellou18, ramatsoku20, condon21, rudnick22}.}. \\
\indent
The GGM-filtered images improve the clarity of surface brightness gradients whose presence would otherwise be assessed from unfiltered images using different contrasts and color scales in the region under investigation \citep{sanders16ggm}. We therefore selected the clearest features (namely, those following a coherent arc-shaped structure extending for $\gtrsim$100 kpc in length; see e.g.,\ \citealt{botteon18edges}) in the filtered images and extracted surface brightness profiles across the identified edges from the discrete source-subtracted \meerkat\ images using a version of \pyproffit\ \citep{eckert20} that was modified to handle radio images with corresponding headers and units.

\section{Results}\label{sec:results}

\subsection{The case of the Bullet Cluster}

The Bullet Cluster (or 1E 0657$-$55.8) is one of the most massive ($\mfive \simeq 1.3 \times 10^{15}$ \msun; \citealt{planck16xxvii}) and hottest ($kT \simeq 17$ keV; \citealt{tucker98}) galaxy clusters observed and hosts the most luminous radio halo known \citep{liang00bullet, shimwell14, shimwell15, sikhosana23}. Due to its brightness, the diffuse radio emission from the ICM is detected at high S/N and is highly resolved in the \meerkat\ image (Fig.~\ref{fig:bullet_meerkat}, top panels), making it possible to search for subtle and small-scale morphological features. Similarly, the X-ray emission of the ICM is observed at high S/N thanks to the presence of deep \chandra\ observations (Fig.~\ref{fig:bullet_chandra}, top panels), which enable a direct comparison between the thermal and nonthermal emission of the system. Due to these high-quality detections, the Bullet Cluster is the obvious choice of system with which to begin our search for radio surface brightness discontinuities in radio halos. 

\subsection{Radio surface brightness discontinuities}

The \meerkat\ image of the Bullet Cluster and the corresponding GGM-filtered images of Fig.~\ref{fig:bullet_meerkat} provide evidence for clear surface brightness discontinuities in the radio halo emission. To the west of the cluster, a strong and arc-shaped gradient marks the boundary of the radio halo. This edge is coincident with the shock front detected in the X-rays in the cluster \citep{markevitch02bullet, markevitch06}, where both radio and X-ray emissions abruptly drop. Additional radio gradients are observed in multiple directions; for example at the location of the cold front outlining the shoulder of the bullet following the western shock \citep{keshet21} and in the northeast region of the cluster. A number of X-ray surface brightness gradients are also present in the \chandra\ GGM-filtered images (Fig.~\ref{fig:bullet_chandra}, bottom panels) \\
\indent
For a better visualization of the discontinuities, in Fig.~\ref{fig:bullet_sb} we report the radio surface brightness profiles extracted from the discrete source-subtracted image across the most prominent features highlighted by the GGM-filtered images. These profiles show rapid surface brightness declines over small scales. In the case of the western edge, which marks the strongest gradient, the radio surface brightness drops by a factor of $\sim$6$-$7 within $\sim$50 kpc (corresponding to $\sim$1.5 resolution element). We note that all the profiles extracted during the analysis were obtained from sectors (whose radial extent and aperture are denoted by the colored segments in the top left panel of Fig.~\ref{fig:bullet_sb}) with a linear binning of 1\arcsec, which is smaller than the resolution of the radio image, meaning that edges are convolved with the beam size ($7.9\arcsec\times7.6\arcsec$) and thus in reality could be even sharper. The modeling of the profiles is beyond the scope of this paper, and will be the focus of a subsequent study.

\subsection{Comparison with the ICM thermal emission}

The shape of the radio surface brightness profiles shown in Fig.~\ref{fig:bullet_sb} is similar to that typically observed in the X-rays for shocks and cold fronts in the ICM \citep[e.g.,][]{markevitch07rev}. It is therefore natural to check whether the radio edges are co-located with X-rays edges. In Fig.~\ref{fig:bullet_comparison} we overlay the position of the radio surface brightness discontinuities onto the \meerkat\ and \chandra\ images and corresponding GGM-filtered images of the cluster, while in Fig.~\ref{fig:ggm_profiles} (left panel) we report an overlay between the radio and X-ray GGM-filtered images. These were obtained by smoothing the \chandra\ image to the resolution of the \meerkat\ image, regridding them to the same pixelation, and applying the GGM filter with $\sigma = 1$ pixel. The correspondence between radio and X-rays edges is striking. To better visualize the correlation between the gradients in the two images, in the right panels of Fig.~\ref{fig:ggm_profiles} we show the profiles extracted from the GGM-filtered images across the four radio discontinuities. Except for the western edge (shown in red in Fig. 5), the profiles are similar and, most importantly, they peak at the same position; that is, the radio and X-ray gradients are correlated. The western radio edge, which is associated with the famous shock front, appears different in radio and X-rays because the X-ray GGM profile across this discontinuity is contaminated at small radii by the strong gradient due to the cold front of the bullet subcluster. Nonetheless, we note that the X-ray GGM profile (which is normalized to its peak value) shows a bump at the radial distance at which we observe the maximum of the radio GGM profile. This bump is due to the strong shock front detected in X-rays by \citet{markevitch02bullet}, and therefore also in this case there is a coincidence between radio and X-ray gradients. \\
\indent
As anticipated, the western discontinuity (shown in red) traces the region of the radio halo that is bounded by the large bow shock \citep{shimwell14}, while the edge downstream (shown in green) is coincident with the cold front surrounding the bullet head \citep{markevitch02bullet}. The northeast radio discontinuity (shown in magenta) is prominent in the \meerkat\ image and profile and is mirrored by an X-ray surface brightness gradient, which is observed more clearly in the \chandra\ GGM-filtered image. The presence of this X-ray edge has not been reported until now and we speculate it traces a cold front given that the thermal gas appears colder in its denser region ($\sim$10 keV, compared to $\sim$12$-$13 keV of the outer region), as inferred from the temperature map of Fig.~\ref{fig:bullet_kt}. The radio edge in the central region of the cluster (blue color) is associated with an X-ray gradient that is probably due to a large Kelvin-Helmholtz roll, where we expect amplified magnetic field due to shear velocity, which may explain the edge in the synchrotron emission. Both the \meerkat\ and \chandra\ GGM-filtered images show many prominent surface brightness structures in the middle of the cluster (Figs.~\ref{fig:bullet_comparison} and \ref{fig:ggm_profiles}). This central region likely traces the remnant of the bigger subcluster that has been disturbed by the high-speed passage of the bullet subcluster \citep{clowe06}.

\subsection{Search for radio edges in the other \mgcls\ radio halos}

The Bullet Cluster is one of the 26 galaxy clusters reported to host a radio halo in the \mgcls\ \citep{knowles22}. We expanded our search for radio surface brightness discontinuities in radio halos by applying the same techniques adopted for the Bullet Cluster to the remaining 25 galaxy clusters in the sample (Table~\ref{tab:sample}). Below we summarize the main results of our analysis. In Appendixes~\ref{app:yes_edges} and \ref{app:no_edges}, we report the images and radio surface brightness profiles for the clusters in which the GGM-filtered images highlight the presence or absence of surface brightness gradients in the radio halo emission, respectively. \\
\indent
In addition to the case of the Bullet Cluster, we find that at least one radio surface brightness gradient is present in another 12 radio halos. That is, $50\%$ of the radio halos in the \mgcls\ may show edges. These additional clusters are: Abell 85, Abell 209, Abell 521, Abell 2744, Abell 3562, MACS J0417.5-1155, RXC J1314.4-2525, J0225.9-4154, J0232.2-4420, J0638.7-5358, J0645.4-5413, and J2023.4-5535 (see Appendix~\ref{app:yes_edges}). Multiple gradients are noted in 4 out of these 12 clusters. Many of the gradients observed show surface brightness profiles similar to that of the Bullet Cluster, while others are more subtle (\ie,\ the surface brightness jump is shallower and/or the profile shows fluctuations), which make any firm claim of an edge less obvious. Surface brightness jumps at the boundary of halos detected with low S/N should be taken with particular care, as in these noisy regions the extended radio emission may not be accurately reconstructed by the \texttt{CLEAN} algorithm. Overall, we extracted radio surface brightness profiles across 16 sectors (20, if we include those in the Bullet Cluster). We anticipate that, in general, the radio halos in which we do not observe surface brightness gradients are those detected with lower S/N in the \meerkat\ images (see Appendix~\ref{app:no_edges}). We return to this point at the end of Section~\ref{sec:discussion}.

\section{Discussion}\label{sec:discussion}

\begin{figure}
  \centering
  \includegraphics[width=\hsize,trim={0cm 0cm 0cm 0cm},clip,valign=c]{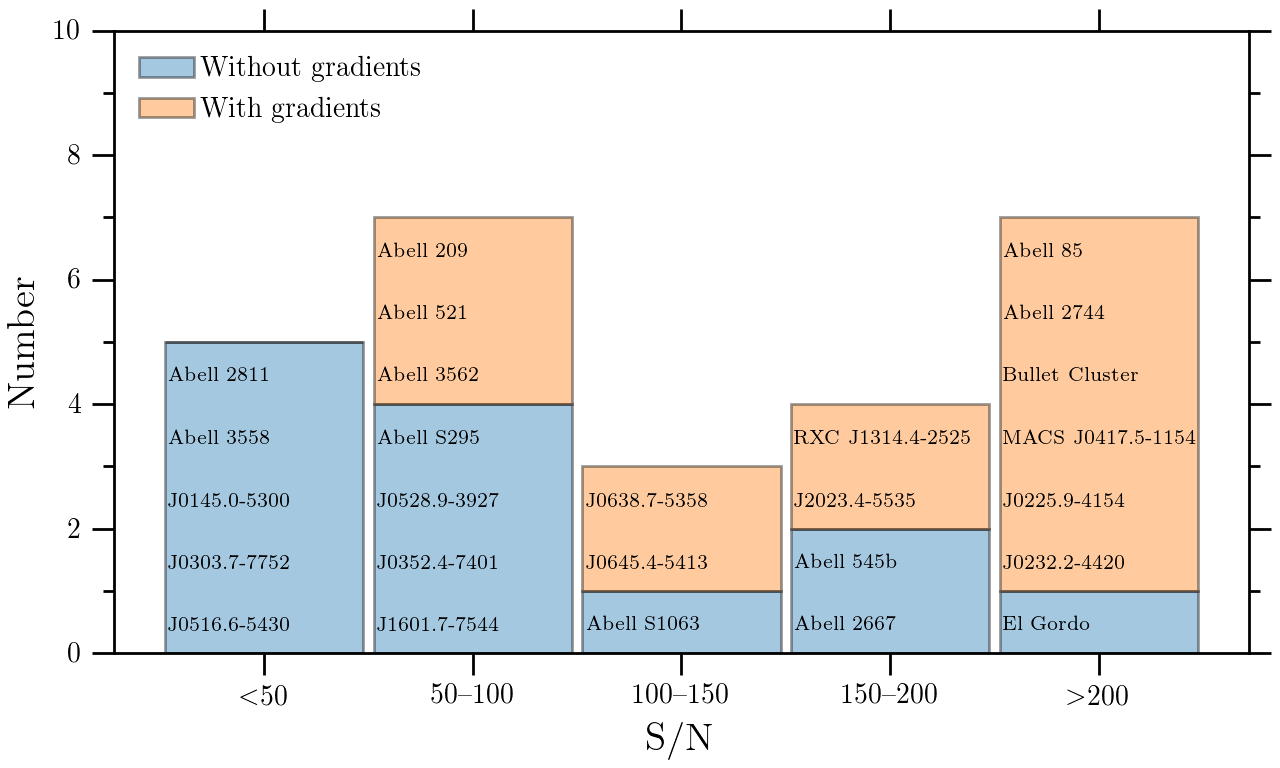}  \caption{Distribution of the S/N for the radio halos in the \mgcls\ grouped according to the presence or absence of radio surface brightness gradients.}
  \label{fig:snr_distribution}
\end{figure}

Surface brightness edges in the cluster X-ray emission have now been observed in numerous systems 
\citep[e.g.,][for collections]{markevitch07rev, owers09sample, ghizzardi10, markevitch10arx, walker16, botteon18edges}. These edges trace density contrasts associated with shocks and cold fronts generated during cluster mergers, sloshing motions, or AGN outbursts. In contrast, only a handful of discontinuities in radio halos have been observed to date. The most striking cases are: Abell 520 \citep{markevitch05, vacca14, wang18a520, hoang19a520}, Coma Cluster \citep{brown11coma, bonafede22}, Bullet Cluster \citep{shimwell14, sikhosana23}, and Toothbrush Cluster \citep{vanweeren16toothbrush, rajpurohit18}. In these clusters, the radio halo emission is bounded by a front that is coincident with an underlying X-ray-confirmed shock \citep[see also][]{markevitch10arx}. Thanks to the high quality of the new \meerkat\ data and the deep archival \chandra\ observations, the Bullet Cluster can be considered a textbook example of a radio halo--shock connection. \\
\indent
In addition to the well-known edge, the \meerkat\ image and corresponding GGM-filtered images of the Bullet Cluster allowed us to pinpoint additional surface brightness discontinuities in the radio halo emission. These gradients are also detected in the inner regions of the cluster, and are associated with discontinuities in the X-ray emission. Clear surface brightness jumps in the diffuse radio emission were found in other galaxy clusters of the studied sample, indicating that the Bullet Cluster is not unique in showing this kind of feature. Although not all the clusters in the sample have been targeted with (deep) X-ray observations, we note that some of the radio edges found are co-located with shocks and cold fronts claimed in previous studies, such as those in Abell 85 \citep{ichinohe15, rahaman22a85}, Abell 2744 \citep{owers11a2744, jauzac16}, MACS J0417.5-1155 \citep{botteon18edges, pandge19macsj0417}, J0225.9-4154 \citep{chon19}, and J0638.7-5358 \citep{botteon18edges}. These results suggest that the dynamical motions in the ICM induced by the ongoing merger activity shape the morphology of both the thermal and nonthermal cluster emissions and that these components are tightly related. This is probably due to the fact that the magnetic field is largely frozen into the thermal plasma and is advected, with gas flows in the ICM dragging the enclosed cosmic rays. \\ %
\indent
In Fig.~\ref{fig:snr_distribution} we grouped the clusters in the sample according to the S/N of their radio halo and to the presence of radio surface brightness gradients. The S/N was evaluated as $S/(\rm{rms}\sqrt{N_{\rm b}})$, where $S$ is the flux density of the radio halo and $N_{\rm b}$ is the number of beams covered by the diffuse emission. This plot highlights that surface brightness gradients in radio halos are preferentially found in the clusters detected at higher S/N in the \meerkat\ images. For example, the five clusters in which we detect multiple radio gradients (Bullet Cluster, Abell 2744, MACS J0417.5-1155, RXC J1314.4-2525, and J0225.9-4154) are among the cases where the radio halo emission is detected at higher S/N, while the 13 clusters where we do not find any radio gradient have generally lower S/N. This could indicate observational bias, in which radio surface brightness discontinuities may be common but may only be detected when the data used have a sufficient S/N.

\subsection{Implications}

Whether or not all radio edges trace shocks and cold fronts in the ICM needs to be confirmed in the future. With the current analysis, we can state that at least some of them are co-located with X-ray discontinuities. Radio edges without an X-ray counterpart may trace specific regions in the ICM that are abundant in seed relativistic electrons or where the acceleration efficiency or magnetic field strength are locally enhanced. For the Bullet Cluster, we pointed out that the northeast radio surface brightness jump is possibly associated with a cold front that was not claimed in past studies. This provides a tantalizing indication that surface brightness gradients in radio halos could be helpful for the discovery of new shocks and cold fronts in the ICM. Moreover, comparisons of the typical duration of the observations in the \mgcls\ ($\sim$6$-$10~h) with the long exposures of the deep \chandra\ observations required to detect X-ray discontinuities in the ICM (\ie,\ $\gtrsim$50~ks $\approx$ 14~h, but more generally $>$100~ks) suggest new radio observations appear more efficient in this task. Radio data have already guided X-ray observations in the search for shock fronts in cluster outskirts associated with previously detected radio relics \citep[e.g.,][]{finoguenov10a3667, macario11, akamatsu12a3376, akamatsu13systematic, bourdin13, shimwell15, botteon16a115, botteon16gordo, eckert16a2744, urdampilleta18, urdampilleta21, digennaro19zwcl0008, ge19a1367}. Thanks to the high-resolution and sensitive imaging enabled by \meerkat, it now appears possible to expand the search for features related to the dynamics of the ICM to the  inner regions of clusters using radio halos. \\
\indent
High-resolution (\ie,\ $\lesssim$10\arcsec) imaging of extended radio sources was not possible until recently because of the requirement for very high surface brightness sensitivity at high angular resolution. Observing radio halos and relics with unprecedented resolution has the potential to transform our understanding of the physical processes leading to the formation of these extended sources. With the discovery of numerous radio edges, it appears that the historical definition of radio halos as ``smooth'' sources with ``regular'' morphology \citep[see][for the definition of radio halo]{feretti12rev, vanweeren19rev} is no longer  valid: our results indicate that radio halos are rich in substructure. \\
\indent
The azimuthal surface brightness profiles of radio halos is usually described with an exponential function \citep[e.g.,][]{murgia09, boxelaar21}. The reasons for the choice of this profile shape are twofold: it is a simple function with only two free parameters, and it provides a reasonable description of many radio halos (despite not being physically motivated). It is clear that once a high S/N is achieved, substructures in the diffuse radio emission become more evident and this model provides poorer fitting statistics. This was recently pointed out by \citet{botteon22dr2}, who fitted 73 radio halos observed with \lofar\ finding that those detected at higher S/N are those that show larger deviations from the exponential profile (see Appendix~\ref{app:bullet_filters} for the residuals obtained for the Bullet Cluster). This is similar to previous findings for X-rays, where the $\beta$-model \citep{cavaliere76} was customarily used to fit the X-ray surface brightness profiles of clusters with low-resolution satellites (e.g.,\ \einstein, \asca, \rosat), but with the advent of higher resolution instruments (such as \chandra\ and \xmm), numerous substructures due to shocks, cold fronts, and cavities became evident, indicating that cluster atmospheres are very dynamic and that their profiles cannot be properly described by a smooth model. The presence of edges in radio halos will naturally lead to large deviations if the azimuthal profiles of these sources are fitted with an exponential model.

\subsection{Prospects}

Here we present the first identification of surface brightness gradients and the extraction of the corresponding profiles in a sample of radio halos. The natural follow-up to this work is the modeling of the profiles and the simultaneous analysis of the thermal and nonthermal emissions. In X-rays, a projection of a 3D spherical discontinuity is generally adopted to describe the underlying density profile for the surface brightness jumps associated with shocks and cold fronts. This model is motivated by the fact that these edges mark contrasts in the thermal electron density $n_{\rm e}$ and that the X-ray emissivity in the hot ($\gtrsim$2.5 keV) ICM is $j_{\rm x} \propto n_{\rm e}^2 (kT)^{1/2}$, which is very weakly dependent on the gas temperature in the soft X-ray band \citep[e.g.,][]{ettori00beta}. In radio, the synchrotron emissivity at frequency $\nu$ for a simple power-law spectrum is $j_{\rm r} \propto n_{\rm CRe} B^{(\delta+1)/2} \nu^{-(\delta-1)/2}$, and therefore depends on the magnetic field strength $B$, the density of the emitting nonthermal electrons $n_{\rm CRe}$, and the slope of the electron energy distribution $\delta$ (which is related to the observed synchrotron spectrum via $\delta = 2\alpha+1$). Therefore, determining what causes the jump is not trivial, because of the degeneracy between $n_{\rm CRe}$ and $B$. Additional effects related to the properties of the magnetic field and processes taking place in the ICM (e.g.,\ stretching, compression, and amplification) have an impact on the emissivity, radiative losses, and reacceleration rate of nonthermal electrons, making the situation even more complex \citep[e.g.,][]{brunetti20, chibueze21, rudnick22}. In addition, the modeling of the extracted profiles must account for the beam size, which artificially smears the surface brightness discontinuities (assuming that edges are sharper than the resolution of the radio images). \\
\indent
Although the count statistics of the available X-ray observations for the clusters in the studied sample is generally lower than that for the Bullet Cluster, previous studies have already reported the presence of shocks and cold fronts at the positions of several of the radio edges we discovered in this study (see discussion above). The majority of the radio discontinuities observed currently lack a counterpart in X-rays and our results can aid future studies to search for the associated edges in the thermal gas distribution in the archival data or eventually to propose targeted follow-up observations.

\section{Conclusions}\label{sec:conclusions}

We used a recently published \meerkat\ image at 1.28 GHz of the Bullet Cluster from the \mgcls\ to detect four surface brightness discontinuities in the radio halo emission. The presence of these edges was highlighted thanks to the GGM filter, which has proven to be helpful in the identification of these features in the diffuse radio emission (previously, this filter was applied to cluster X-ray images to facilitate the detection of shocks and cold fronts). The surface brightness profiles extracted across the radio gradients are very similar to those obtained in X-rays for cluster shocks and cold fronts reported in the literature. The four radio edges observed in the Bullet Cluster are coincident with X-ray gradients. This suggests that the magnetic field frozen into the ICM mediates the coupling between nonthermal and thermal particles and that the morphology of the radio and X-ray cluster emissions is shaped by the dynamical motions of the gas during the merger. \\
\indent
By applying the same techniques adopted for the Bullet Cluster to the other 25 radio halos in the \mgcls, we find that half of the clusters in the sample show at least one radio surface brightness gradient. The fact that edges are more frequent and more clear in halos detected at higher S/N (such as the Bullet Cluster and Abell 2744) indicates that high-quality images are necessary to characterize the discontinuities, and also that these edges may be common in radio halos but their presence could be hindered when the radio emission is observed with low S/N. \\
\indent
In general, the results of this work imply that the emission of radio halos is not smooth, but rich in substructure arising from complex motions in the ICM. We are able to make this claim thanks to the high resolution and sensitivity provided by the \meerkat\ images. In the future, we expect that radio surface brightness discontinuities in radio halos will be routinely observed thanks to the new generation of instruments, which may provide a more efficient means to detect shocks and cold fronts in the ICM   than
X-ray observations.

\begin{acknowledgements}
We thank the anonymous referee for providing constructive comments to improve the presentation of the results. RJvW acknowledges support from the ERC Starting Grant ClusterWeb 804208. MGCLS data products were provided by the South African Radio Astronomy Observatory and the MGCLS team and were derived from observations with the MeerKAT radio telescope. The MeerKAT telescope is operated by the South African Radio Astronomy Observatory, which is a facility of the National Research Foundation, an agency of the Department of Science and Innovation.
This research made use of APLpy, an open-source plotting package for Python \citep{robitaille12}, and Astropy, a community-developed core Python package and an ecosystem of tools and resources for astronomy \citep{astropy13, astropy18}. Parts of the results in this work make use of the colormaps in the CMasher package \citep{vandervelden20}.
\end{acknowledgements}

\bibliographystyle{aa}
\bibliography{library.bib}

\begin{appendix}

\onecolumn

\section{Additional images of the Bullet Cluster}\label{app:bullet_filters}

\begin{figure}[h]
  \centering
  \includegraphics[width=\hsize,trim={0cm 0cm 0cm 0cm},clip,valign=c]{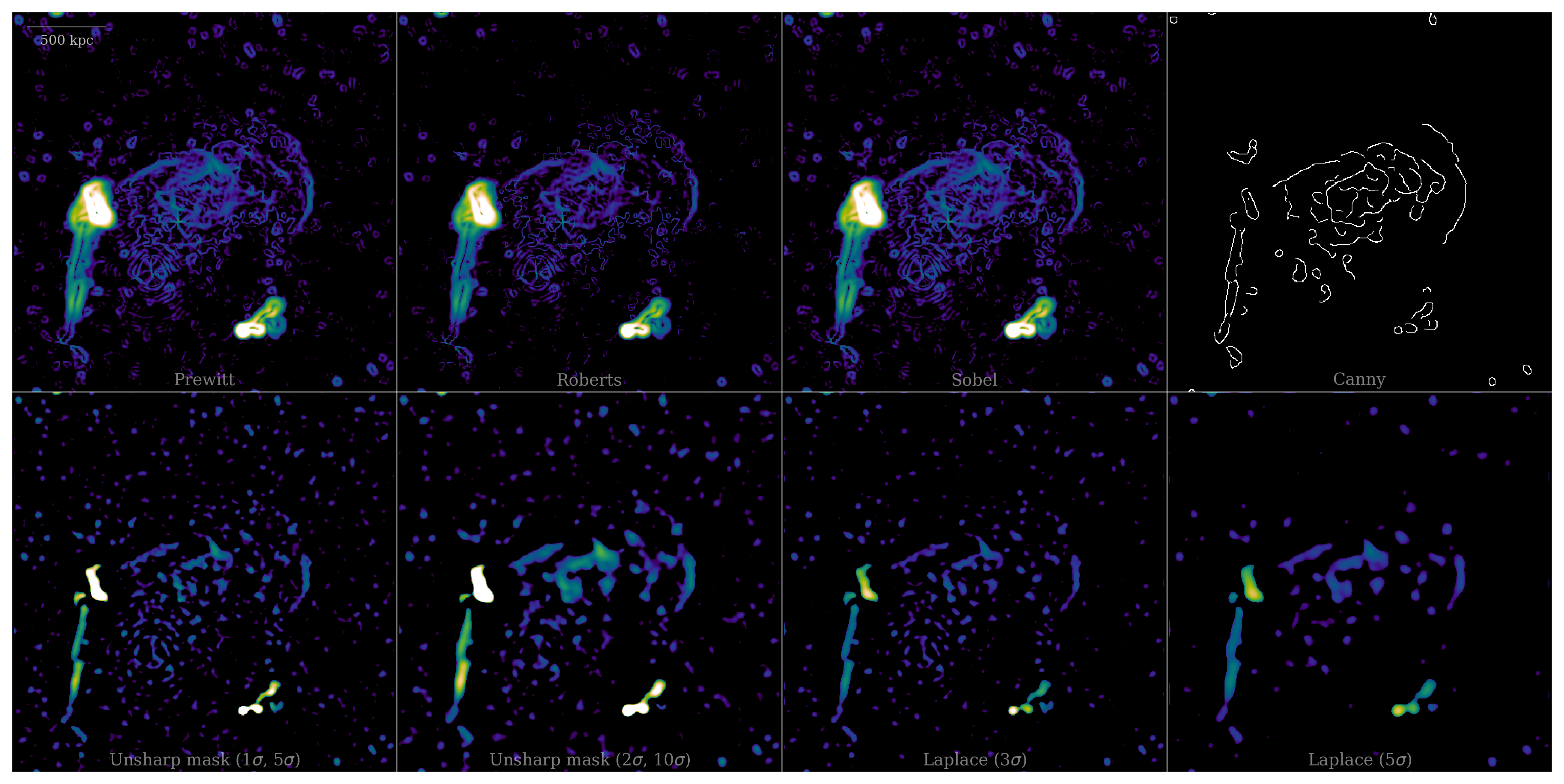}
  \caption{Different edge-detection algorithms (see labels) applied to the \meerkat\ image of the Bullet Cluster.}
  \label{fig:bullet_others}
\end{figure}

\begin{figure}
  \centering
  \includegraphics[width=.54\hsize,trim={0cm 0cm 0cm 0cm},clip,valign=c]{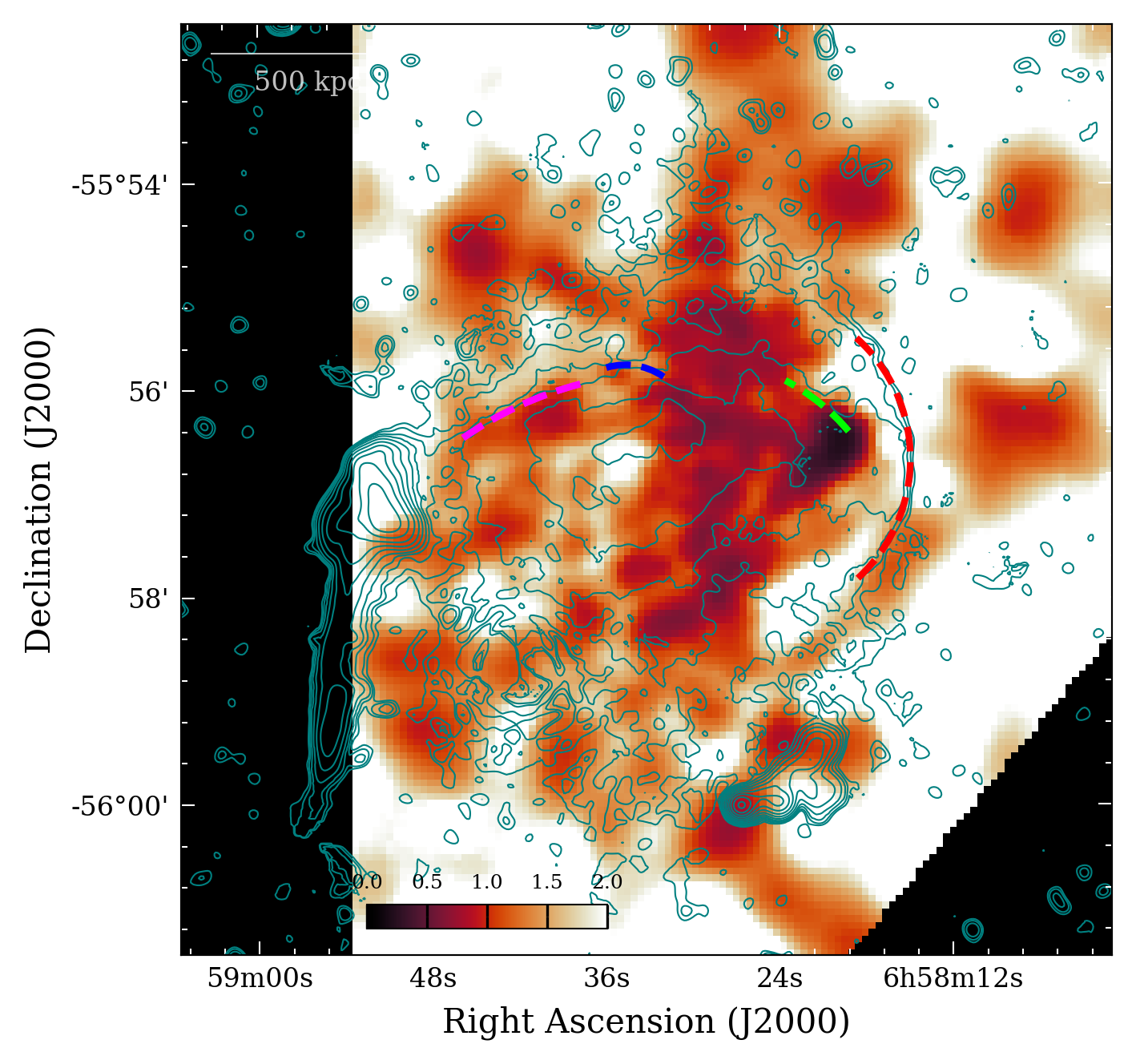}
  \caption{Temperature error map of the Bullet cluster (\cf\ with Fig.~\ref{fig:bullet_kt}).}
  \label{fig:bullet_kt_error}
\end{figure}

\begin{figure}
  \centering
  \includegraphics[width=\hsize,trim={0cm 0cm 0cm 0cm},clip,valign=c]{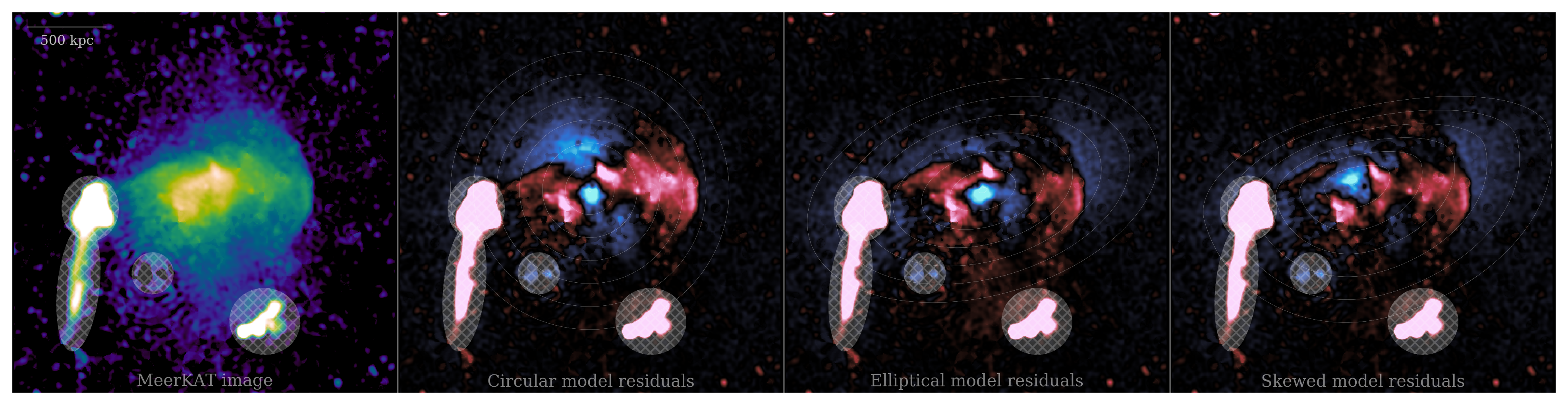}
  \caption{\meerkat\ image of the Bullet Cluster and residuals of the best-fit exponential models (red is positive, blue is negative). Isocontour levels of the corresponding model are reported. Hatched regions were masked during the fit with \halofdca.}
  \label{fig:bullet_residuals}
\end{figure}

In Fig.~\ref{fig:bullet_others} we collect the \meerkat\ images of the Bullet Cluster filtered with the edge-detection methods outlined in Section~\ref{sec:techniques}. \\
In Fig.~\ref{fig:bullet_kt_error} we show the temperature error map corresponding to Fig.~\ref{fig:bullet_kt}. \\
In Fig.~\ref{fig:bullet_residuals} we report the residuals obtained by subtracting the best-fit exponential models (circular, elliptical, and skewed) describing the surface brightness profile of the radio halo in the Bullet Cluster derived with the \halofdcaE\ \citep[\halofdca;][]{boxelaar21} from the \meerkat\ image. Deviations from these smooth models are due to substructures (such as edges) which are highlighted in the residual maps. 

\FloatBarrier

\section{Radio halos in MGCLS with radio surface brightness gradients}\label{app:yes_edges}

\paragraph{Abell 85} The surface brightness drops at the boundary of the radio halo emission. This edge is coincident with a cold front observed in the X-rays \citep{ichinohe15, rahaman22a85}. %

\paragraph{Abell 209} The surface brightness gradient is detected at the boundary of the diffuse emission.

\paragraph{Abell 521} A mild gradient located in a low surface brightness region is found at the northern border of the radio halo. Deeper data will help to determine the significance of this jump.

\paragraph{Abell 2744} Multiple radio gradients are identified in the radio halo. The southeastern edge (green color) is located in a region where a shock front has been detected with X-ray observations \citep{owers11a2744, jauzac16}.

\paragraph{Abell 3562} A possible gradient is observed in a region where the radio halo has low surface brightness; confirmation will require deeper observations.

\paragraph{MACS J0417.5-1155} Two radio edges co-located with two cold fronts observed in the X-rays \citep{botteon18edges, pandge19macsj0417} are detected.

\paragraph{RXC J1314.4-2525} Two gradients are observed at the boundary of the radio halo. 
 
\paragraph{J0225.9-4154} (Abell 3017) Two surface brightness gradients are identified. The southern jump is located in the direction where a cold front and a shock front have been detected in X-rays \citep{chon19}.

\paragraph{J0232.2-4420} (PSZ2 G259.98-63.43) A surface brightness gradient is present in the northern region of the radio halo. The subtraction of the bright AGN at the center of the diffuse emission introduced some artifacts in the image that were avoided during the extraction of the surface brightness profile.

\paragraph{J0638.7-5358} (Abell S592) A significant surface brightness jump is found in coincidence with an X-ray-detected shock front \citep{botteon18edges}.
 
\paragraph{J0645.4-5413} (Abell 3404) The gradient in the inner region of the cluster is mild and a better characterization will require deeper observations.

\paragraph{J2023.4-5535} (PSZ1 G342.33-34.92) The radio edge is located at the eastern boundary of the radio halo.

\begin{figure}
  \centering
  \includegraphics[width=\hsize,trim={0cm 0cm 0cm 0cm},clip,valign=c]{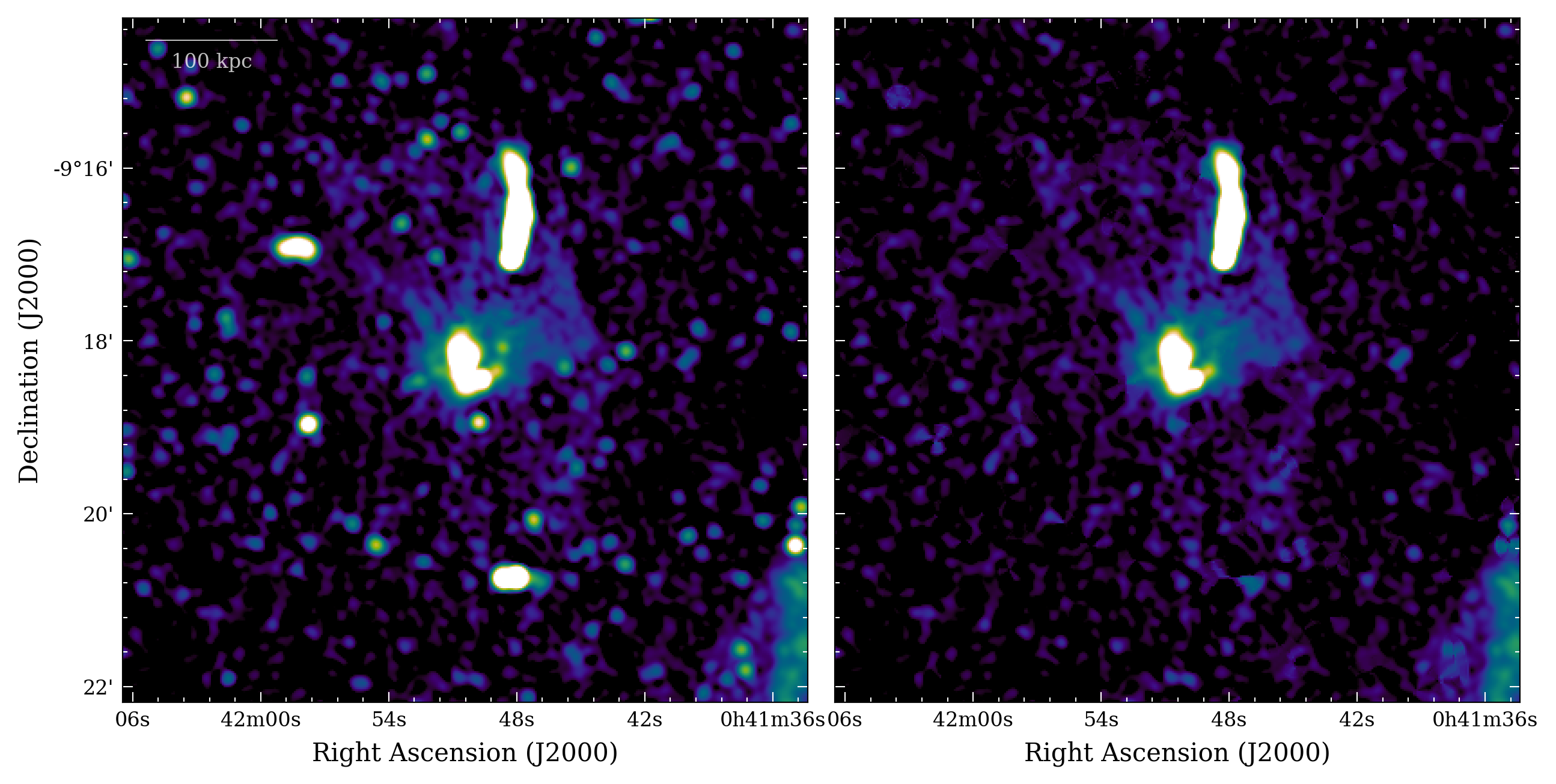}
  \includegraphics[width=\hsize,trim={0cm 0cm 0cm 0cm},clip,valign=c]{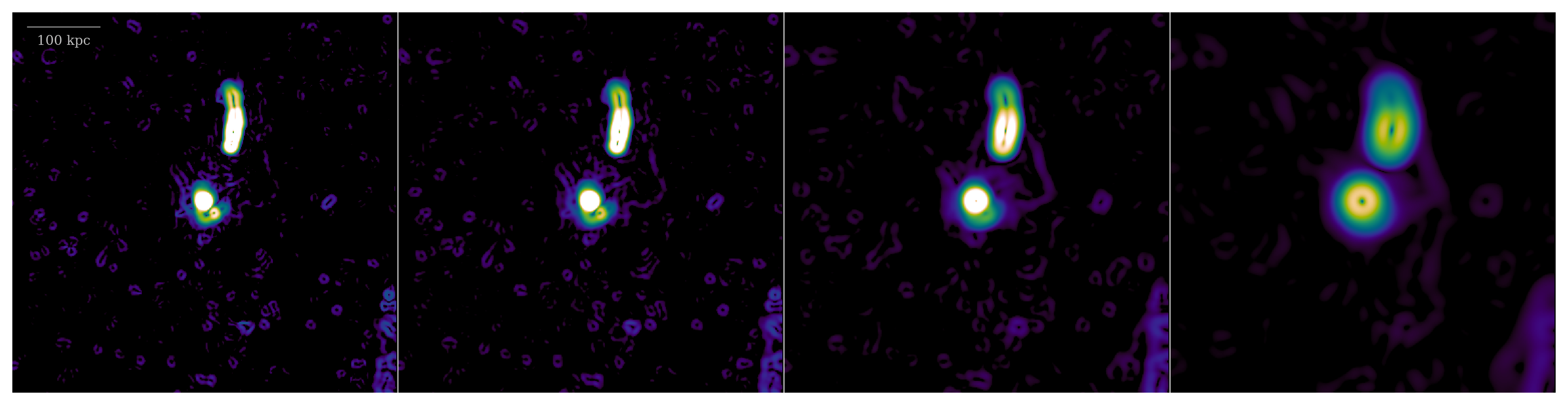}
  \caption{Same as Fig.~\ref{fig:bullet_meerkat} but for Abell 85.}
  \label{fig:Abell_85_meerkat}
\end{figure}

\begin{figure}
  \centering
  \includegraphics[width=.348\hsize,trim={0cm 0cm 0cm 0cm},clip,valign=c]{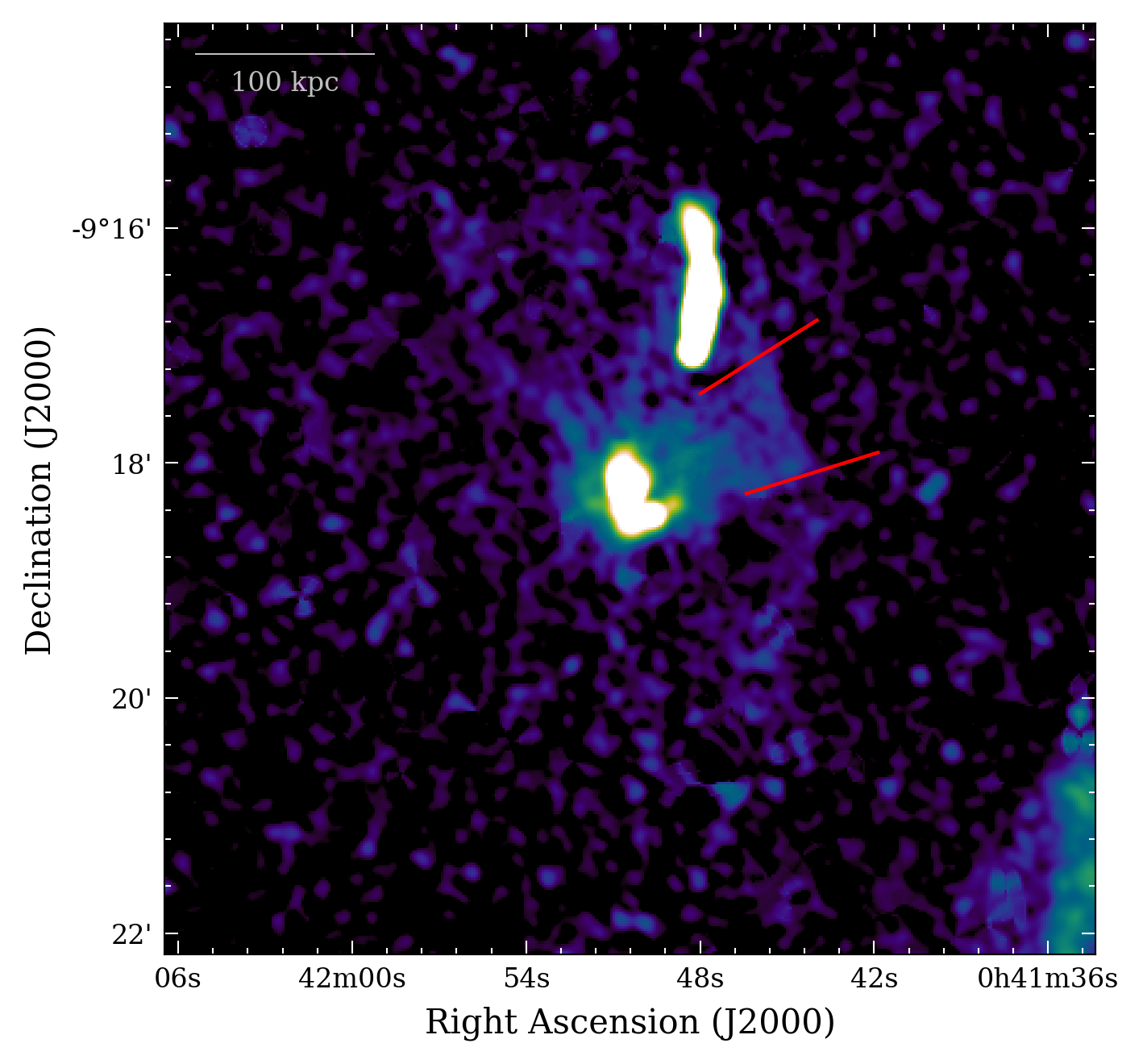}
  \includegraphics[width=.32\hsize,trim={0cm 0cm 0cm 0cm},clip,valign=c]{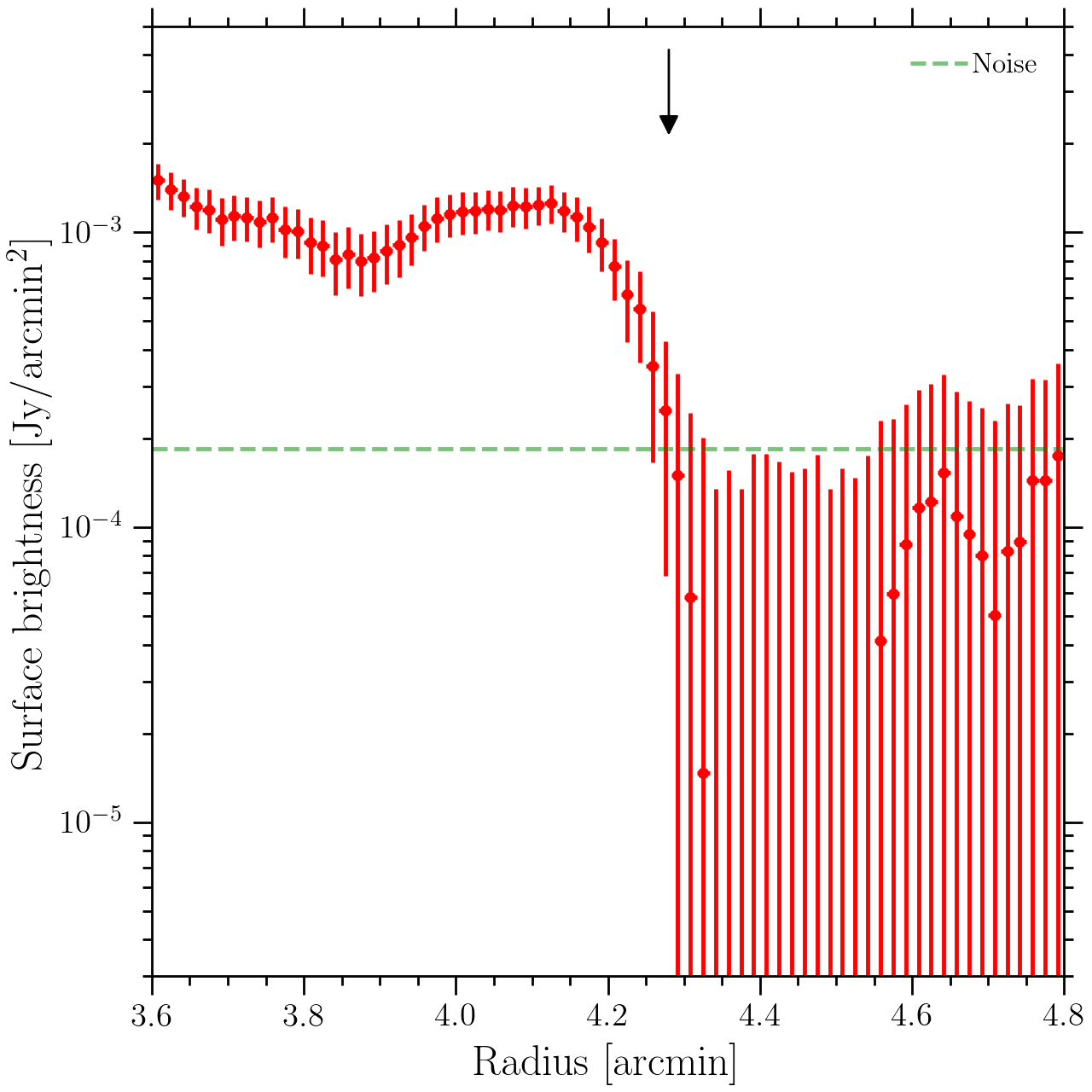}
  \caption{Same as Fig.~\ref{fig:bullet_sb} but for Abell 85.}
  \label{fig:Abell_85_sb}
\end{figure}

\begin{figure}
  \centering
  \includegraphics[width=\hsize,trim={0cm 0cm 0cm 0cm},clip,valign=c]{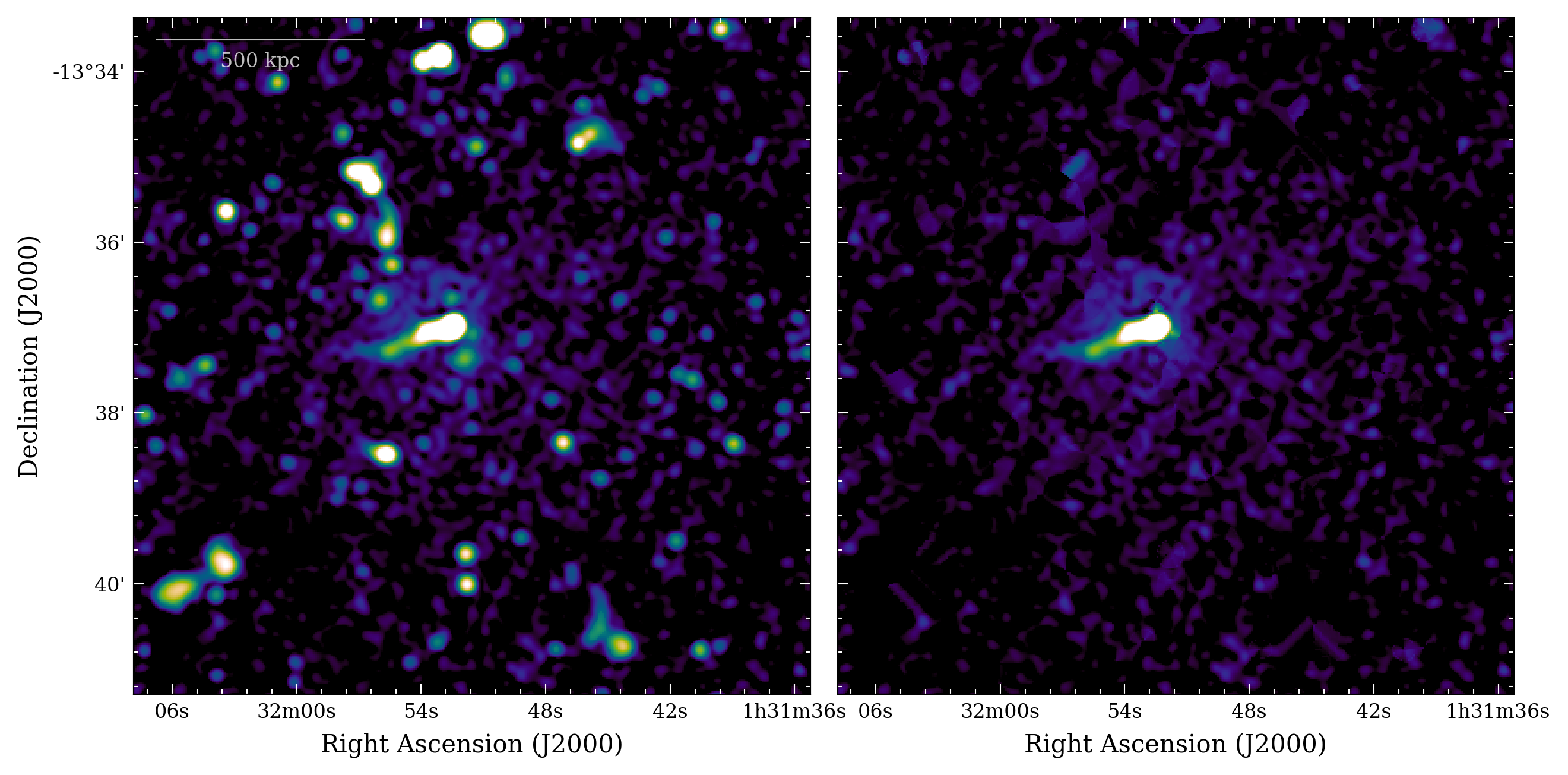}
  \includegraphics[width=\hsize,trim={0cm 0cm 0cm 0cm},clip,valign=c]{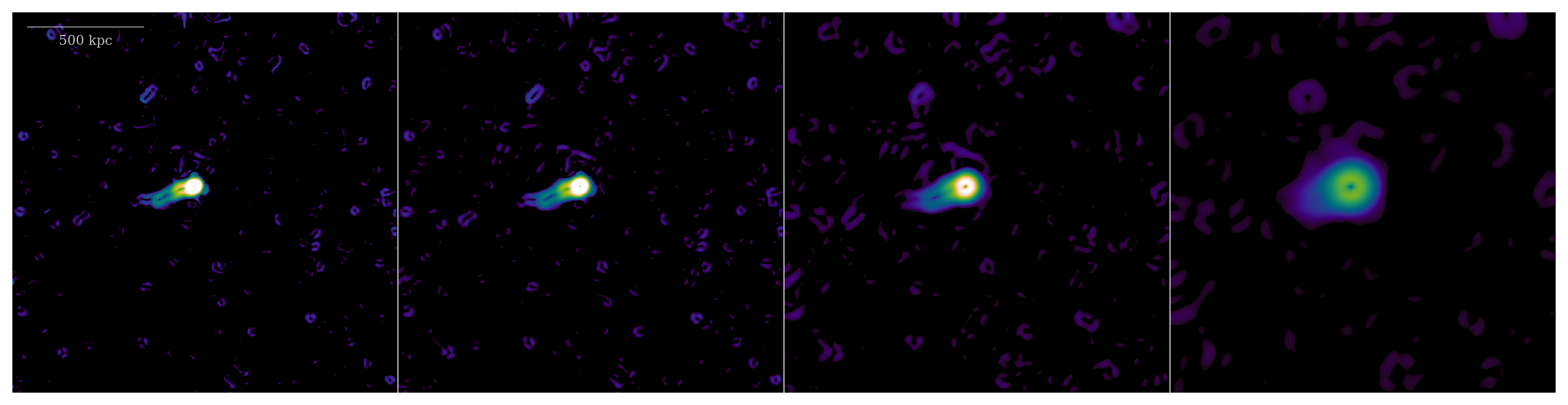}
  \caption{Same as Fig.~\ref{fig:bullet_meerkat} but for Abell 209.}
  \label{fig:Abell_209_meerkat}
\end{figure}

\begin{figure}
  \centering
  \includegraphics[width=.348\hsize,trim={0cm 0cm 0cm 0cm},clip,valign=c]{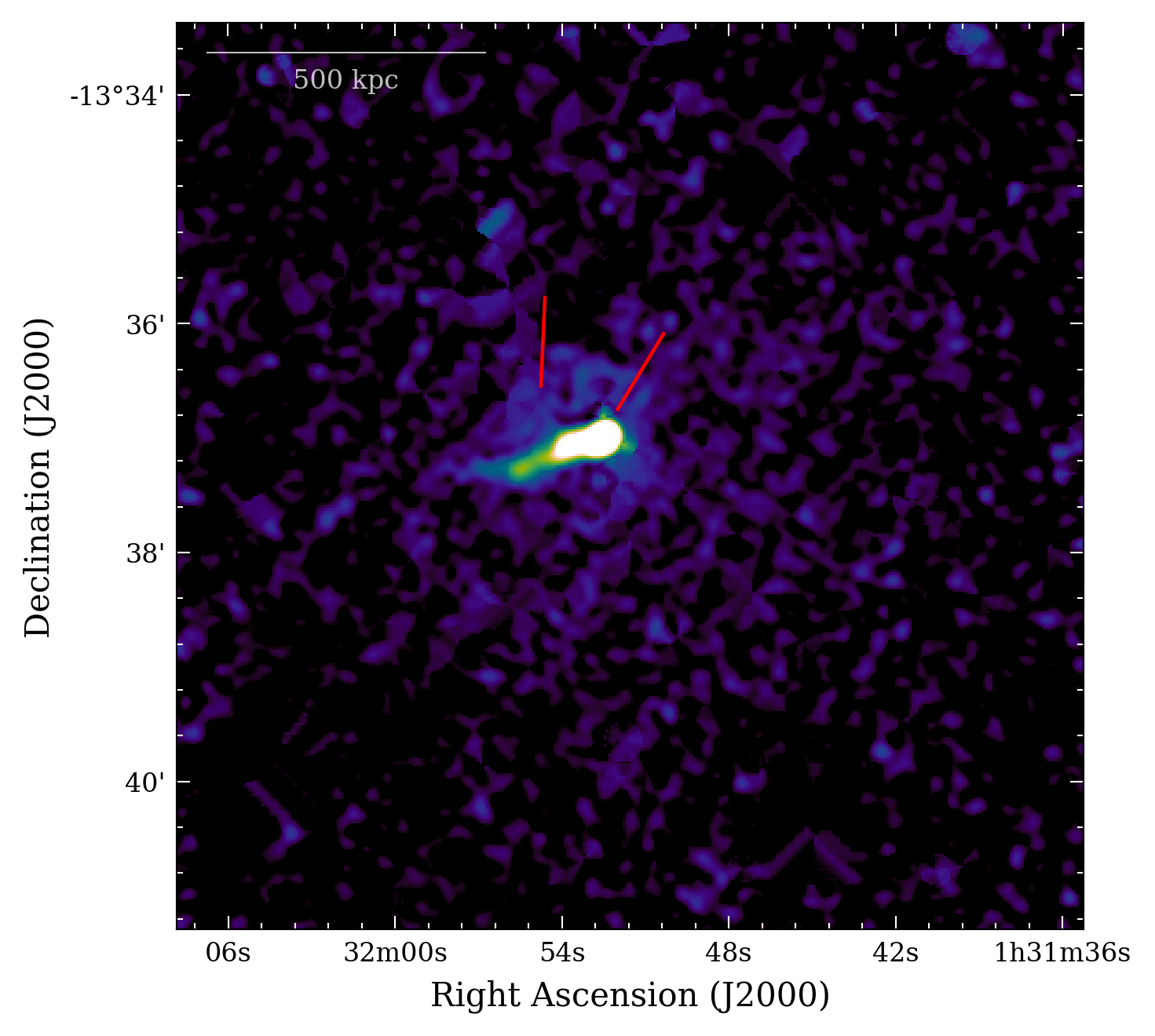}
  \includegraphics[width=.32\hsize,trim={0cm 0cm 0cm 0cm},clip,valign=c]{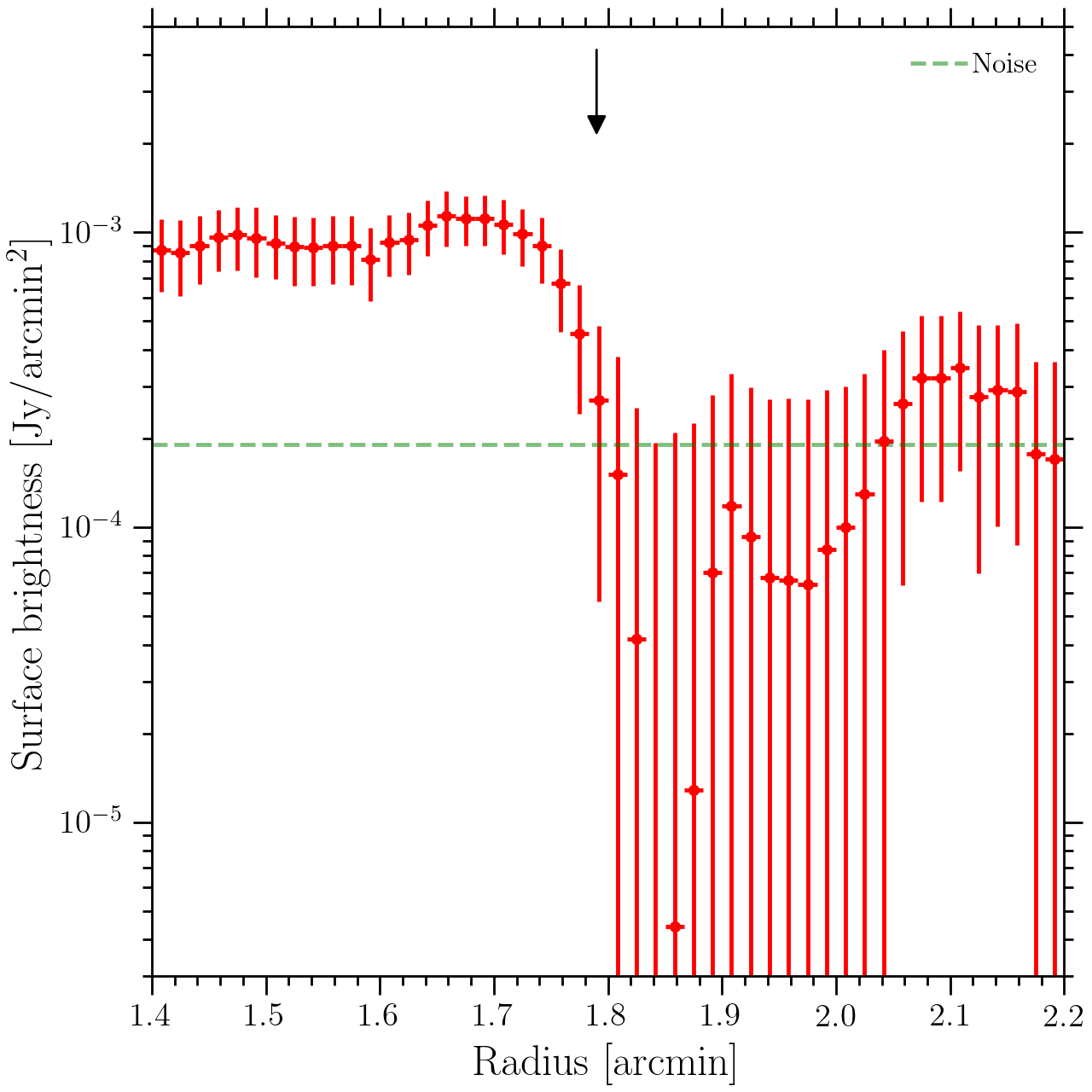}
  \caption{Same as Fig.~\ref{fig:bullet_sb} but for Abell 209.}
  \label{fig:Abell_209_sb}
\end{figure}

\begin{figure}
  \centering
  \includegraphics[width=\hsize,trim={0cm 0cm 0cm 0cm},clip,valign=c]{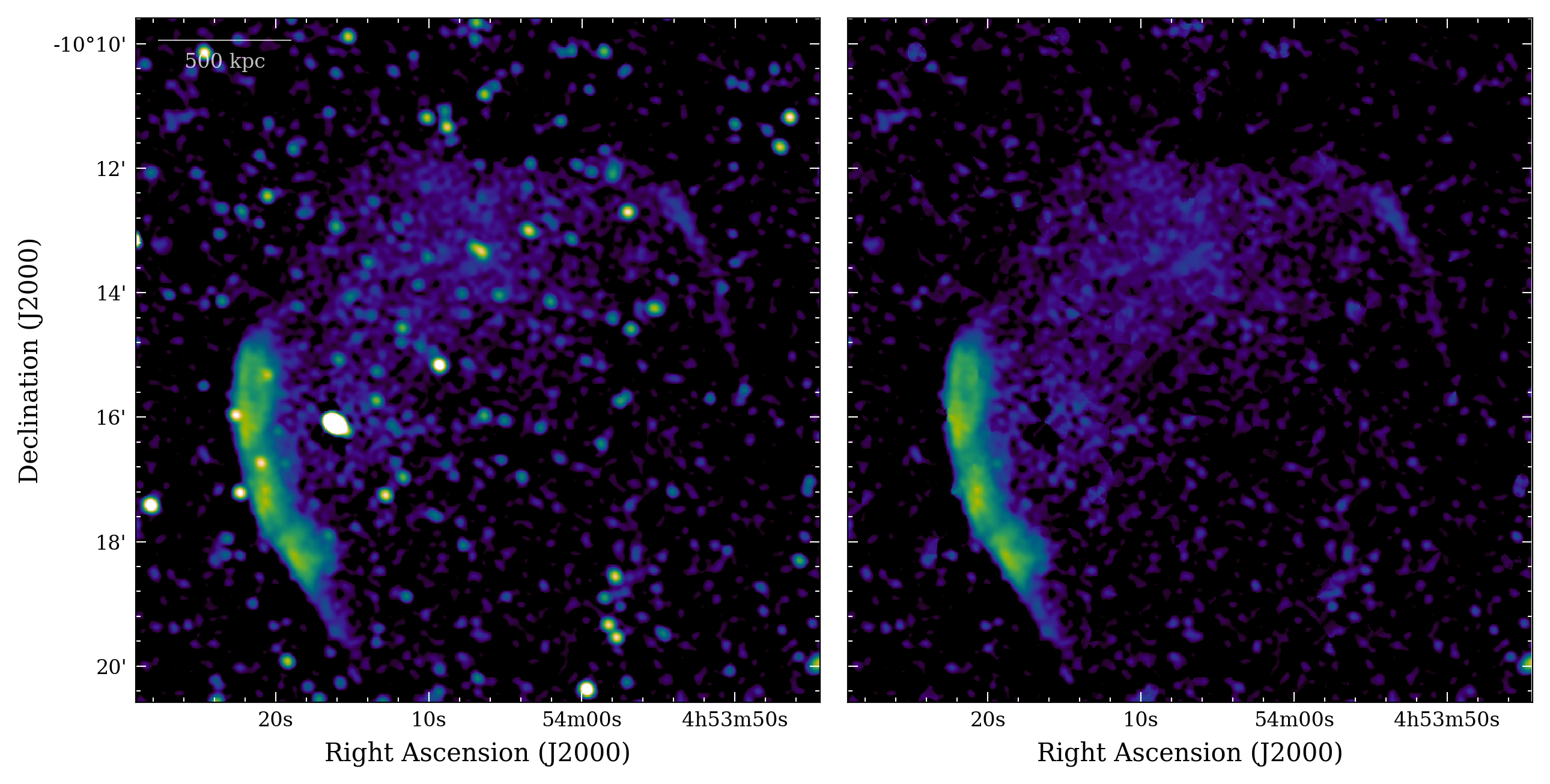}
  \includegraphics[width=\hsize,trim={0cm 0cm 0cm 0cm},clip,valign=c]{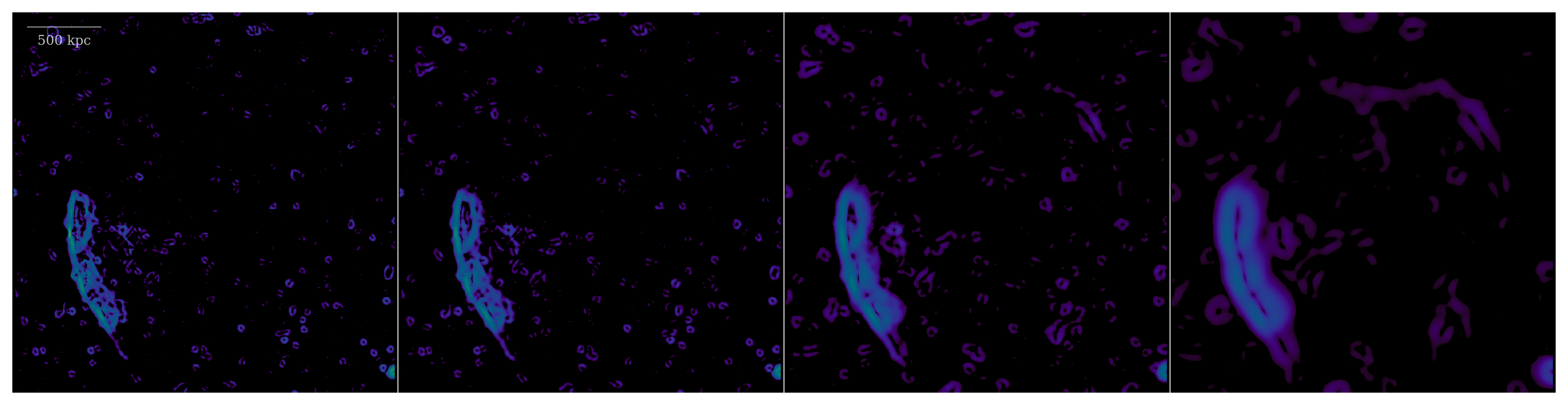}
  \caption{Same as Fig.~\ref{fig:bullet_meerkat} but for Abell 521.}
  \label{fig:Abell_521_meerkat}
\end{figure}

\begin{figure}
  \centering
  \includegraphics[width=.348\hsize,trim={0cm 0cm 0cm 0cm},clip,valign=c]{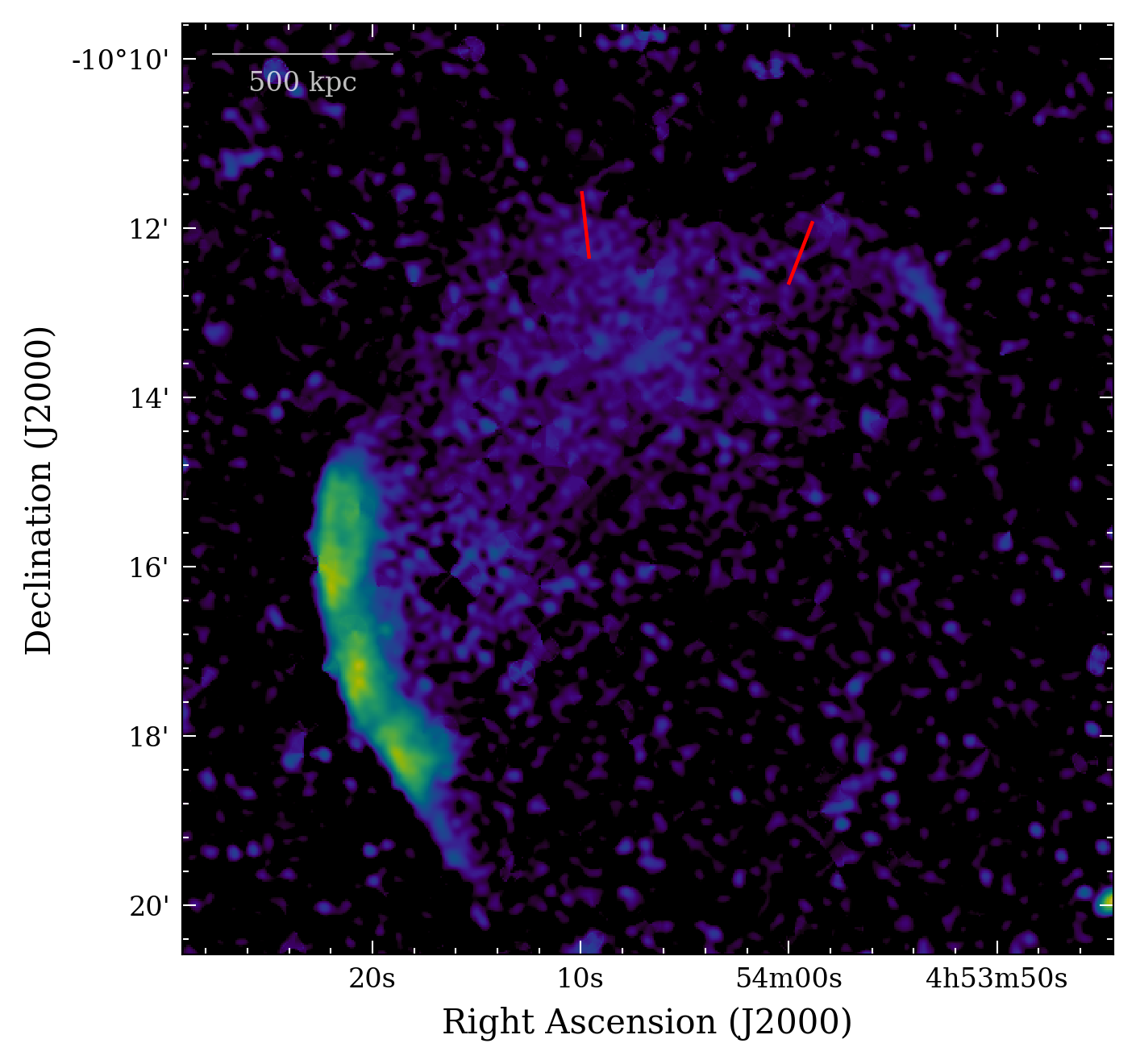}
  \includegraphics[width=.32\hsize,trim={0cm 0cm 0cm 0cm},clip,valign=c]{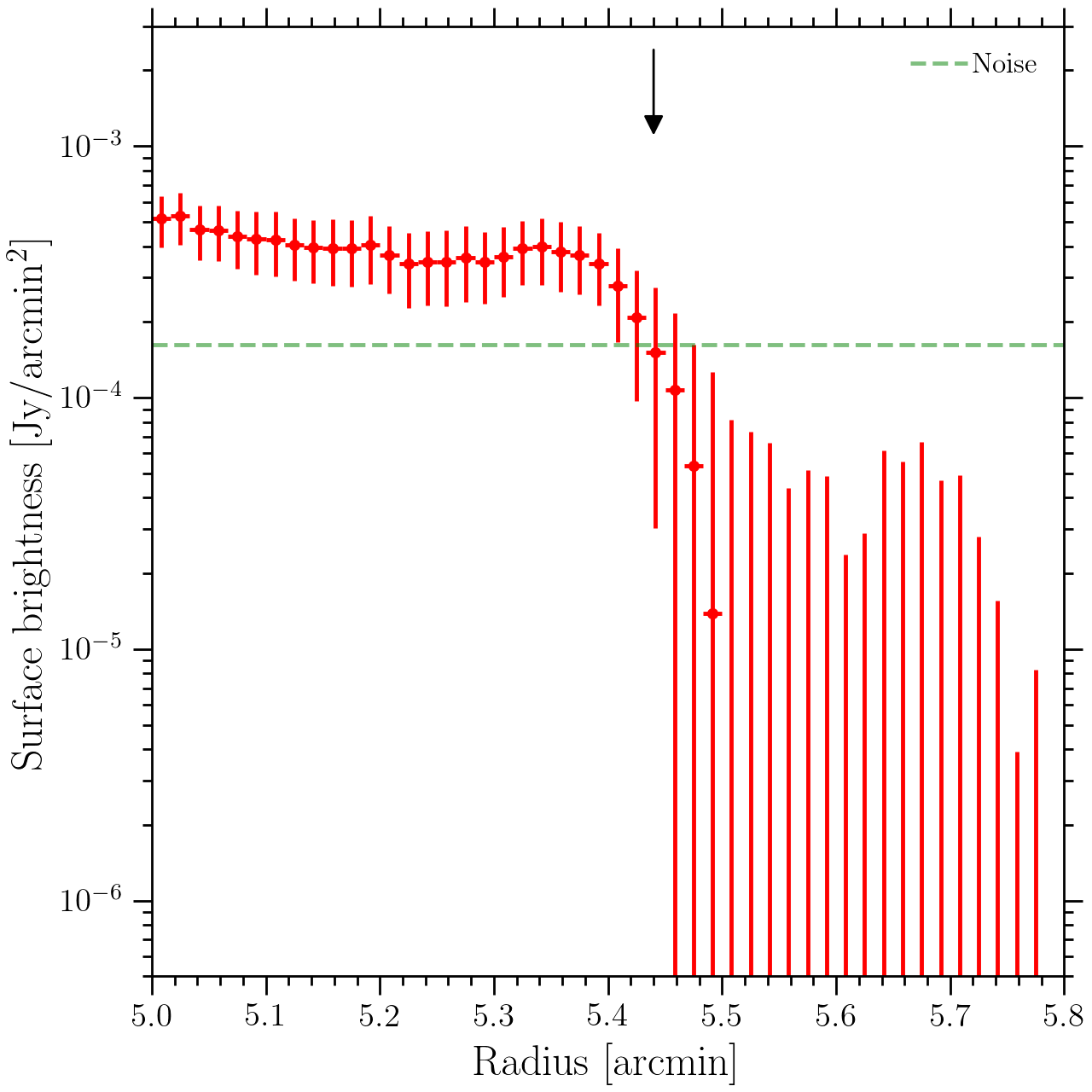}
  \caption{Same as Fig.~\ref{fig:bullet_sb} but for Abell 521.}
  \label{fig:Abell_521_sb}
\end{figure}

\begin{figure}
  \centering
  \includegraphics[width=\hsize,trim={0cm 0cm 0cm 0cm},clip,valign=c]{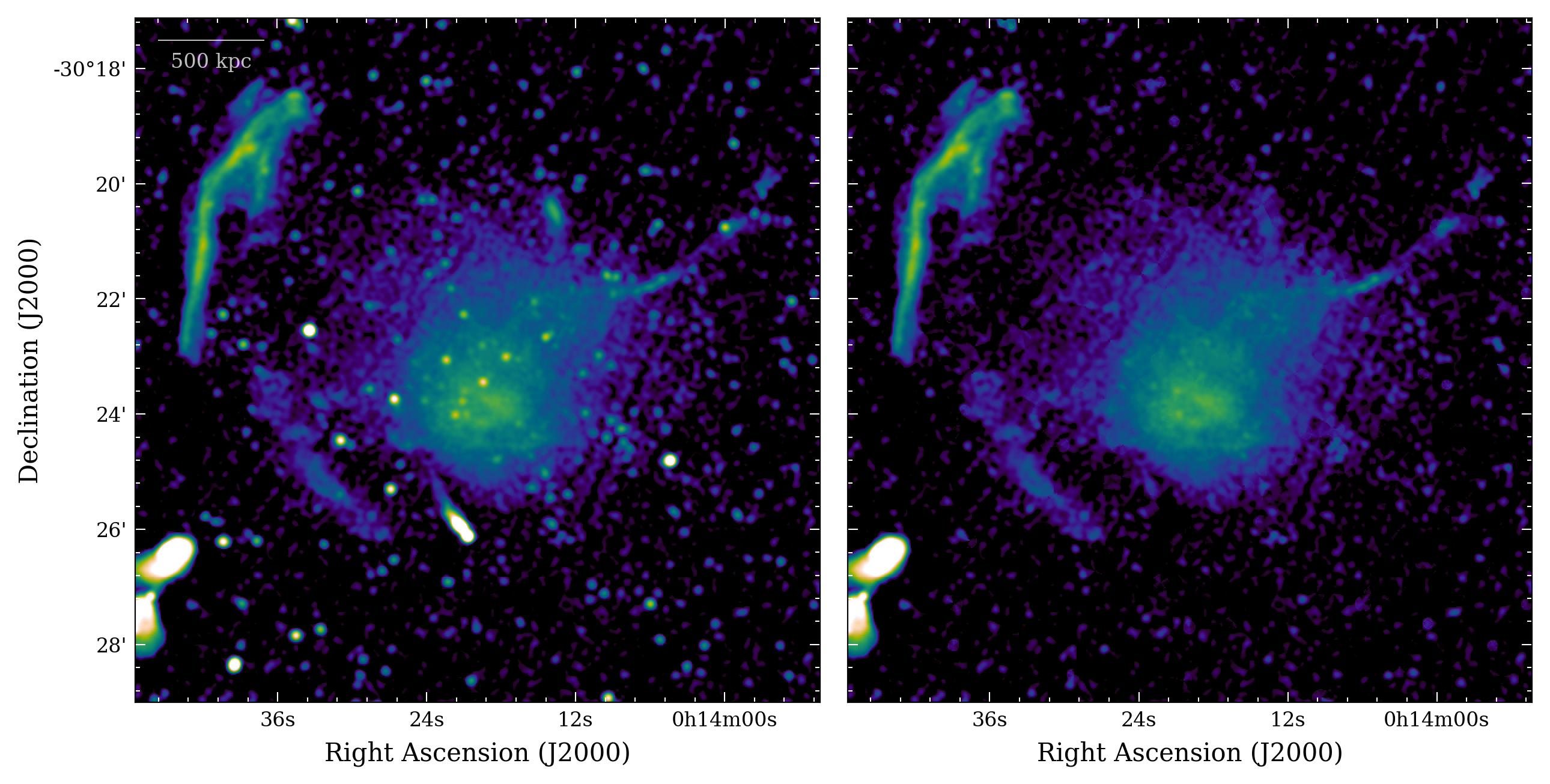}
  \includegraphics[width=\hsize,trim={0cm 0cm 0cm 0cm},clip,valign=c]{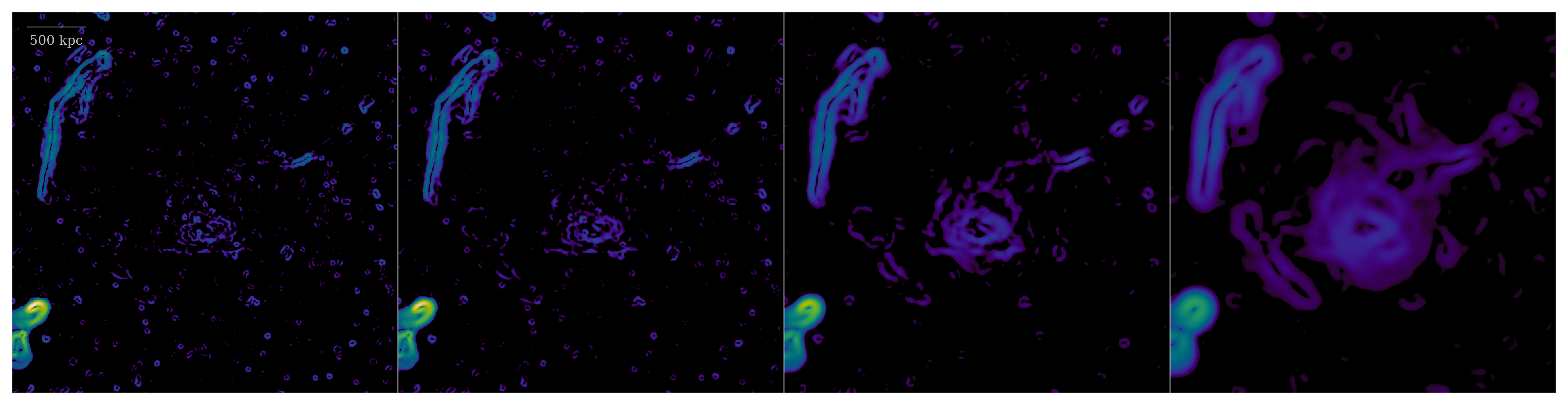}
  \caption{Same as Fig.~\ref{fig:bullet_meerkat} but for Abell 2744.}
  \label{fig:Abell_2744_meerkat}
\end{figure}

\begin{figure}
  \centering
  \includegraphics[width=.348\hsize,trim={0cm 0cm 0cm 0cm},clip,valign=c]{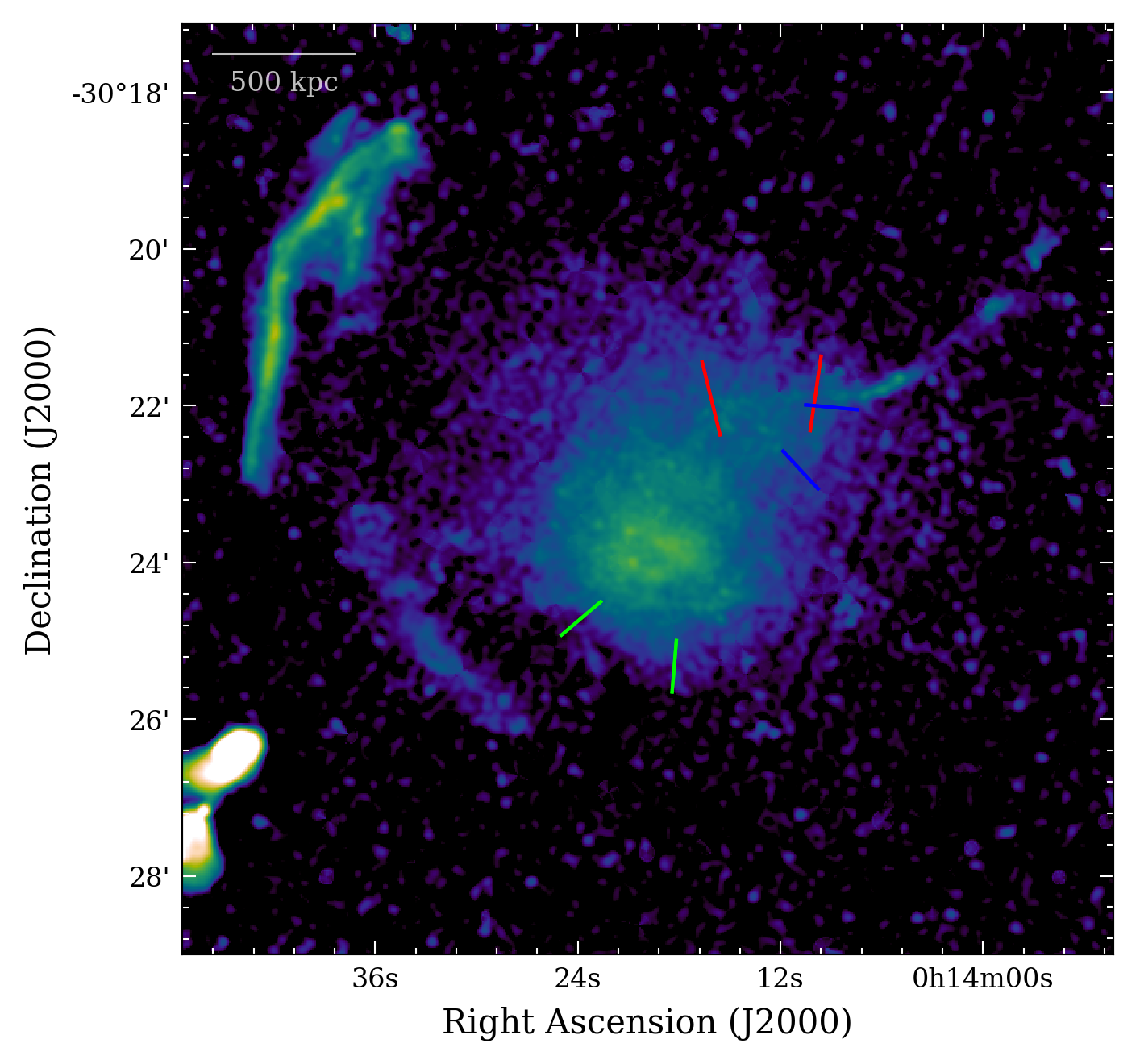}
  \includegraphics[width=.32\hsize,trim={0cm 0cm 0cm 0cm},clip,valign=c]{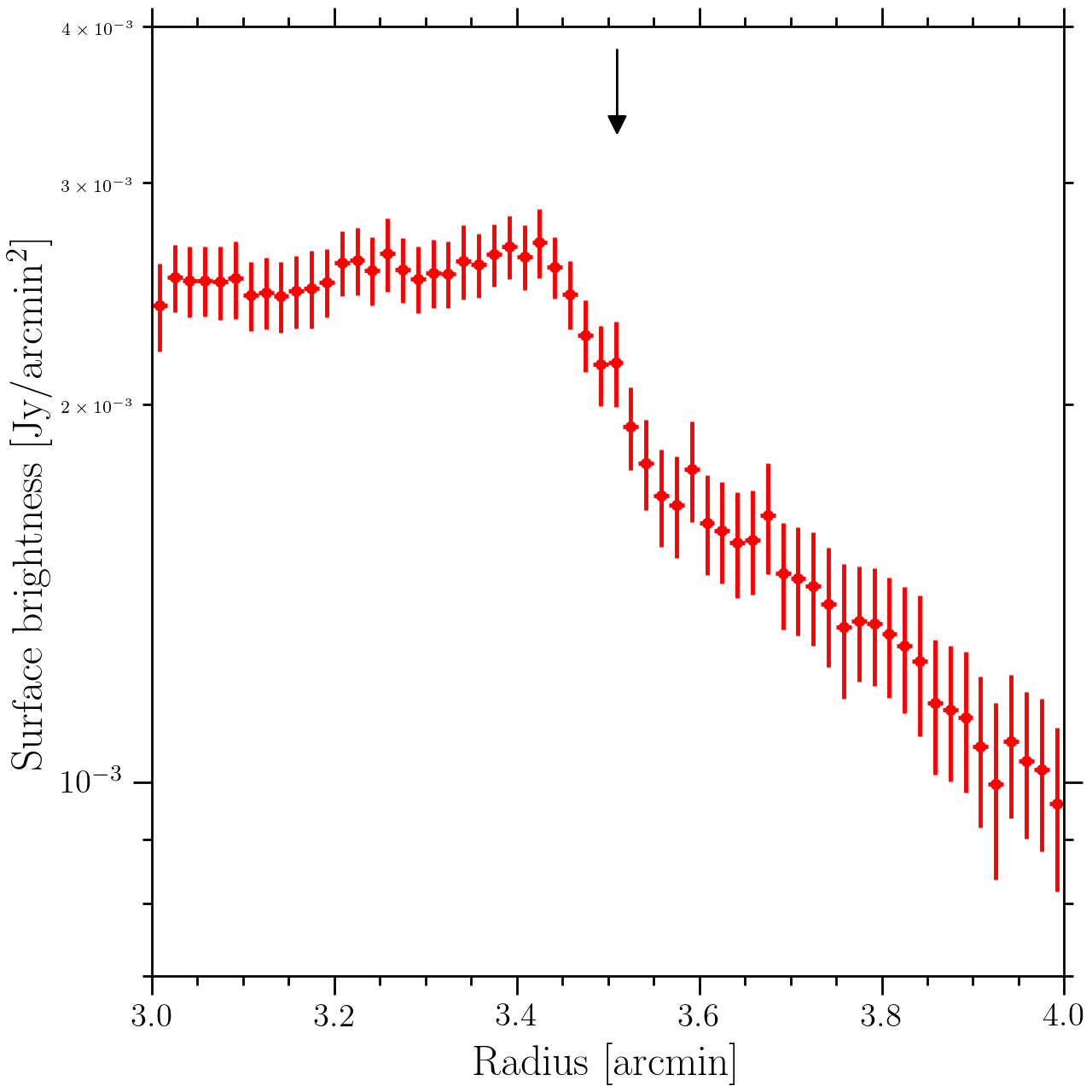}
  \includegraphics[width=.32\hsize,trim={0cm 0cm 0cm 0cm},clip,valign=c]{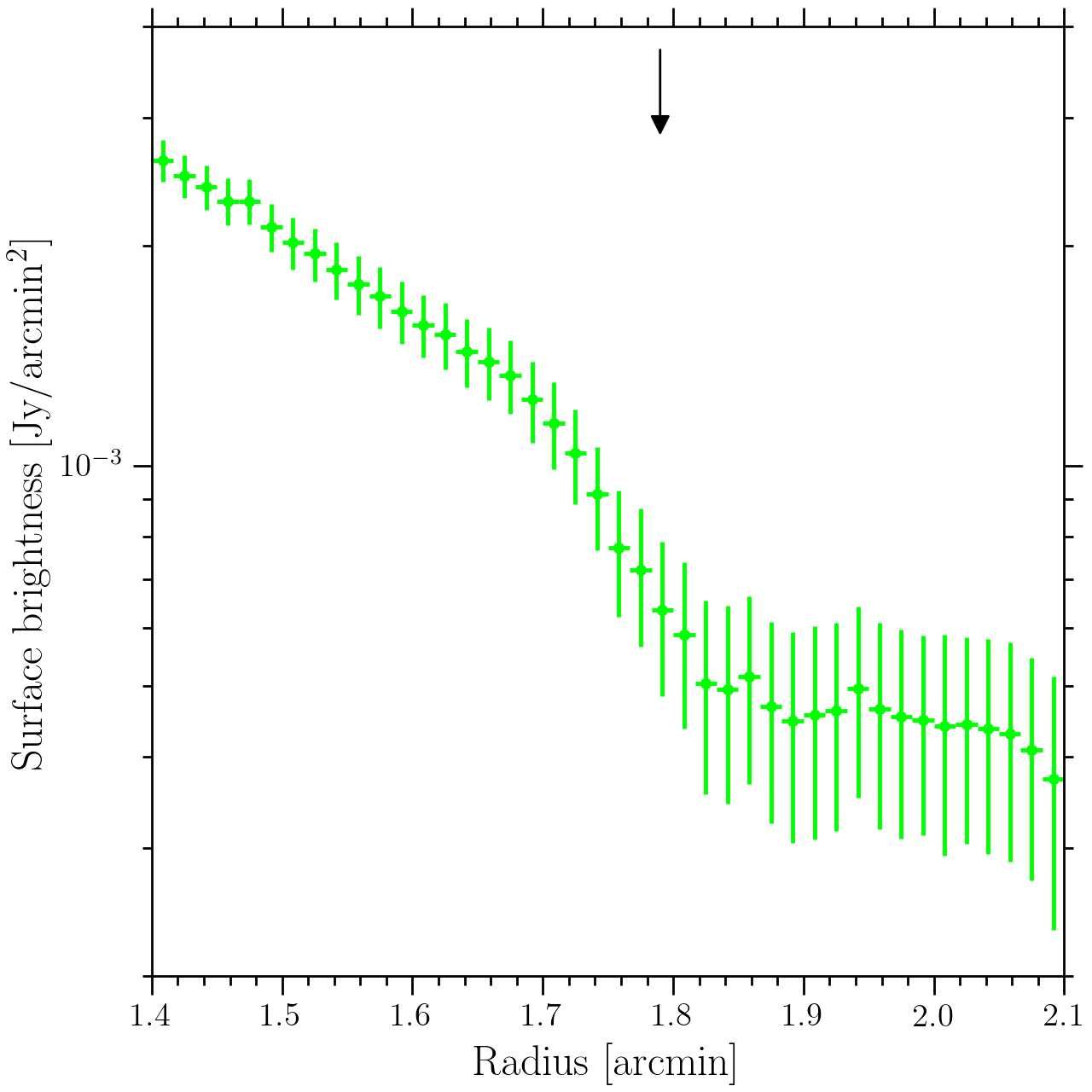}
  \hspace*{0.425cm} \includegraphics[width=.32\hsize,trim={0cm 0cm 0cm 0cm},clip,valign=c]{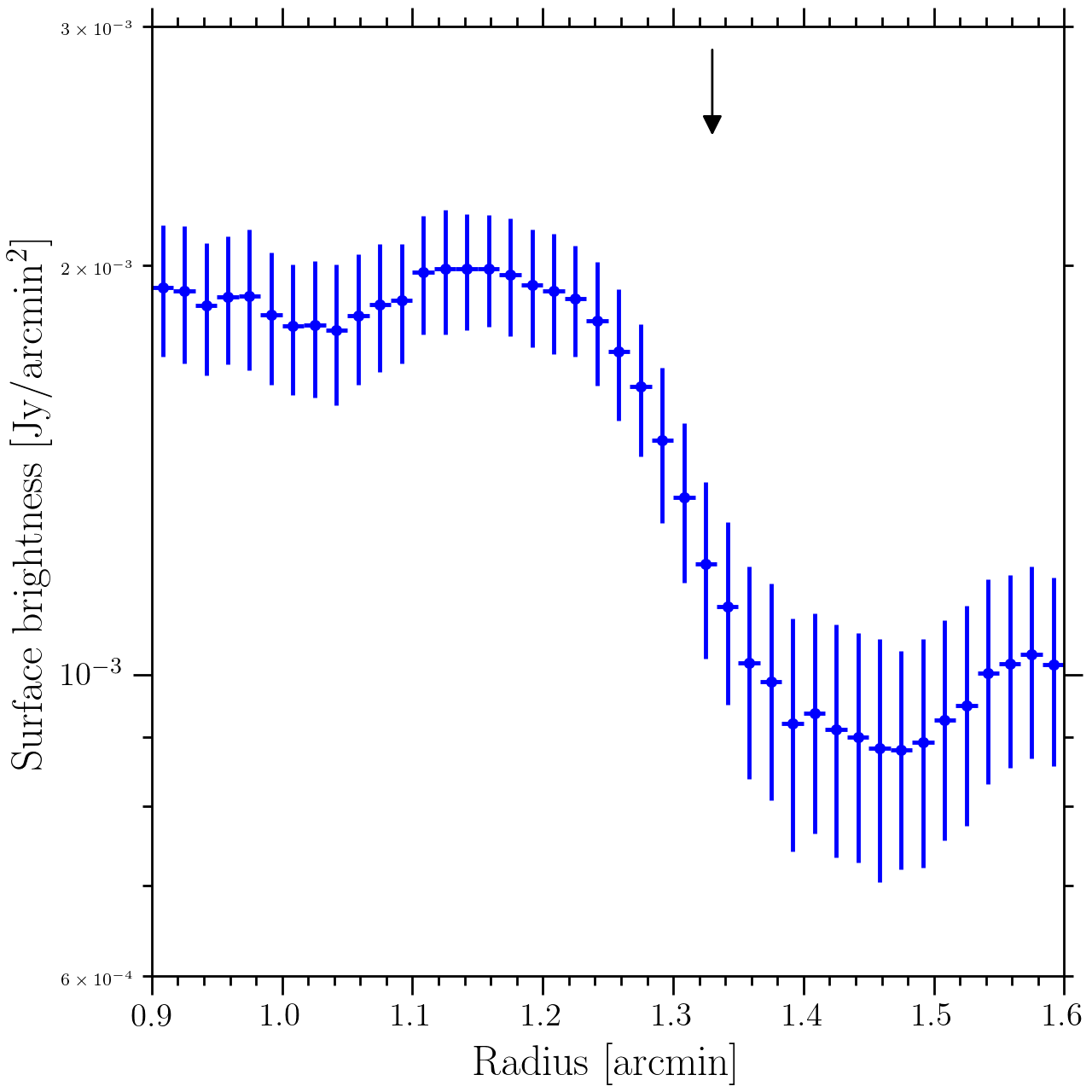}
  \caption{Same as Fig.~\ref{fig:bullet_sb} but for Abell 2744.}
  \label{fig:Abell_2744_sb}
\end{figure}

\begin{figure}
  \centering
  \includegraphics[width=\hsize,trim={0cm 0cm 0cm 0cm},clip,valign=c]{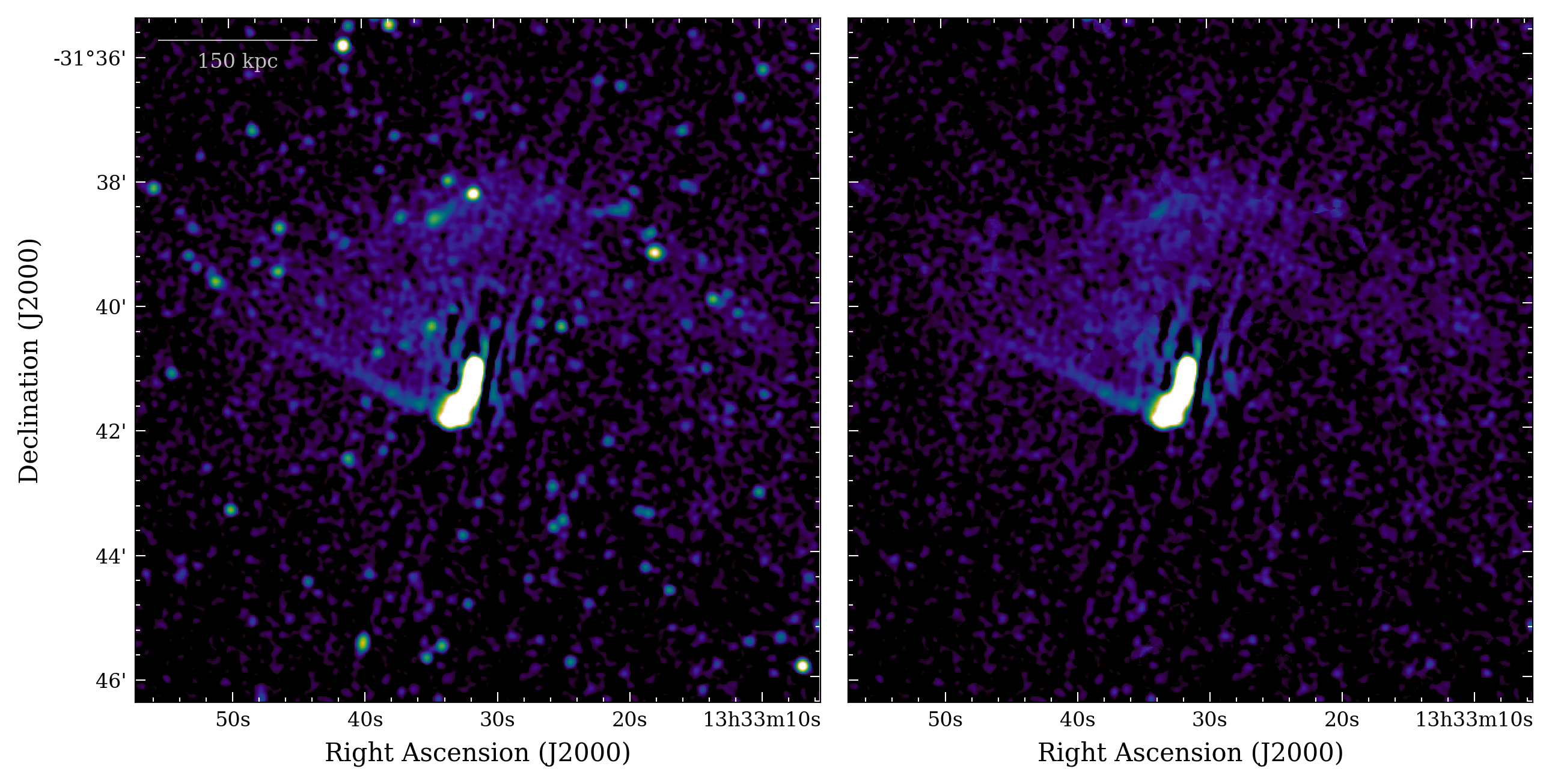}
  \includegraphics[width=\hsize,trim={0cm 0cm 0cm 0cm},clip,valign=c]{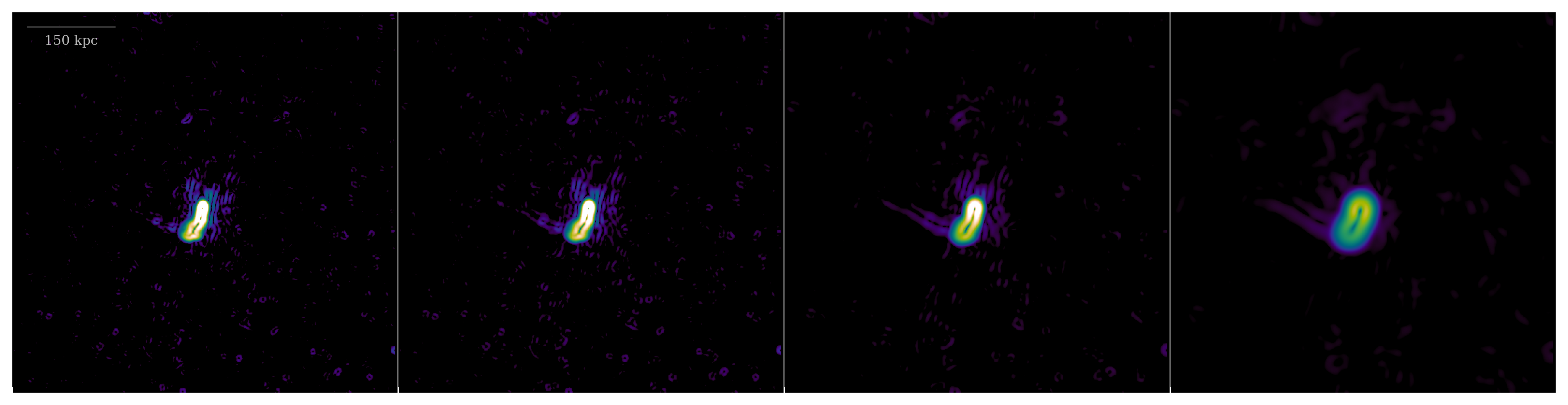}
  \caption{Same as Fig.~\ref{fig:bullet_meerkat} but for Abell 3562.}
  \label{fig:Abell_3562_meerkat}
\end{figure}

\begin{figure}
  \centering
  \includegraphics[width=.348\hsize,trim={0cm 0cm 0cm 0cm},clip,valign=c]{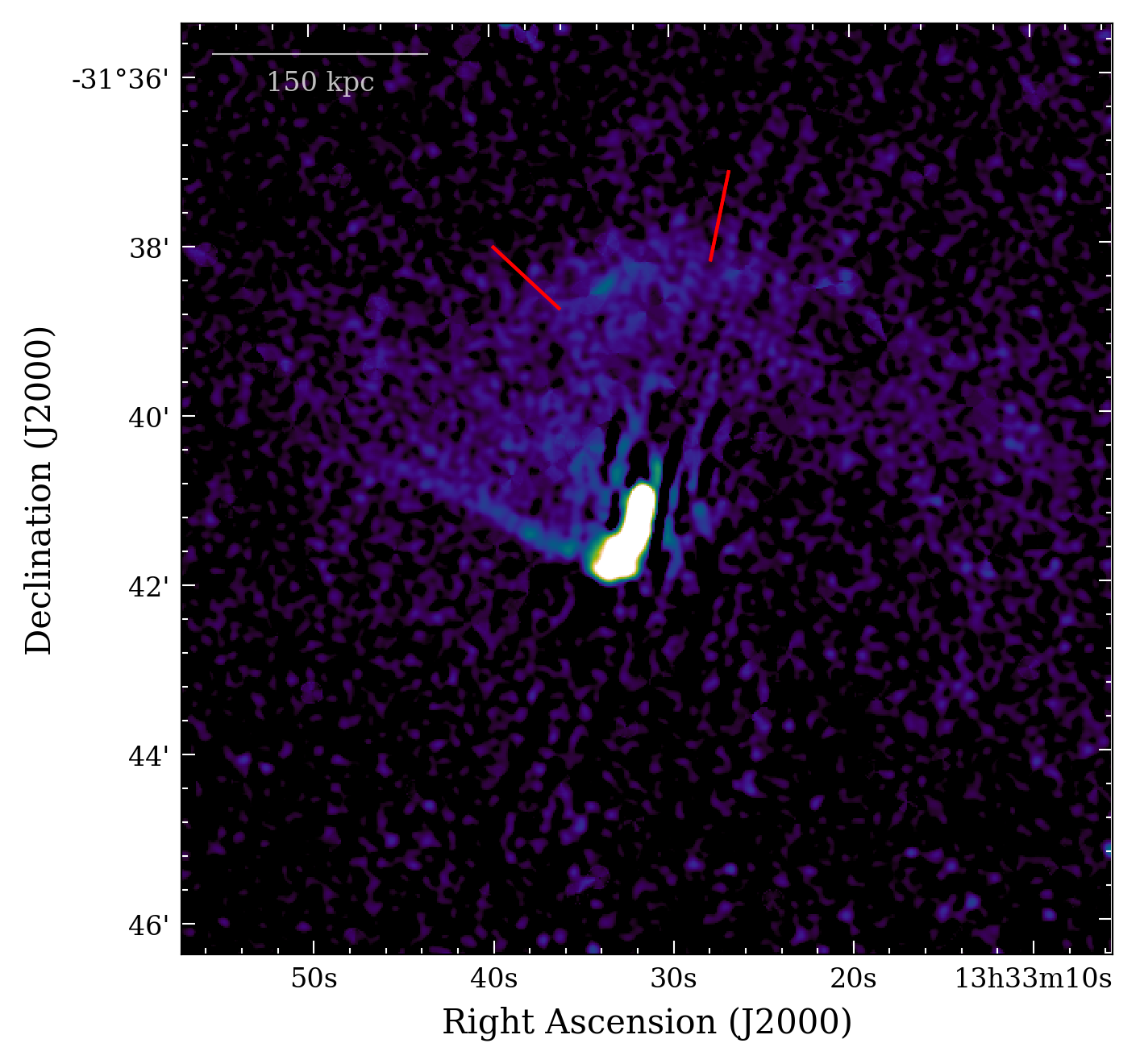}
  \includegraphics[width=.32\hsize,trim={0cm 0cm 0cm 0cm},clip,valign=c]{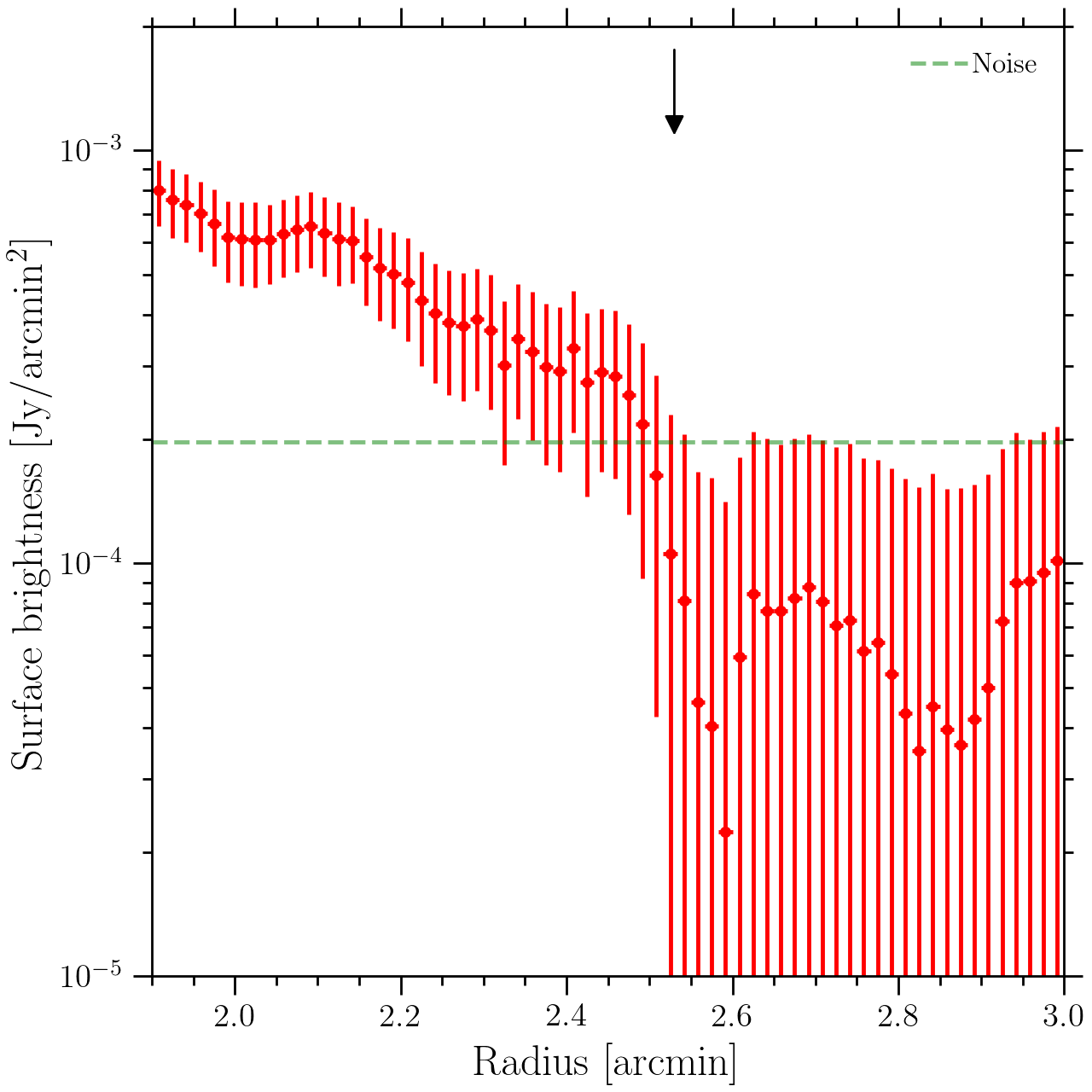}
  \caption{Same as Fig.~\ref{fig:bullet_sb} but for Abell 3562.}
  \label{fig:Abell_3562_sb}
\end{figure}

\begin{figure}
  \centering
  \includegraphics[width=\hsize,trim={0cm 0cm 0cm 0cm},clip,valign=c]{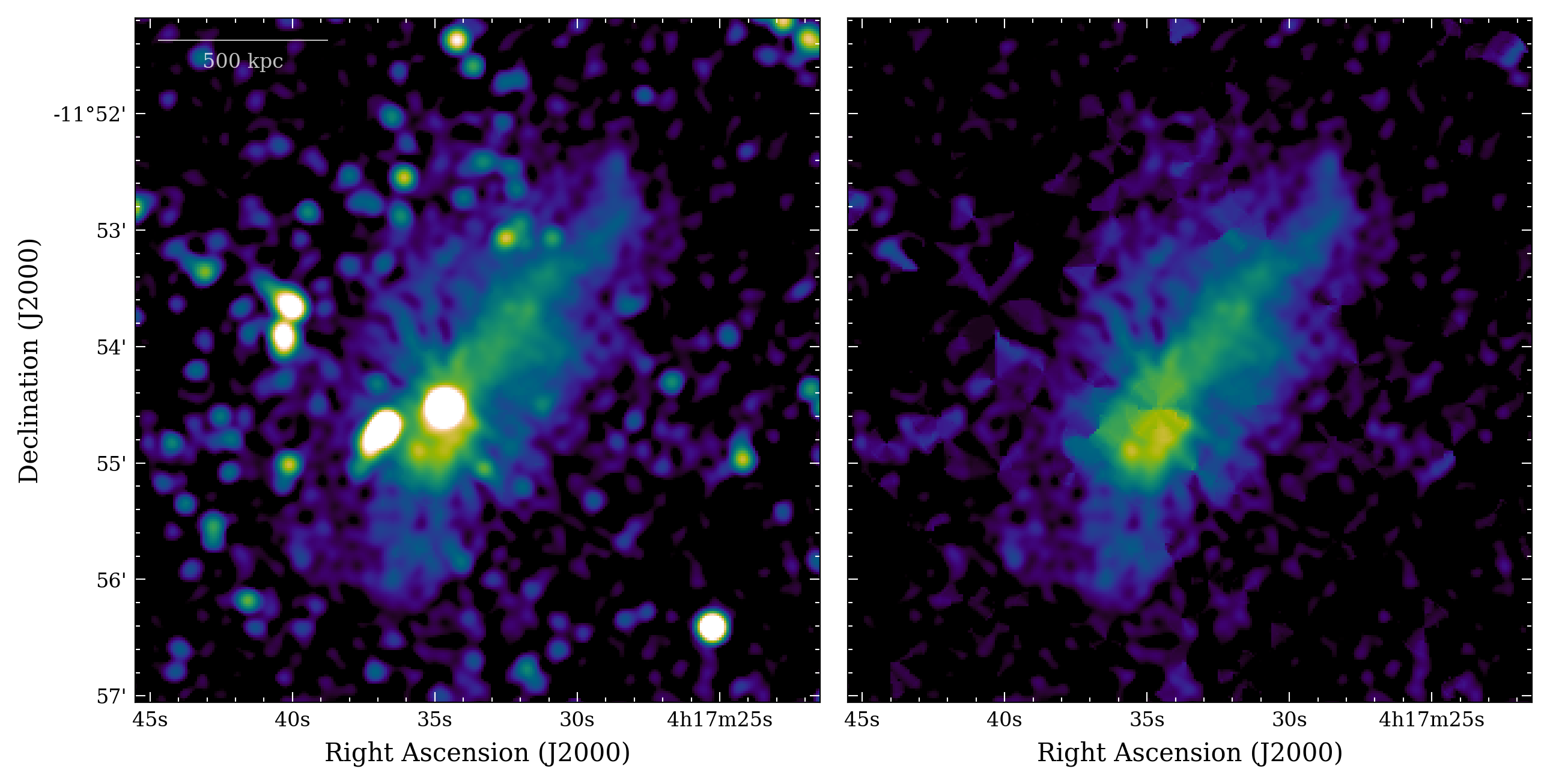}
  \includegraphics[width=\hsize,trim={0cm 0cm 0cm 0cm},clip,valign=c]{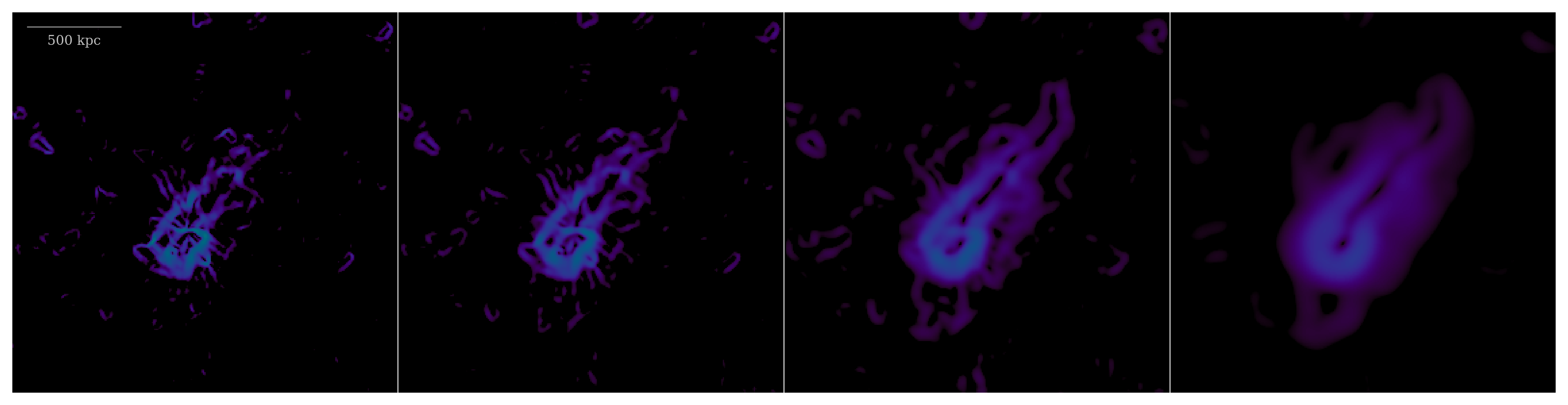}
  \caption{Same as Fig.~\ref{fig:bullet_meerkat} but for MACS J0417.5-1155.}
  \label{fig:MACSJ0417_meerkat}
\end{figure}

\begin{figure}
  \centering
  \includegraphics[width=.348\hsize,trim={0cm 0cm 0cm 0cm},clip,valign=c]{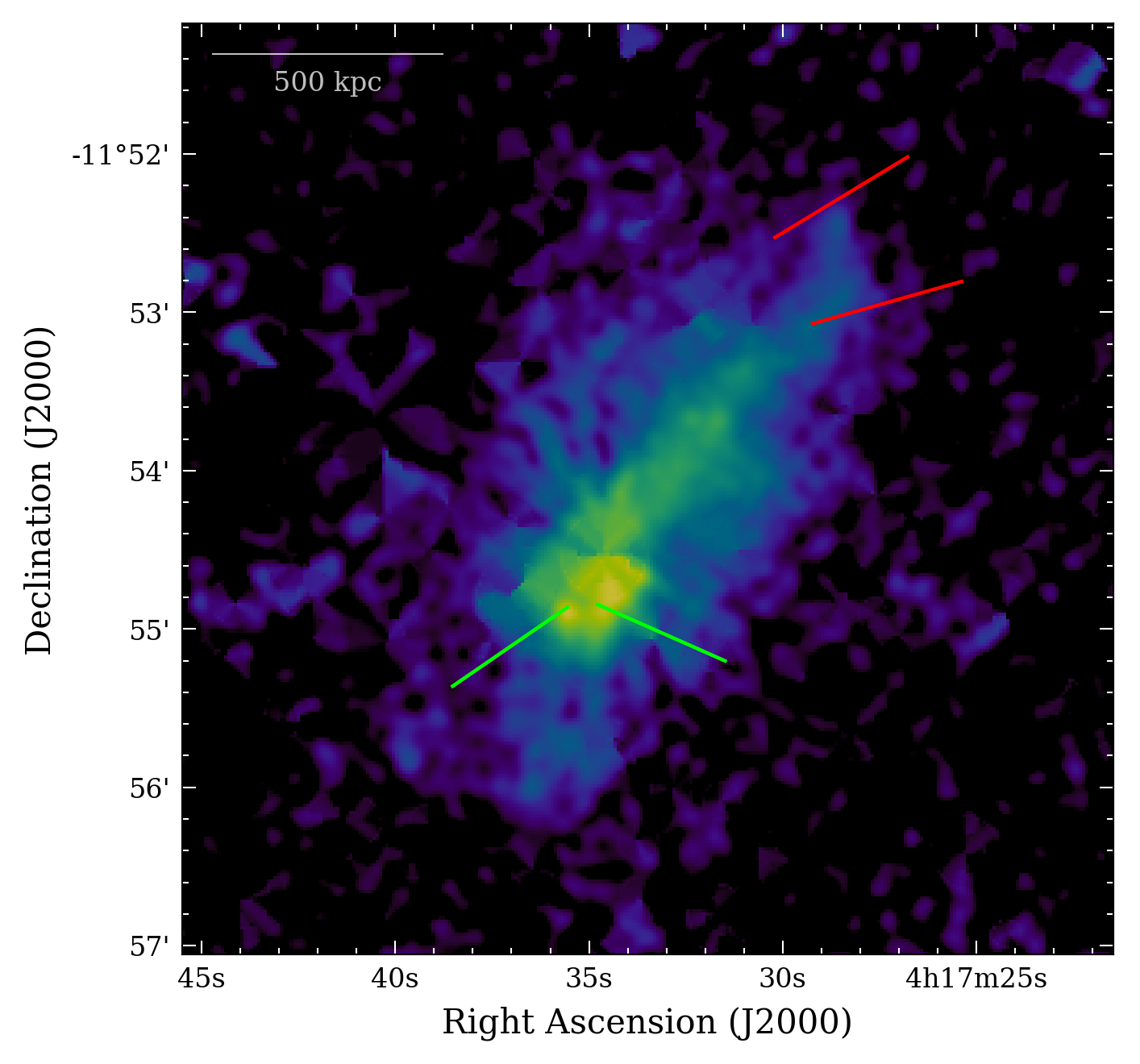}
  \includegraphics[width=.32\hsize,trim={0cm 0cm 0cm 0cm},clip,valign=c]{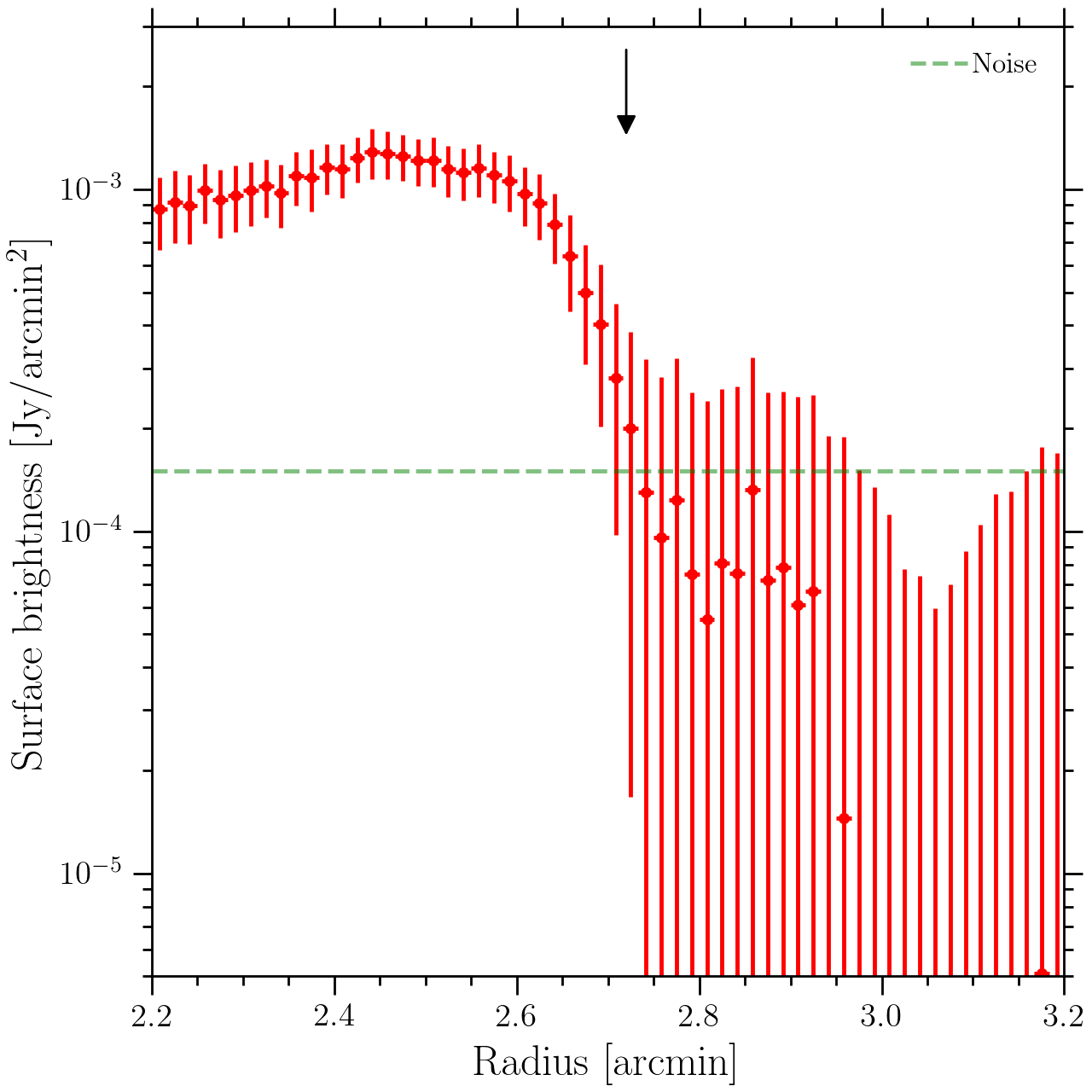}
  \includegraphics[width=.32\hsize,trim={0cm 0cm 0cm 0cm},clip,valign=c]{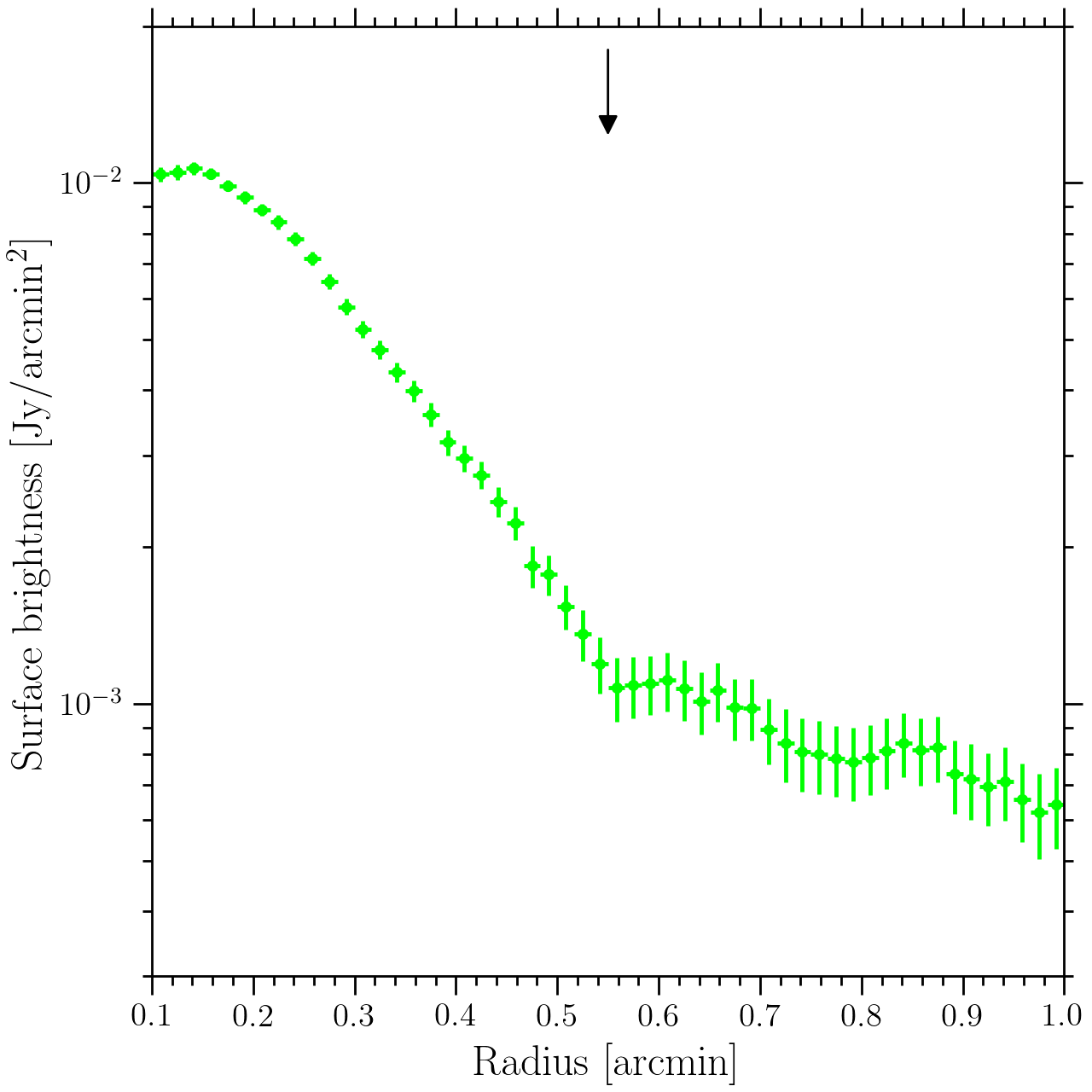}
  \caption{Same as Fig.~\ref{fig:bullet_sb} but for MACS J0417.5-1155.}
  \label{fig:MACSJ0417_sb}
\end{figure}

\begin{figure}
  \centering
  \includegraphics[width=\hsize,trim={0cm 0cm 0cm 0cm},clip,valign=c]{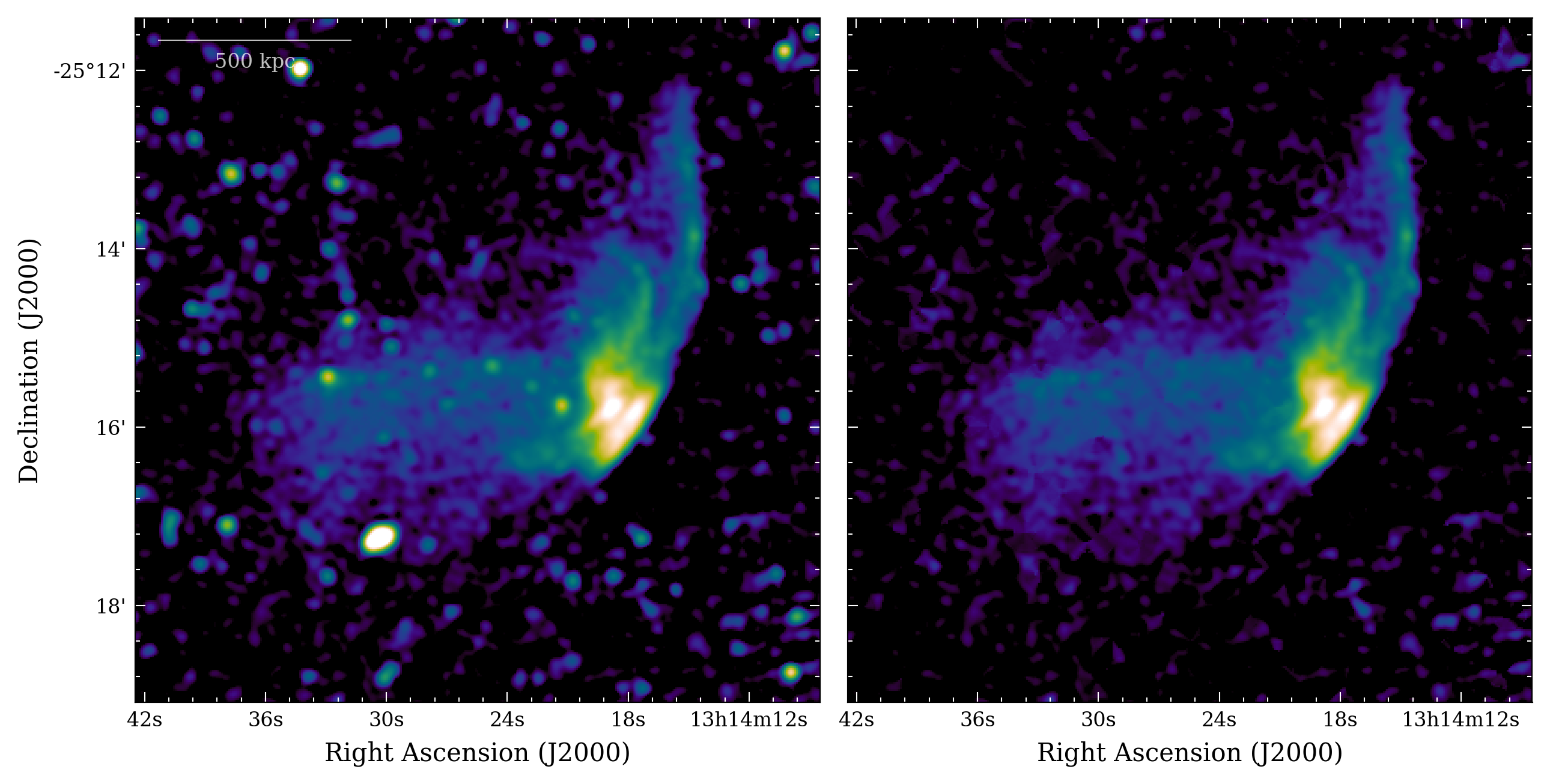}
  \includegraphics[width=\hsize,trim={0cm 0cm 0cm 0cm},clip,valign=c]{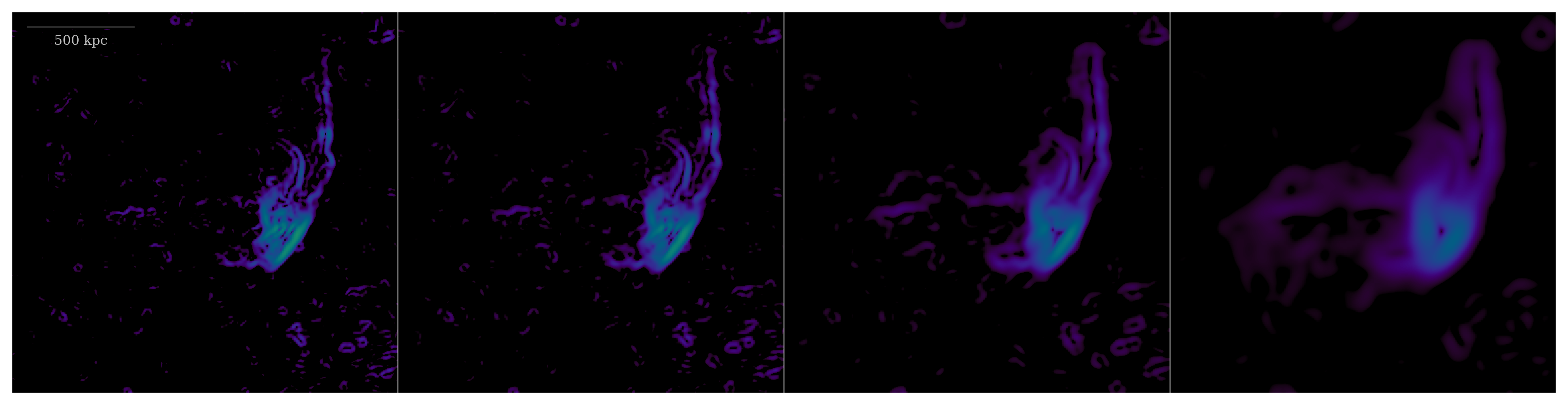}
  \caption{Same as Fig.~\ref{fig:bullet_meerkat} but for RXC J1314.4-2525.}
  \label{fig:RXCJ1314_meerkat}
\end{figure}

\begin{figure}
  \centering
  \includegraphics[width=.348\hsize,trim={0cm 0cm 0cm 0cm},clip,valign=c]{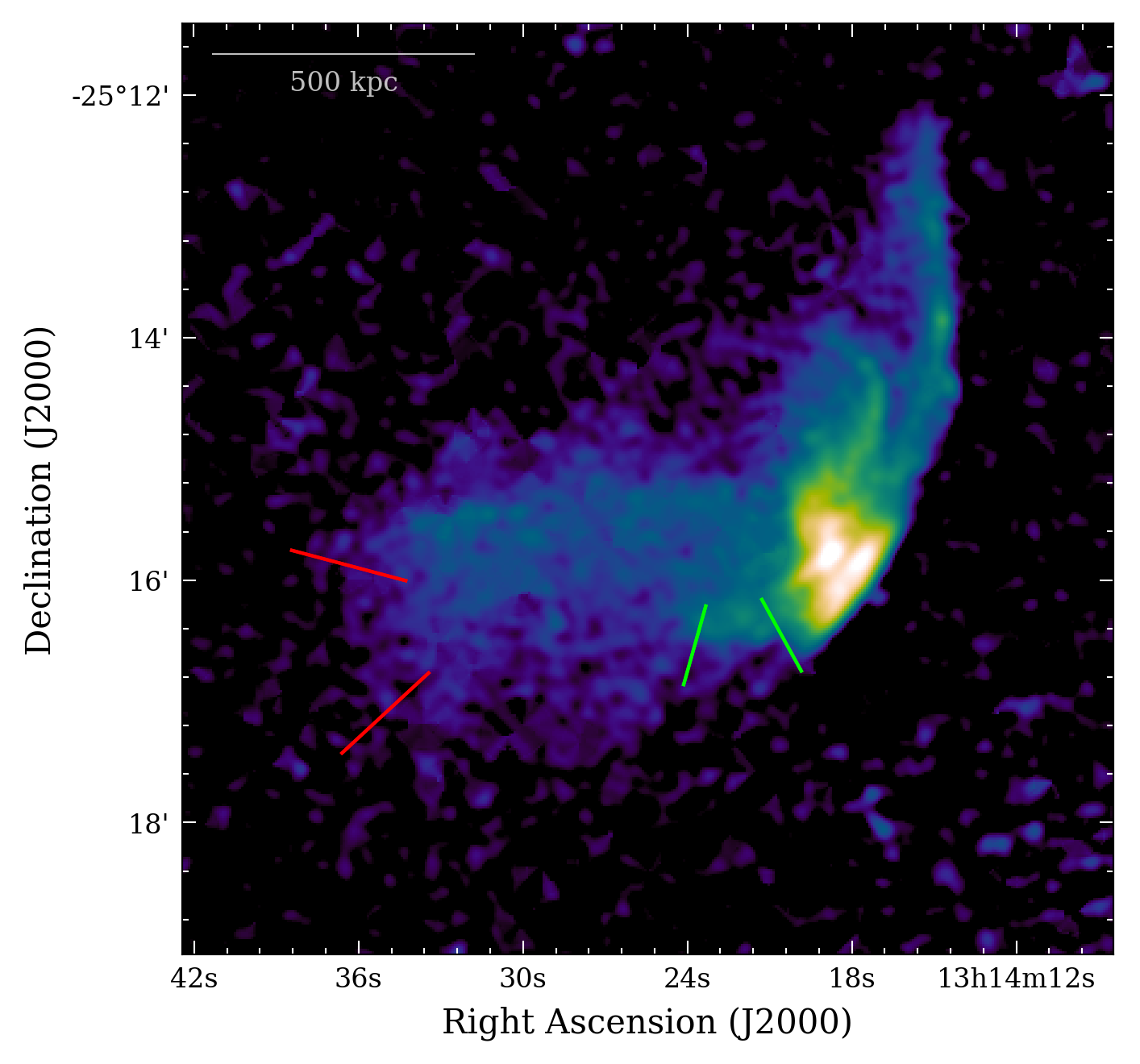}
  \includegraphics[width=.32\hsize,trim={0cm 0cm 0cm 0cm},clip,valign=c]{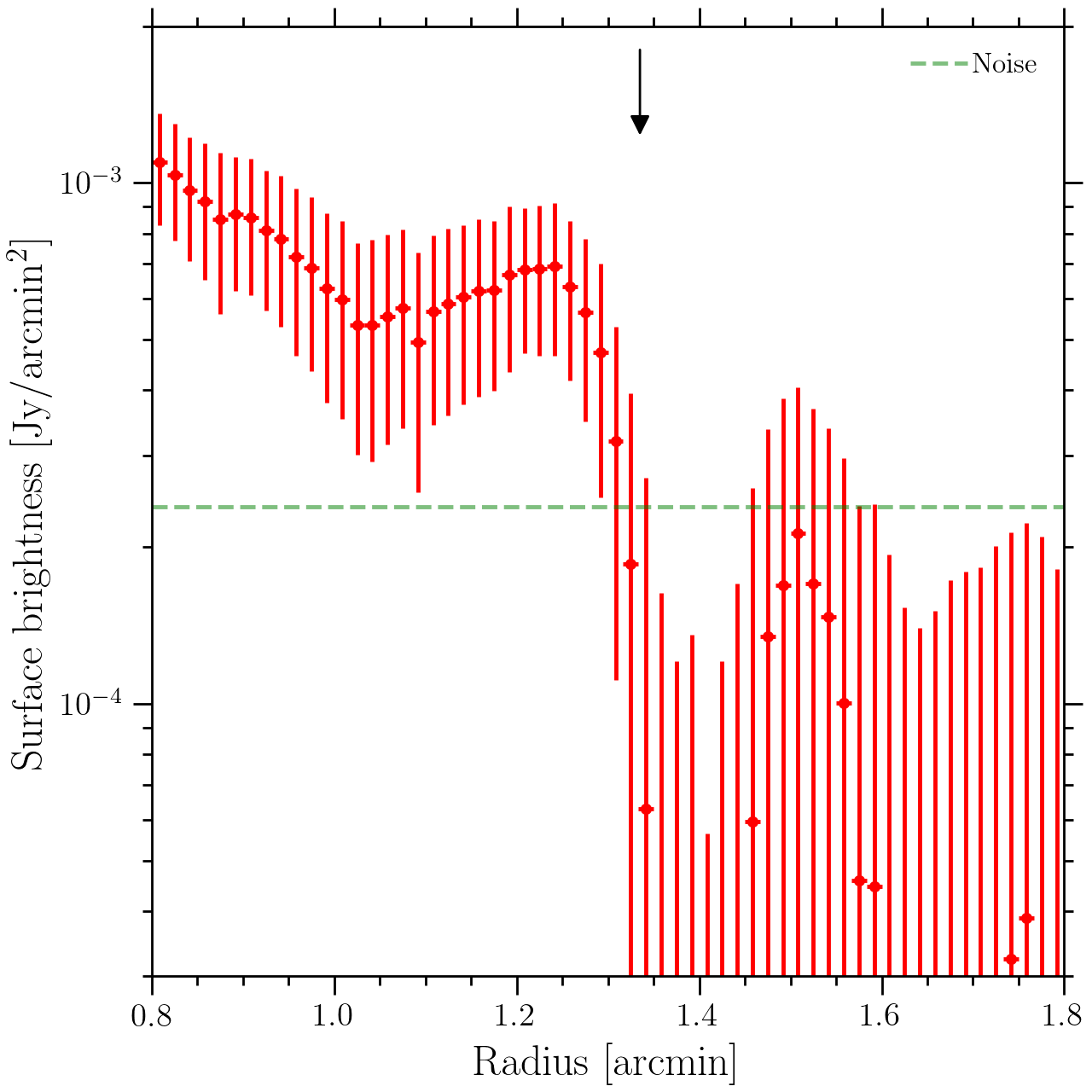}
  \includegraphics[width=.32\hsize,trim={0cm 0cm 0cm 0cm},clip,valign=c]{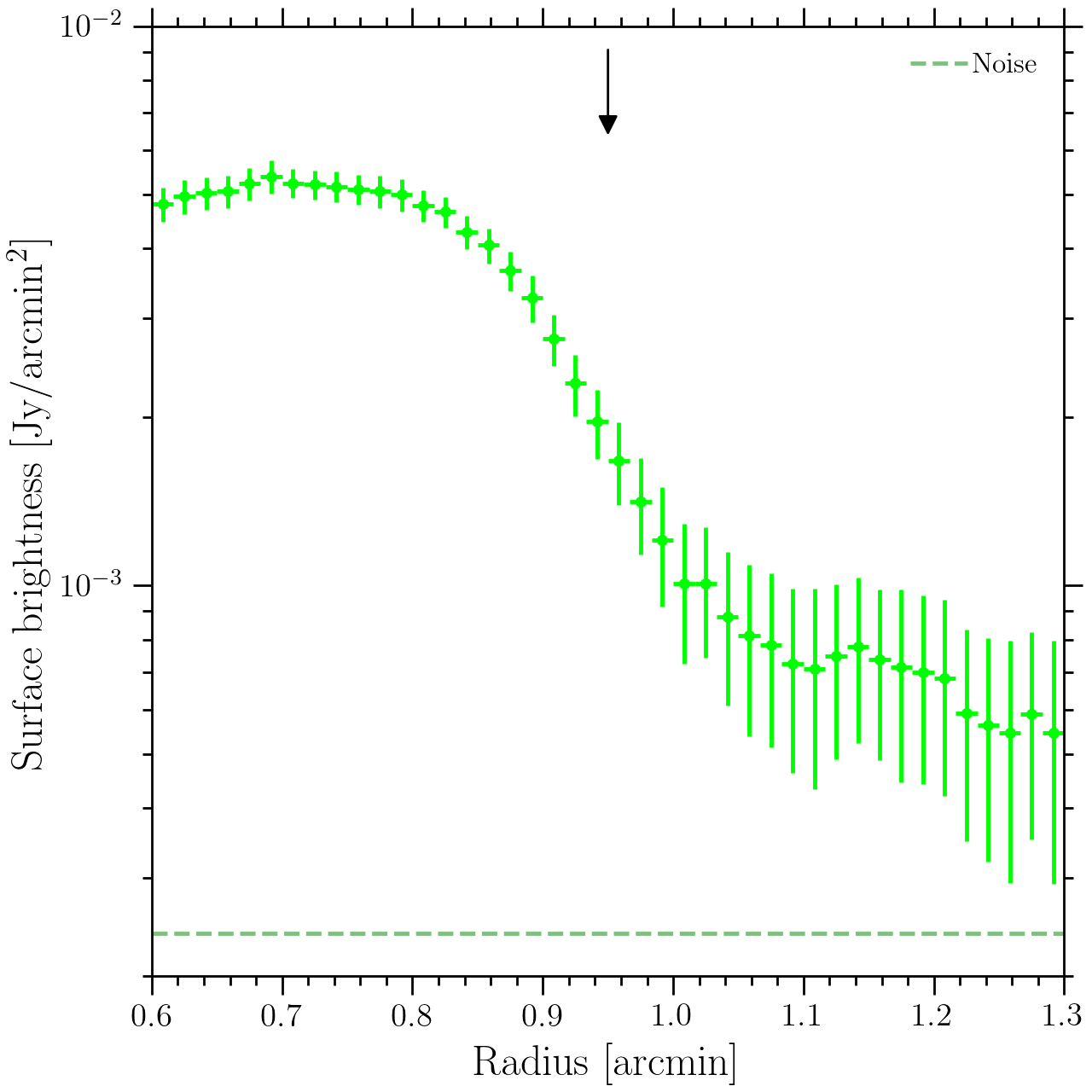}
    \caption{Same as Fig.~\ref{fig:bullet_sb} but for RXC J1314.4-2525.}
  \label{fig:RXCJ1314_sb}
\end{figure}

\begin{figure}
  \centering
  \includegraphics[width=\hsize,trim={0cm 0cm 0cm 0cm},clip,valign=c]{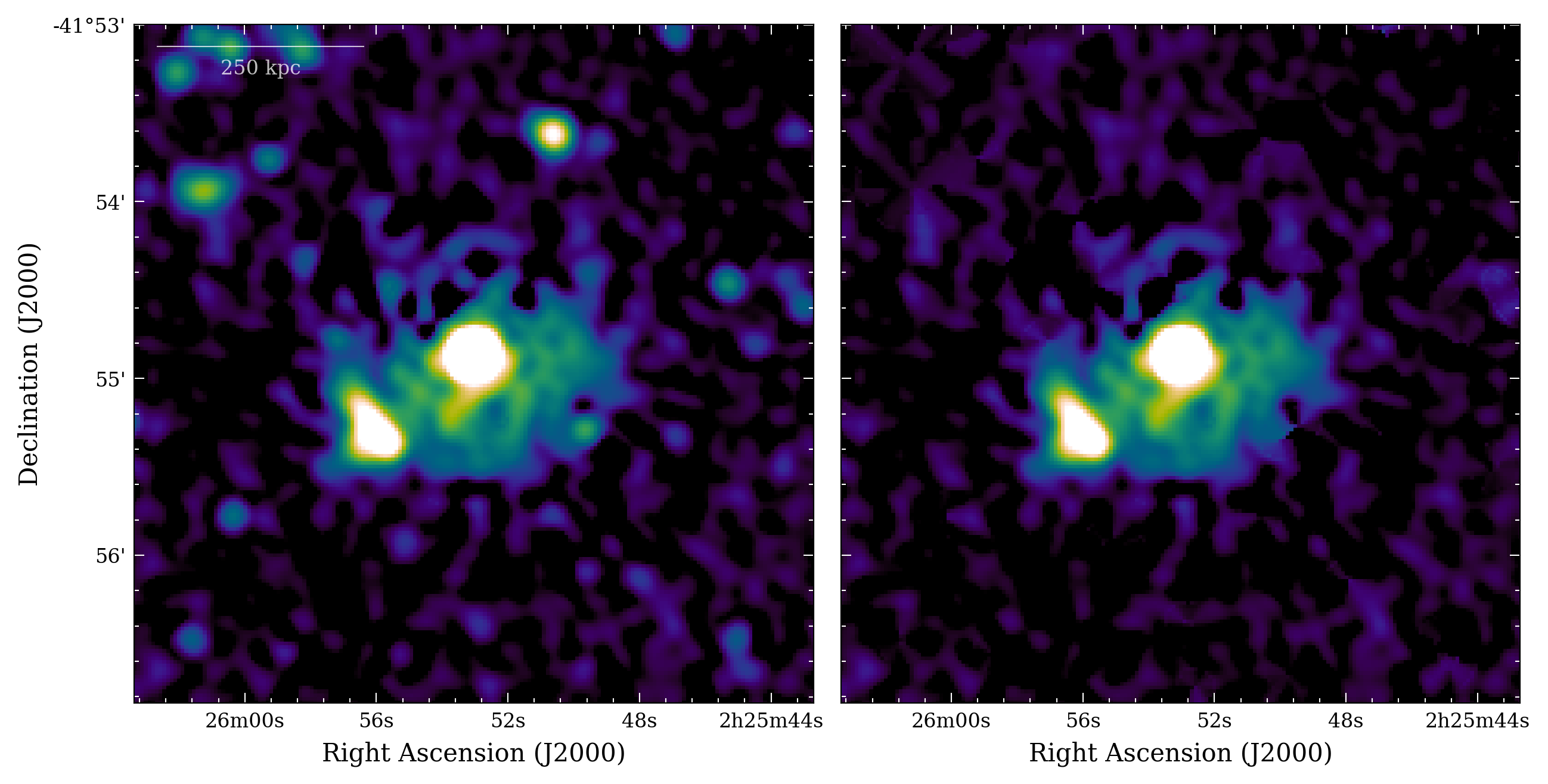}
  \includegraphics[width=\hsize,trim={0cm 0cm 0cm 0cm},clip,valign=c]{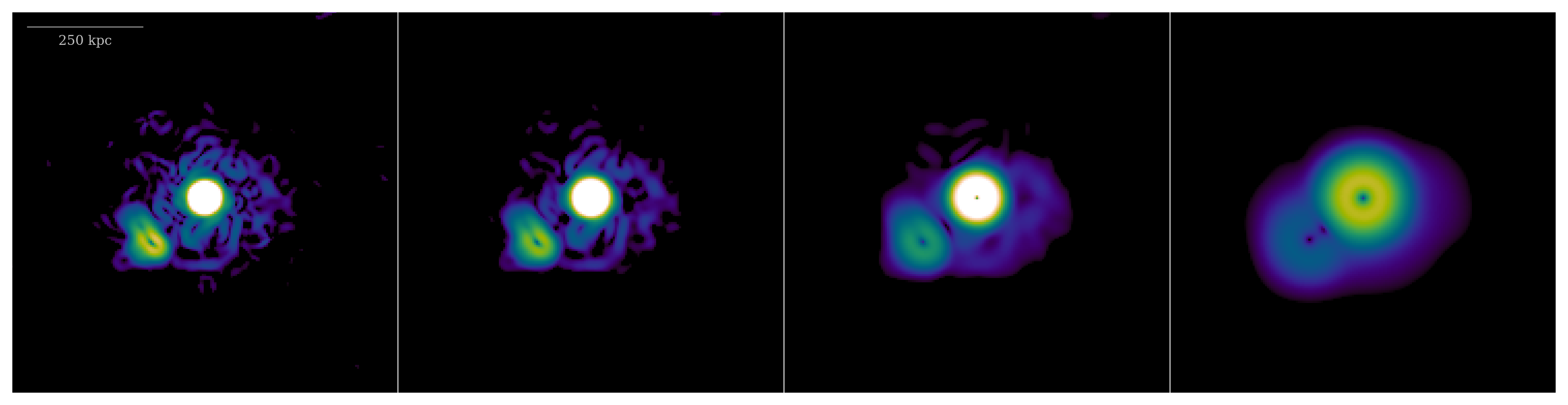}
  \caption{Same as Fig.~\ref{fig:bullet_meerkat} but for J0225.9-4154.}
  \label{fig:J0225.9-4154_meerkat}
\end{figure}

\begin{figure}
  \centering
  \includegraphics[width=.348\hsize,trim={0cm 0cm 0cm 0cm},clip,valign=c]{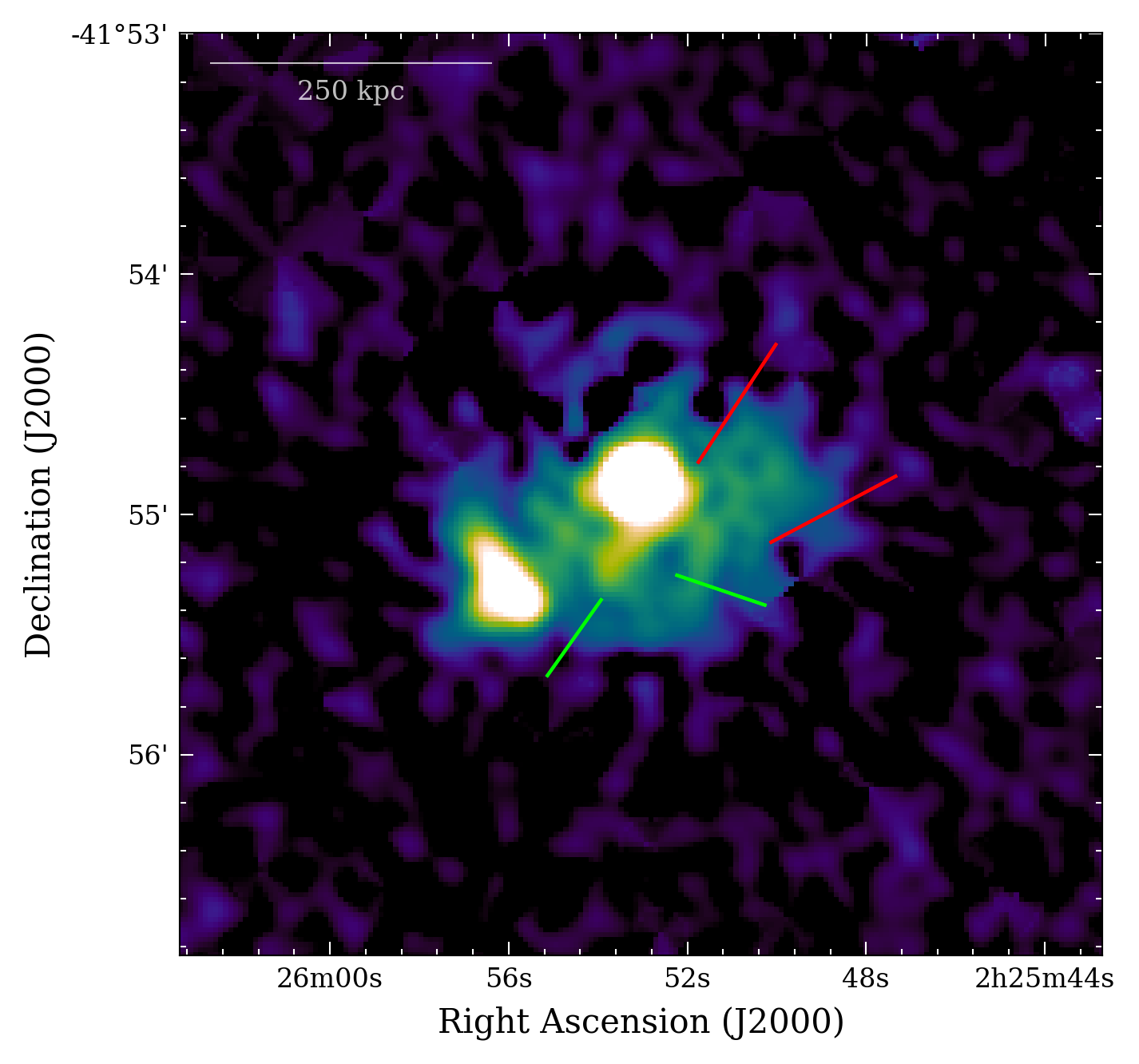}
  \includegraphics[width=.32\hsize,trim={0cm 0cm 0cm 0cm},clip,valign=c]{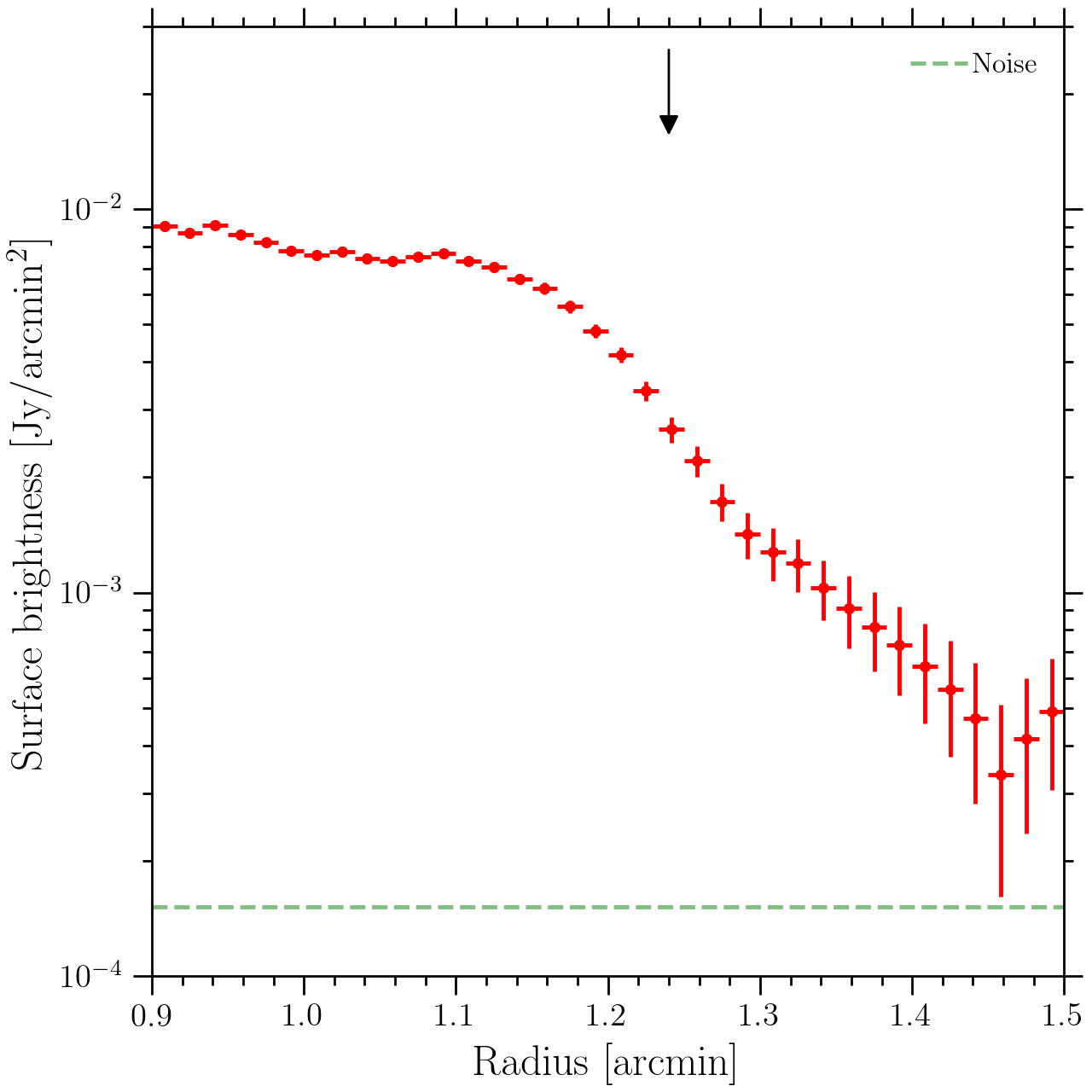}
  \includegraphics[width=.32\hsize,trim={0cm 0cm 0cm 0cm},clip,valign=c]{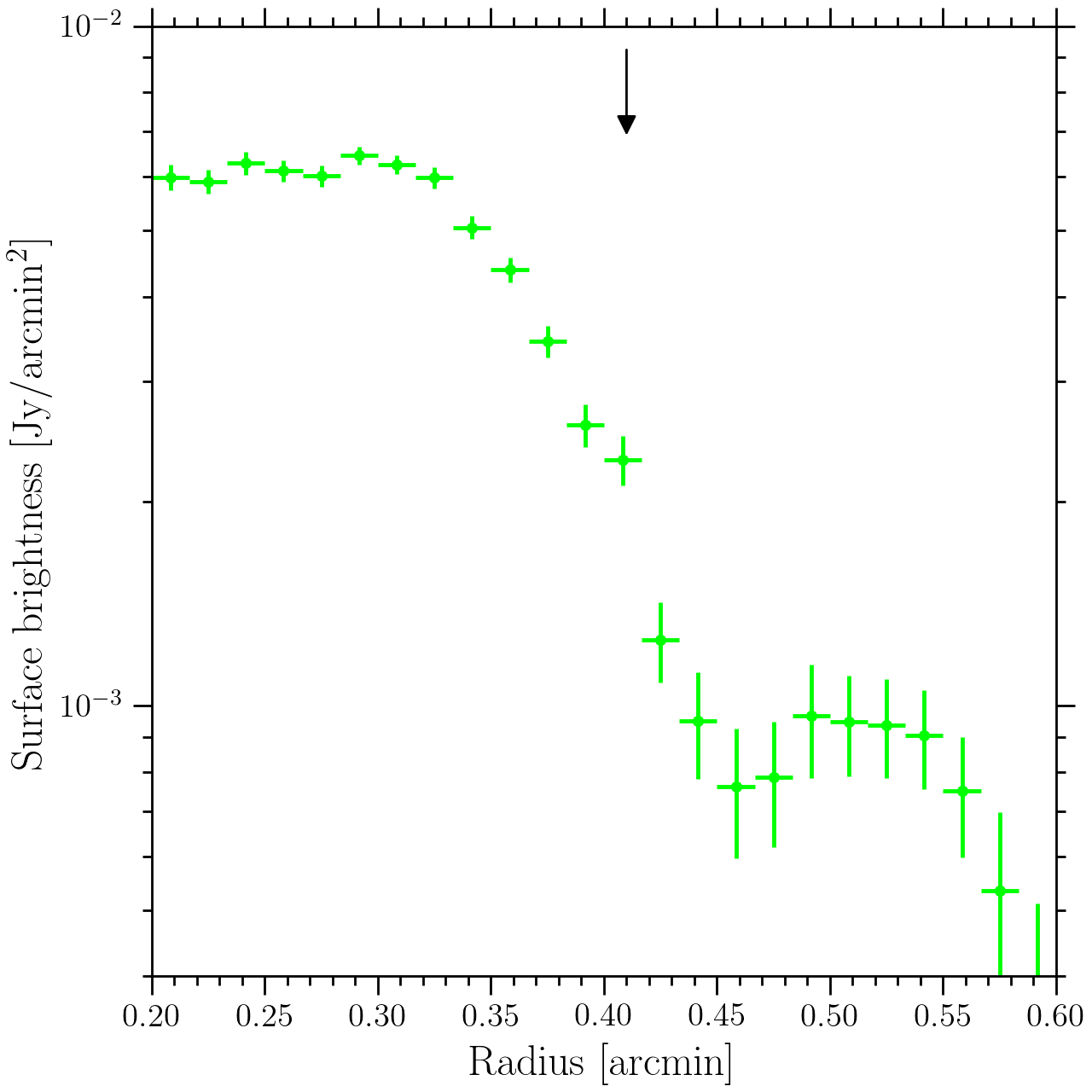}
    \caption{Same as Fig.~\ref{fig:bullet_sb} but for J0225.9-4154.}
  \label{fig:J0225.9-4154_sb}
\end{figure}

\begin{figure}
  \centering
  \includegraphics[width=\hsize,trim={0cm 0cm 0cm 0cm},clip,valign=c]{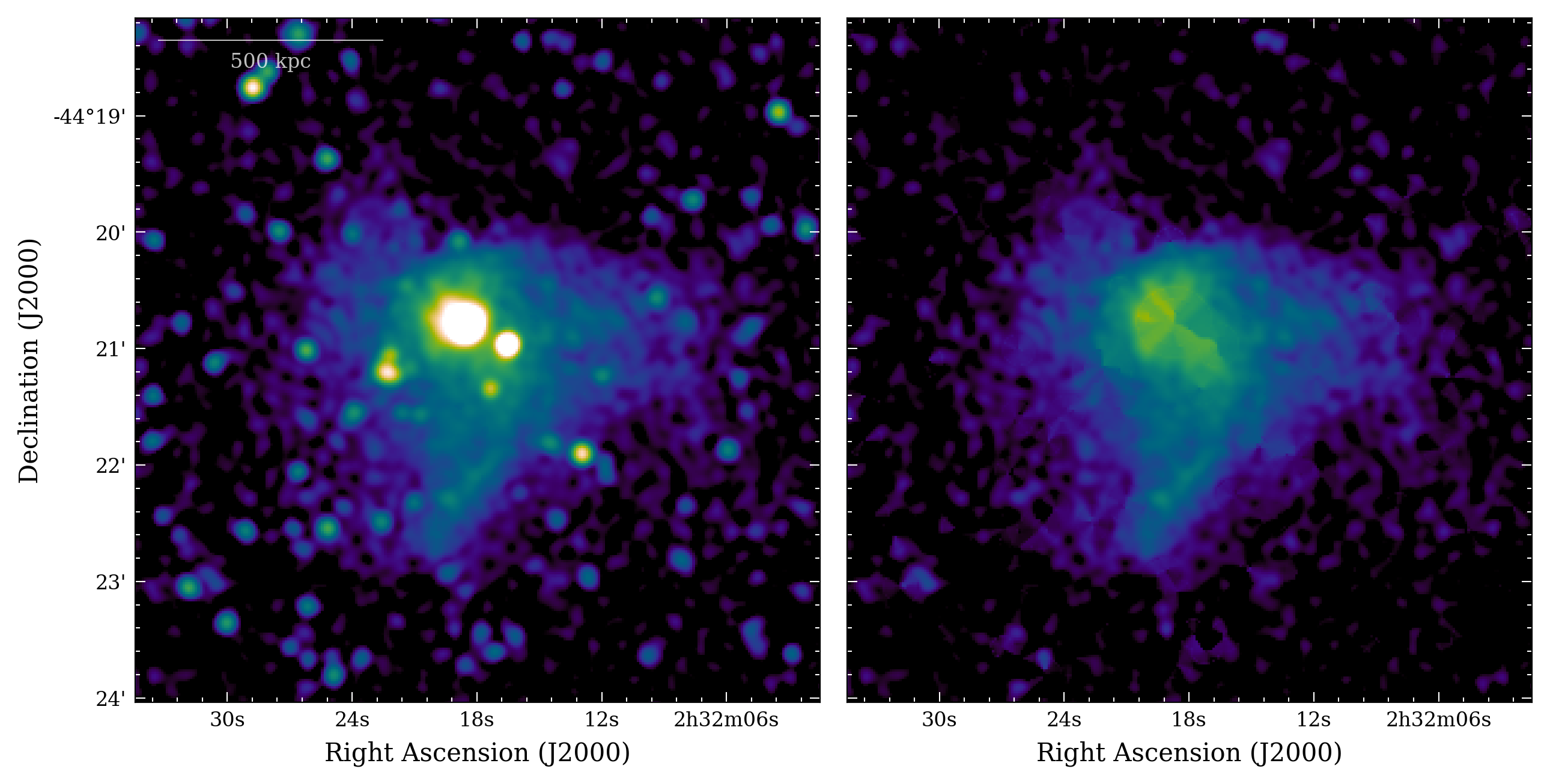}
  \includegraphics[width=\hsize,trim={0cm 0cm 0cm 0cm},clip,valign=c]{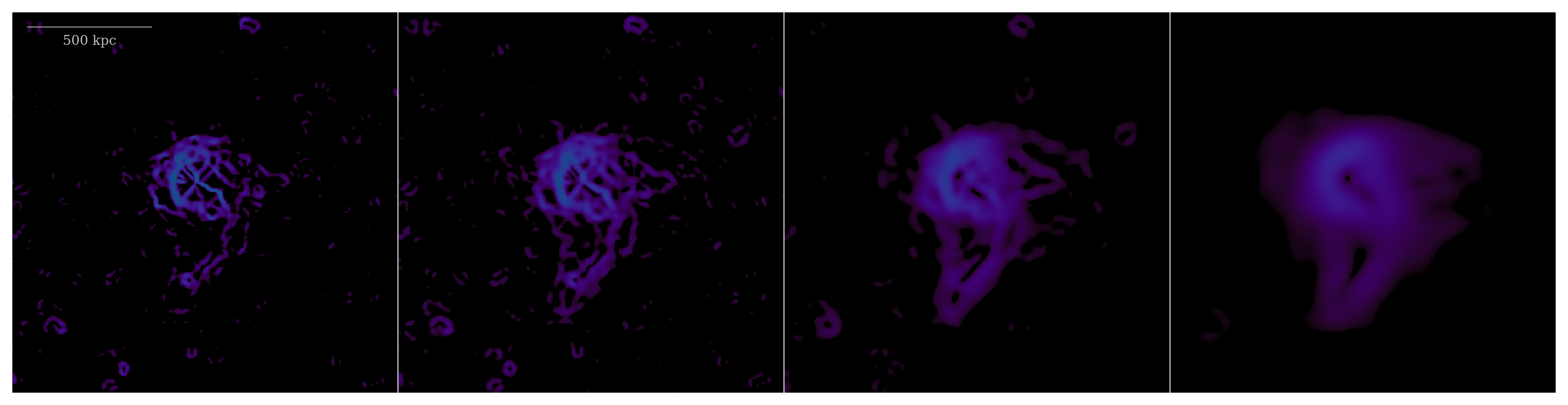}
  \caption{Same as Fig.~\ref{fig:bullet_meerkat} but for J0232.2-4420.}
  \label{fig:J0232.2-4420_meerkat}
\end{figure}

\begin{figure}
  \centering
  \includegraphics[width=.348\hsize,trim={0cm 0cm 0cm 0cm},clip,valign=c]{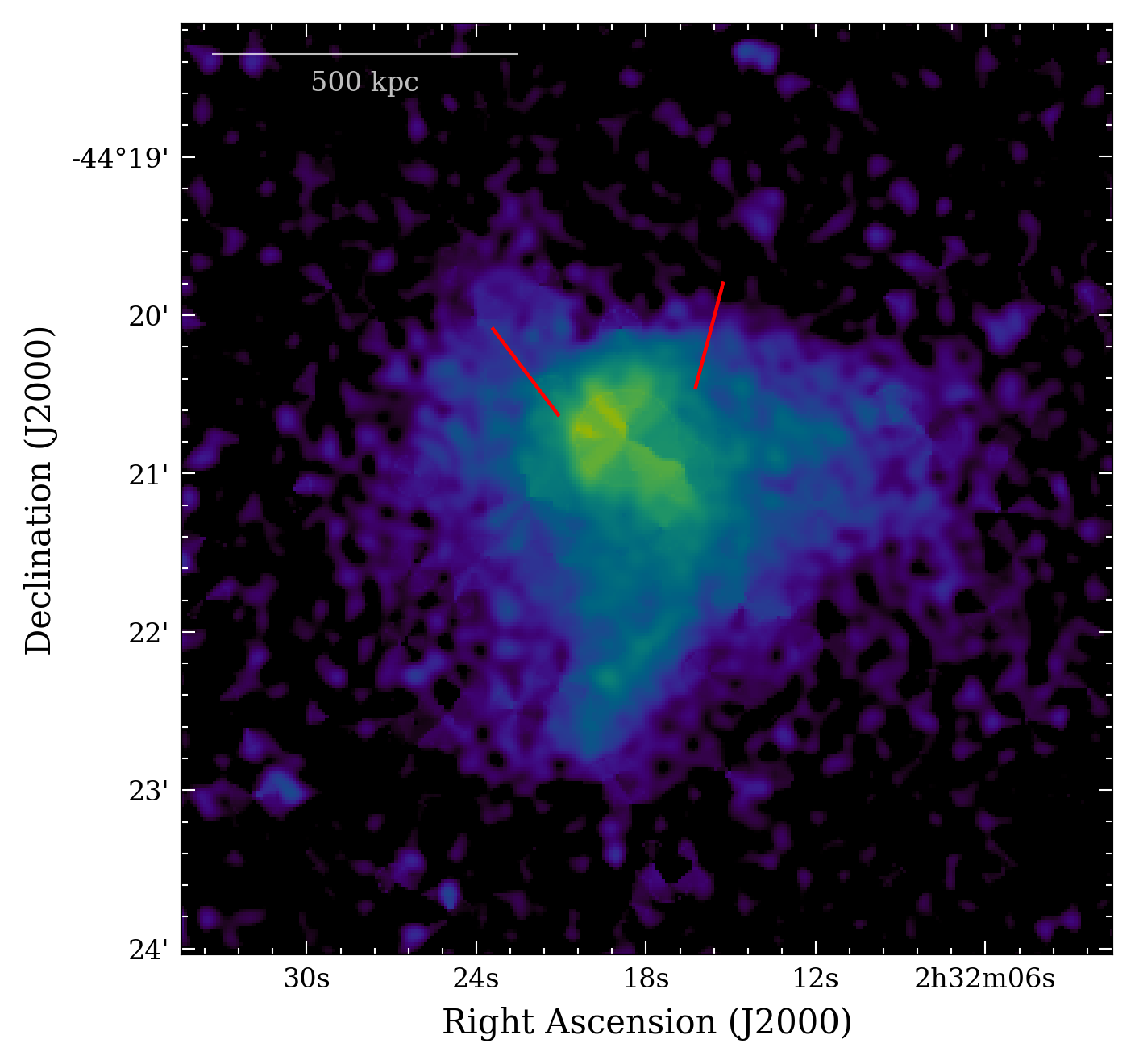}
  \includegraphics[width=.32\hsize,trim={0cm 0cm 0cm 0cm},clip,valign=c]{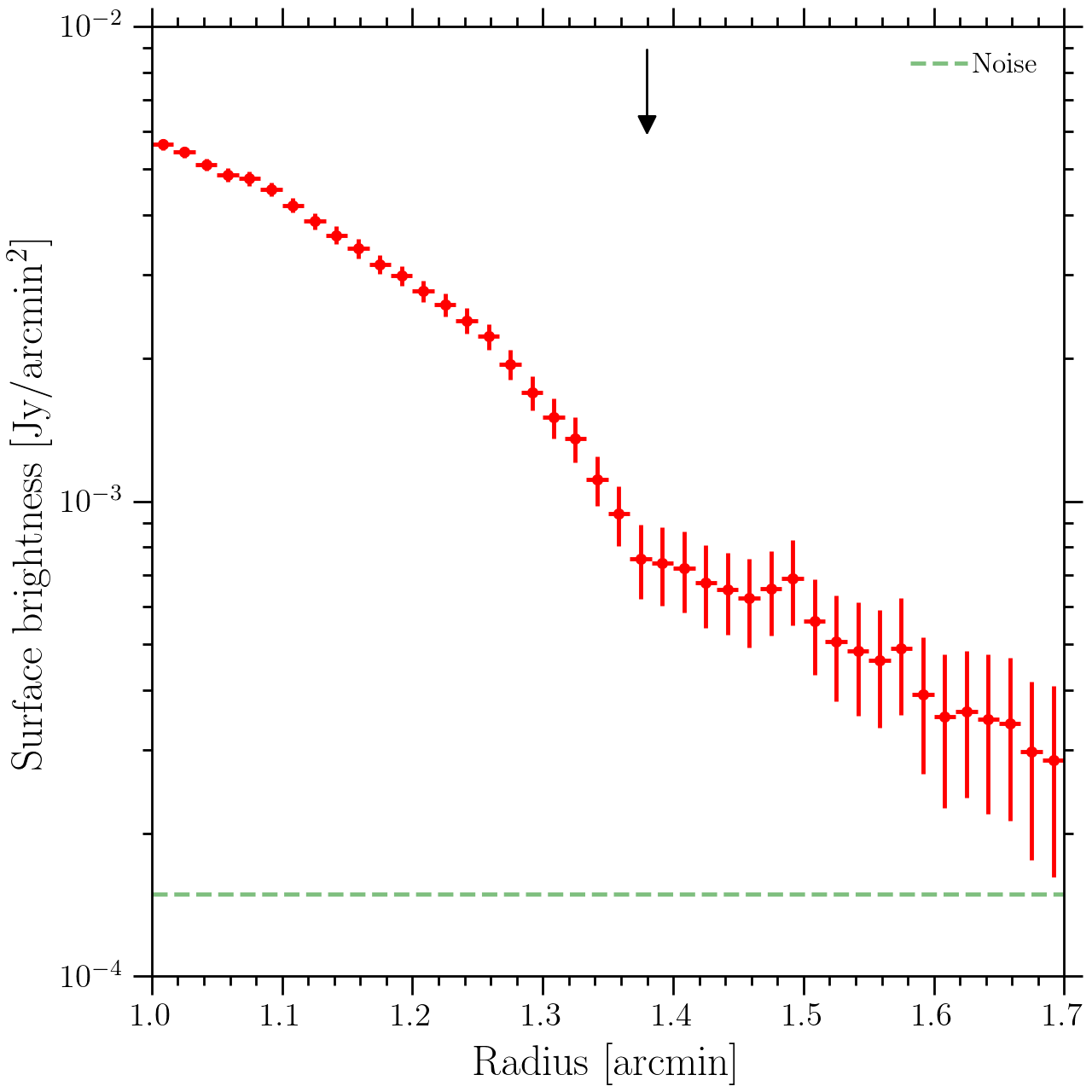}
    \caption{Same as Fig.~\ref{fig:bullet_sb} but for J0232.2-4420.}
  \label{fig:J0232.2-4420_sb}
\end{figure}

\begin{figure}
  \centering
  \includegraphics[width=\hsize,trim={0cm 0cm 0cm 0cm},clip,valign=c]{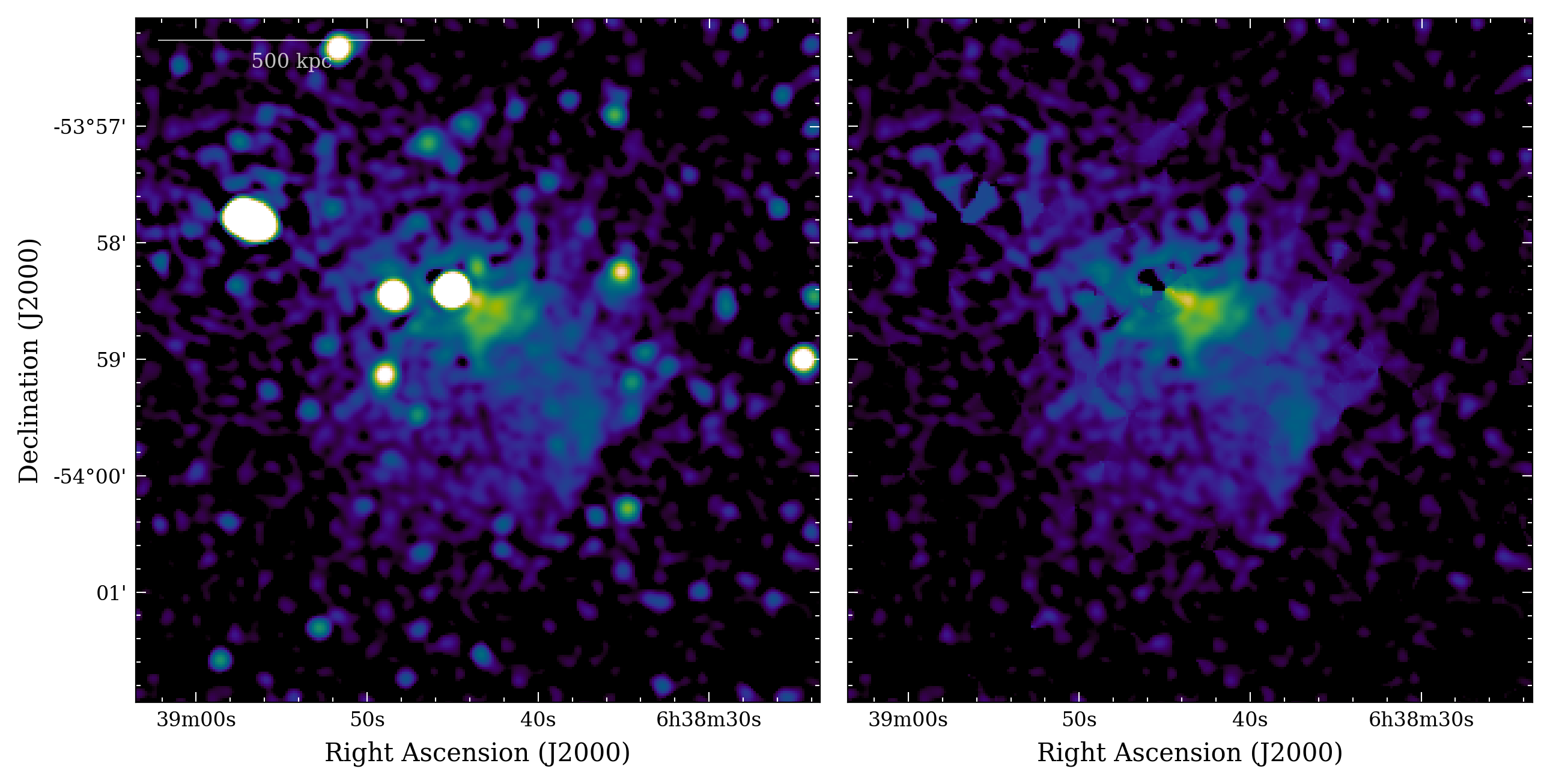}
  \includegraphics[width=\hsize,trim={0cm 0cm 0cm 0cm},clip,valign=c]{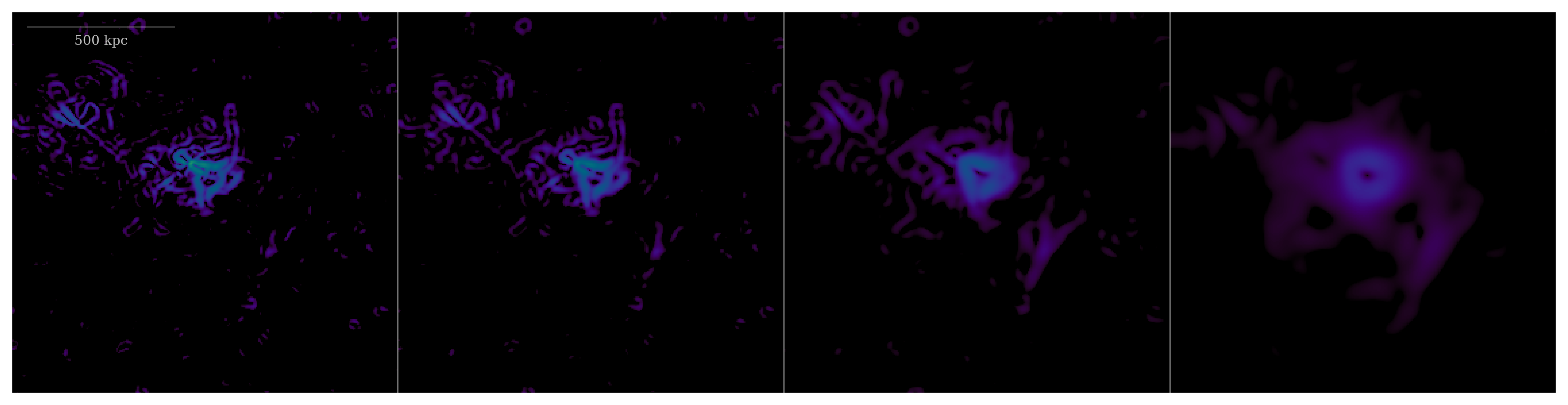}
  \caption{Same as Fig.~\ref{fig:bullet_meerkat} but for J0638.7-5358.}
  \label{fig:J0638.7-5358_meerkat}
\end{figure}

\begin{figure}
  \centering
  \includegraphics[width=.348\hsize,trim={0cm 0cm 0cm 0cm},clip,valign=c]{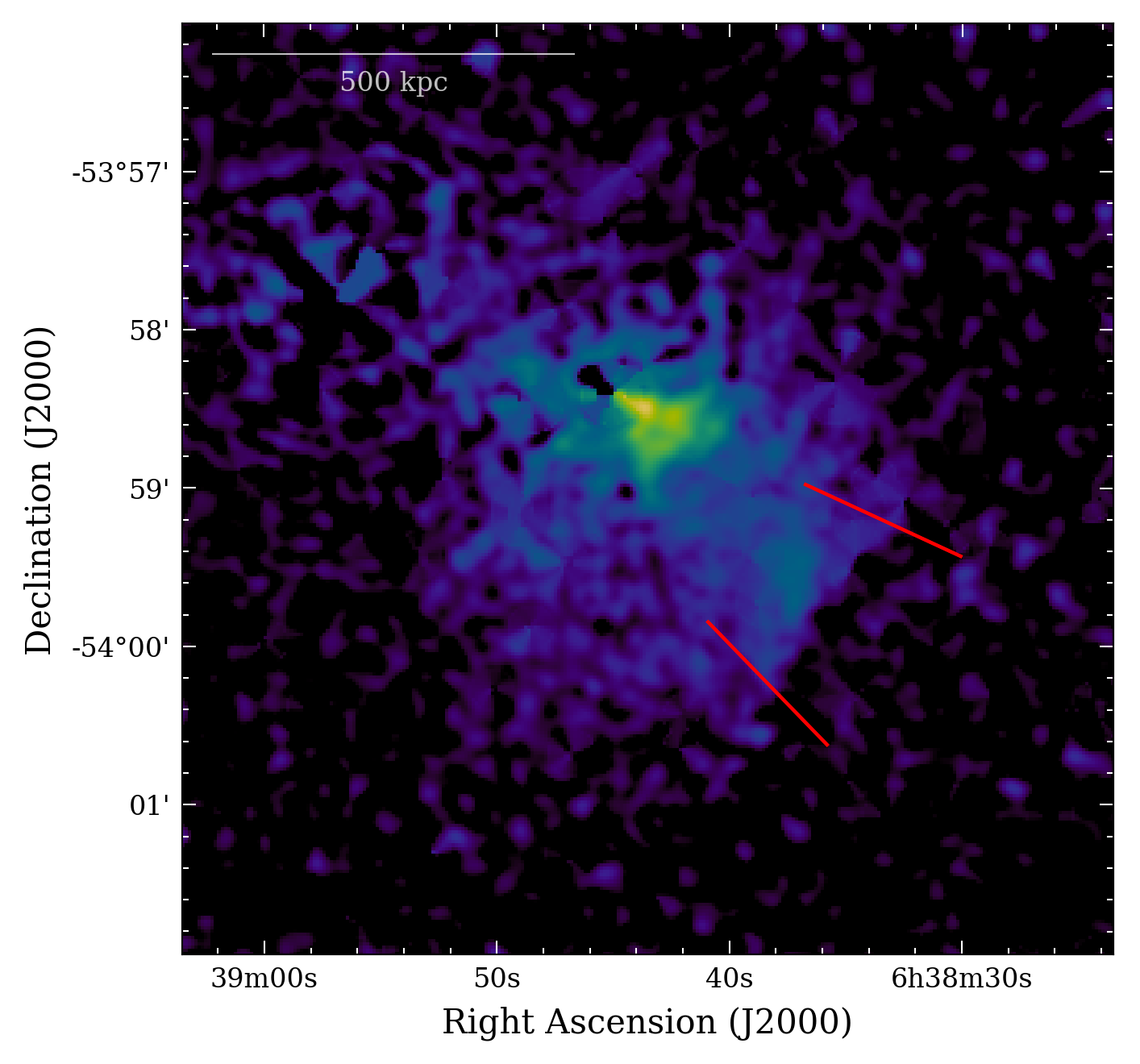}
  \includegraphics[width=.32\hsize,trim={0cm 0cm 0cm 0cm},clip,valign=c]{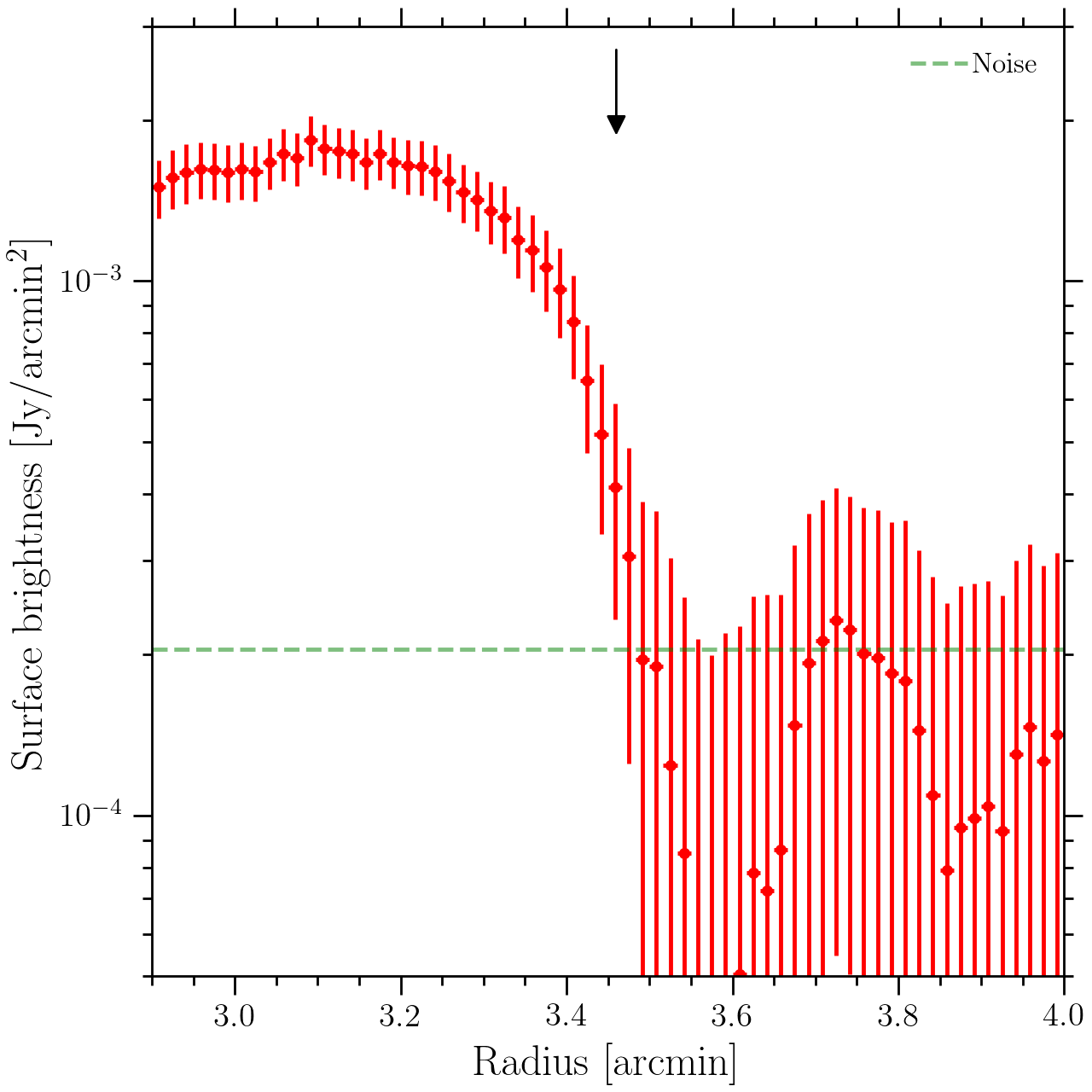}
  \caption{Same as Fig.~\ref{fig:bullet_sb} but for J0638.7-5358.}
  \label{fig:J0638.7-5358_sb}
\end{figure}

\begin{figure}
  \centering
  \includegraphics[width=\hsize,trim={0cm 0cm 0cm 0cm},clip,valign=c]{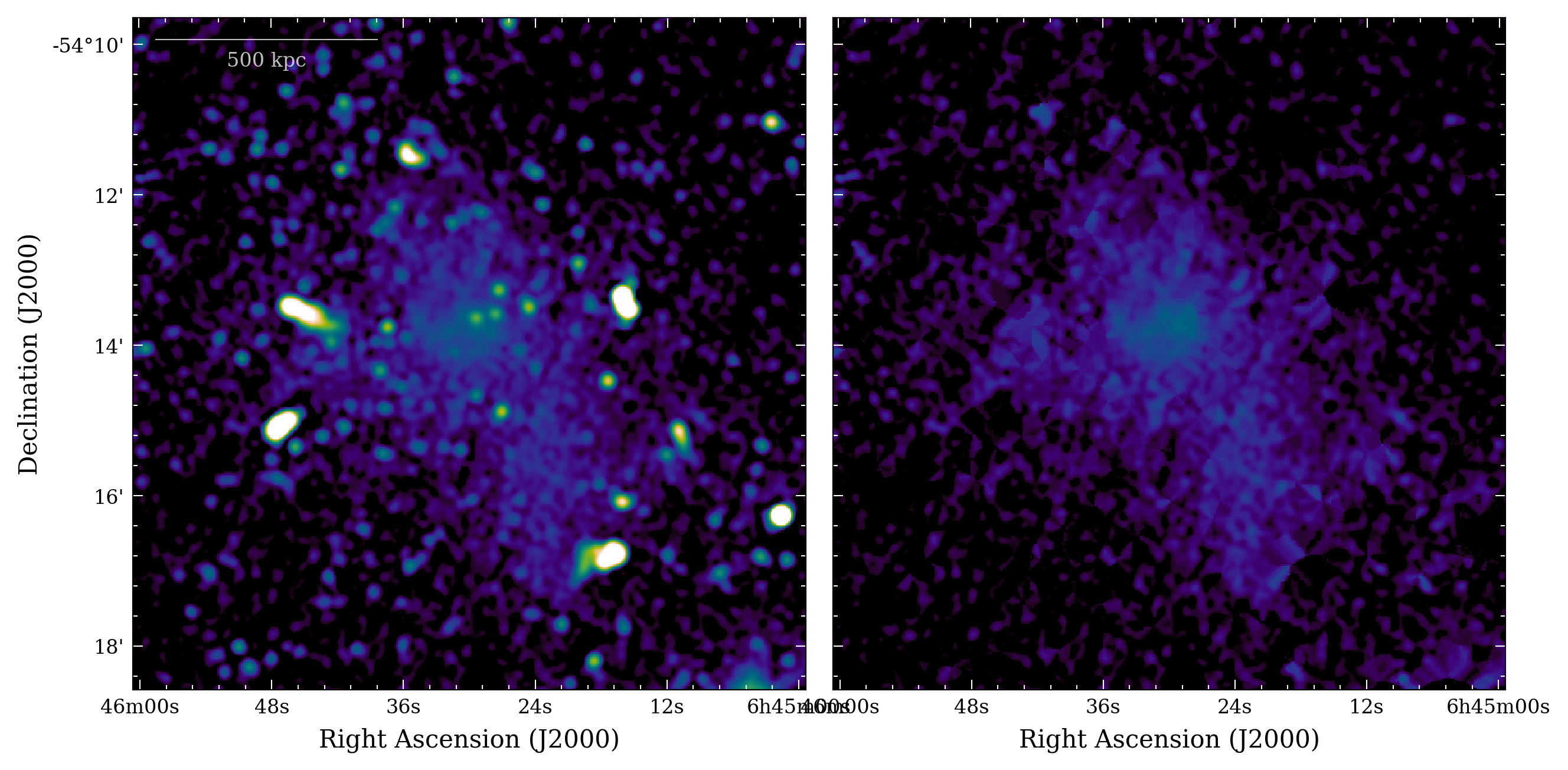}
  \includegraphics[width=\hsize,trim={0cm 0cm 0cm 0cm},clip,valign=c]{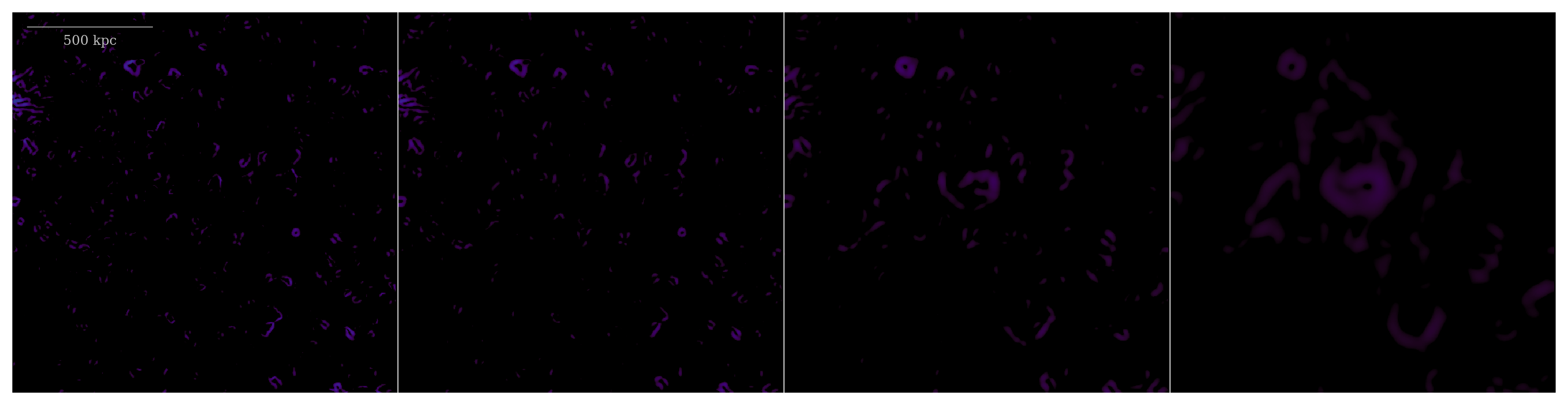}
  \caption{Same as Fig.~\ref{fig:bullet_meerkat} but for J0645.4-5413.}
  \label{fig:J0645.4-5413_meerkat}
\end{figure}

\begin{figure}
  \centering
  \includegraphics[width=.348\hsize,trim={0cm 0cm 0cm 0cm},clip,valign=c]{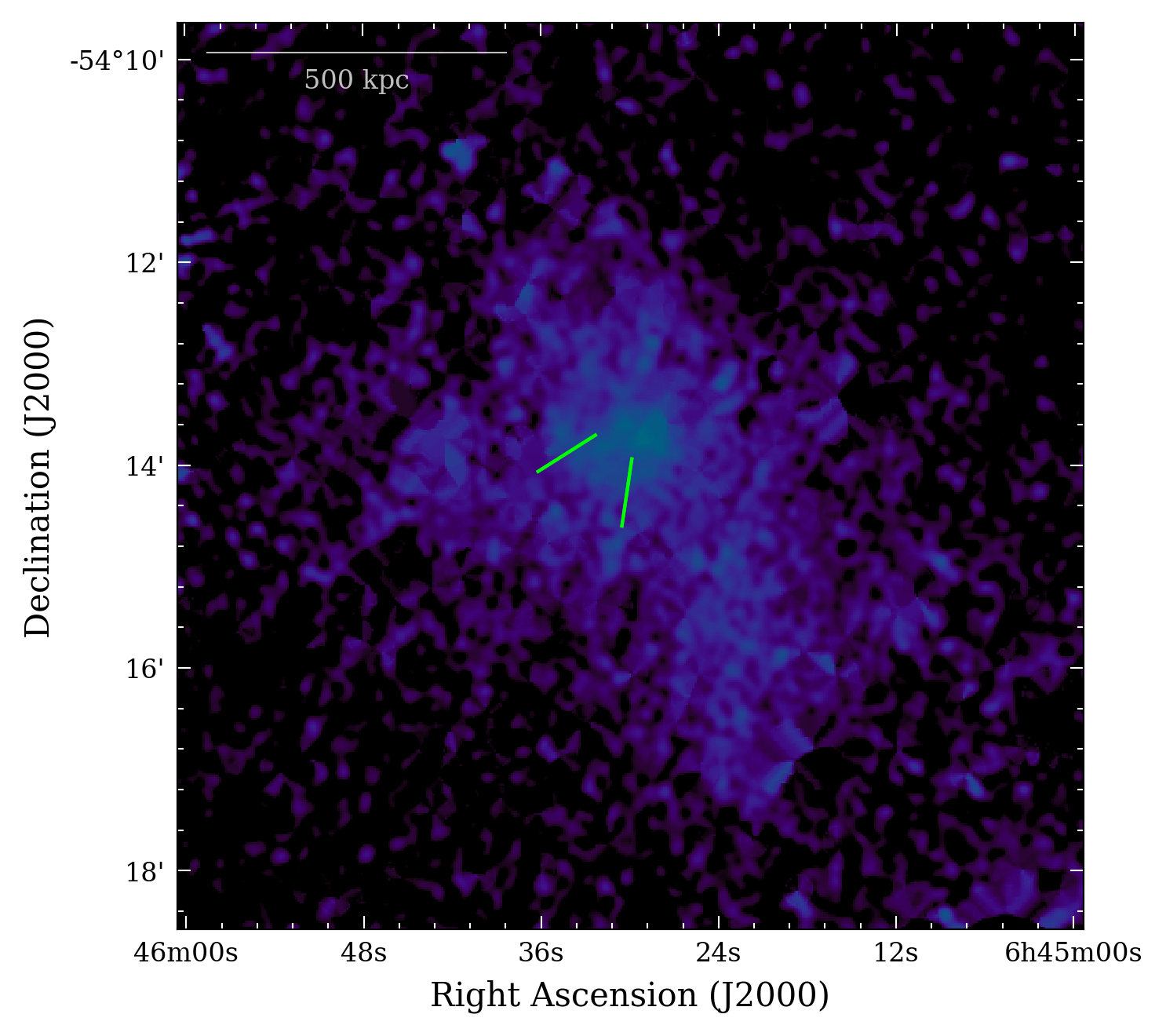}
  \includegraphics[width=.32\hsize,trim={0cm 0cm 0cm 0cm},clip,valign=c]{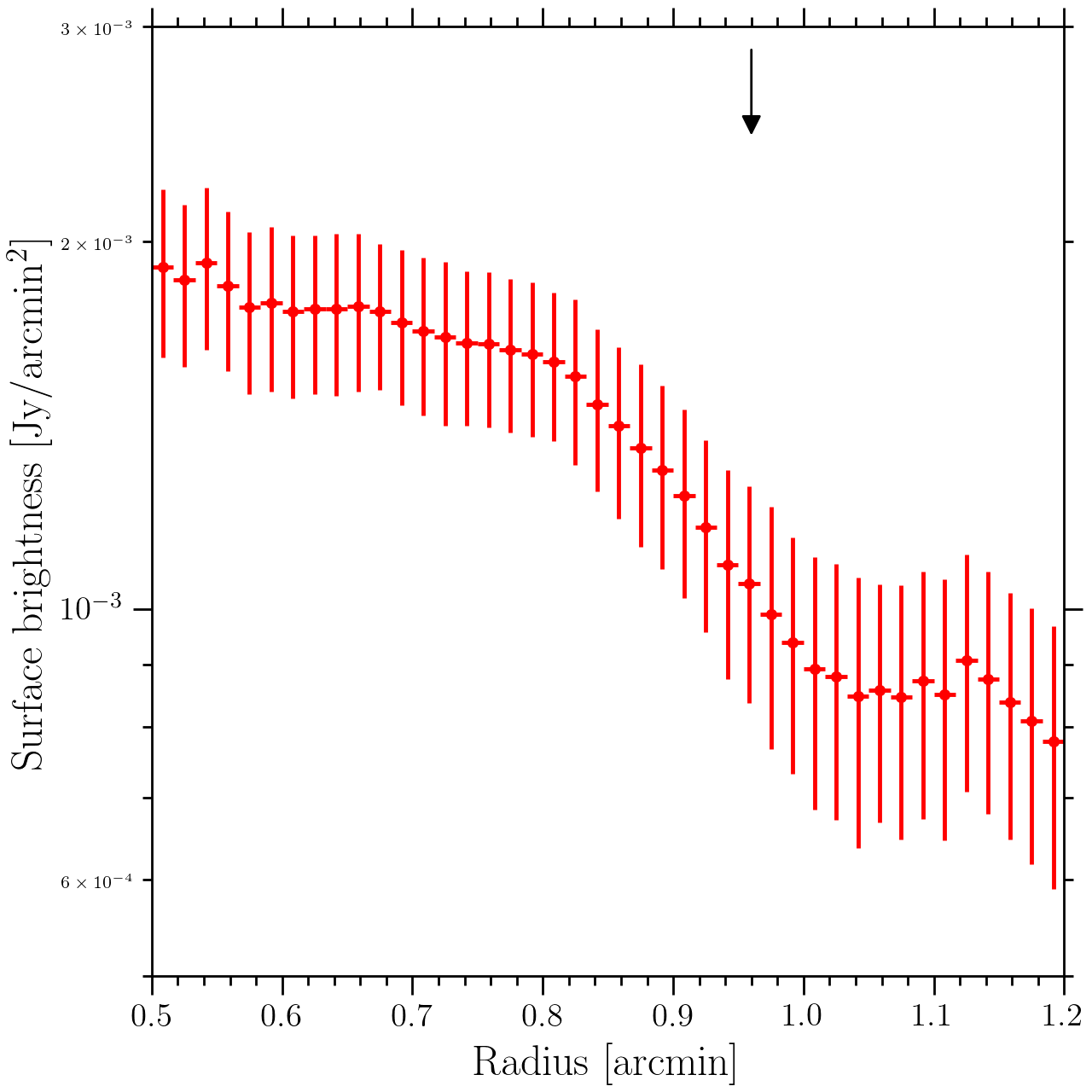}
  \caption{Same as Fig.~\ref{fig:bullet_sb} but for J0645.4-5413.}
  \label{fig:J0645.4-5413_sb}
\end{figure}

\begin{figure}
  \centering
  \includegraphics[width=\hsize,trim={0cm 0cm 0cm 0cm},clip,valign=c]{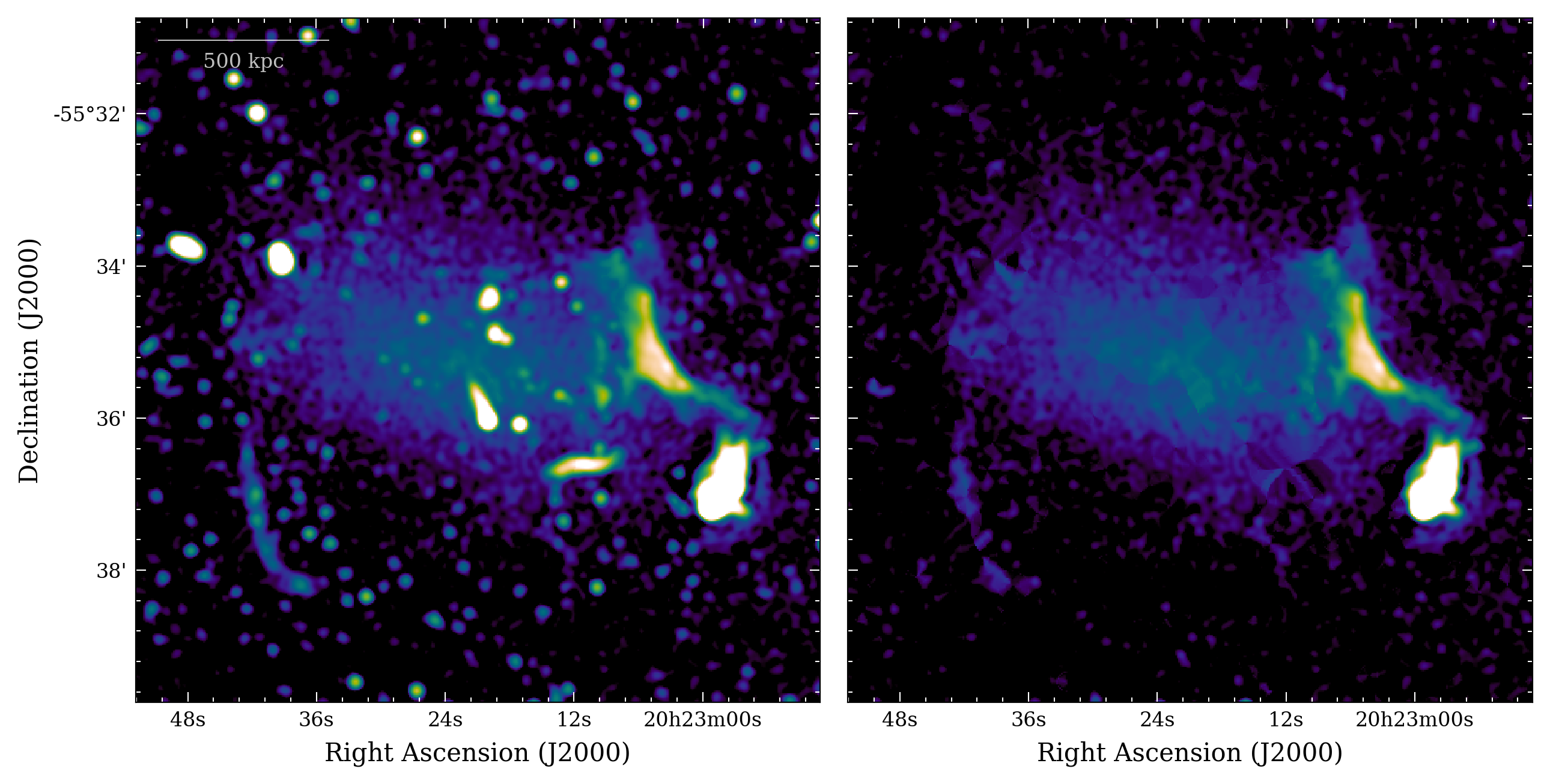}
  \includegraphics[width=\hsize,trim={0cm 0cm 0cm 0cm},clip,valign=c]{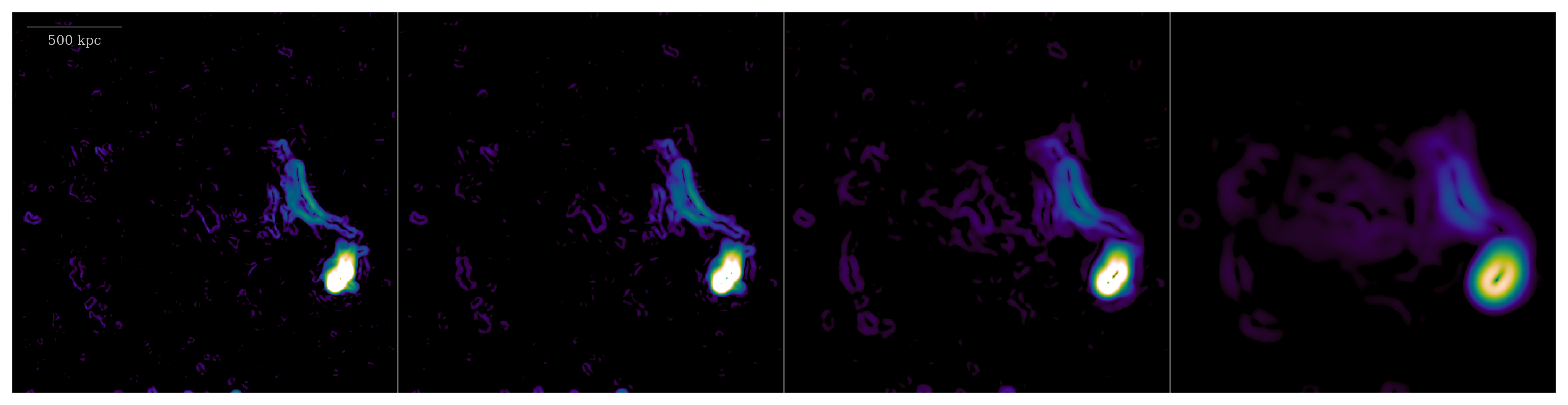}
  \caption{Same as Fig.~\ref{fig:bullet_meerkat} but for J2023.4-5535.}
  \label{fig:J2023.4-5535_meerkat}
\end{figure}

\begin{figure}
  \centering
  \includegraphics[width=.348\hsize,trim={0cm 0cm 0cm 0cm},clip,valign=c]{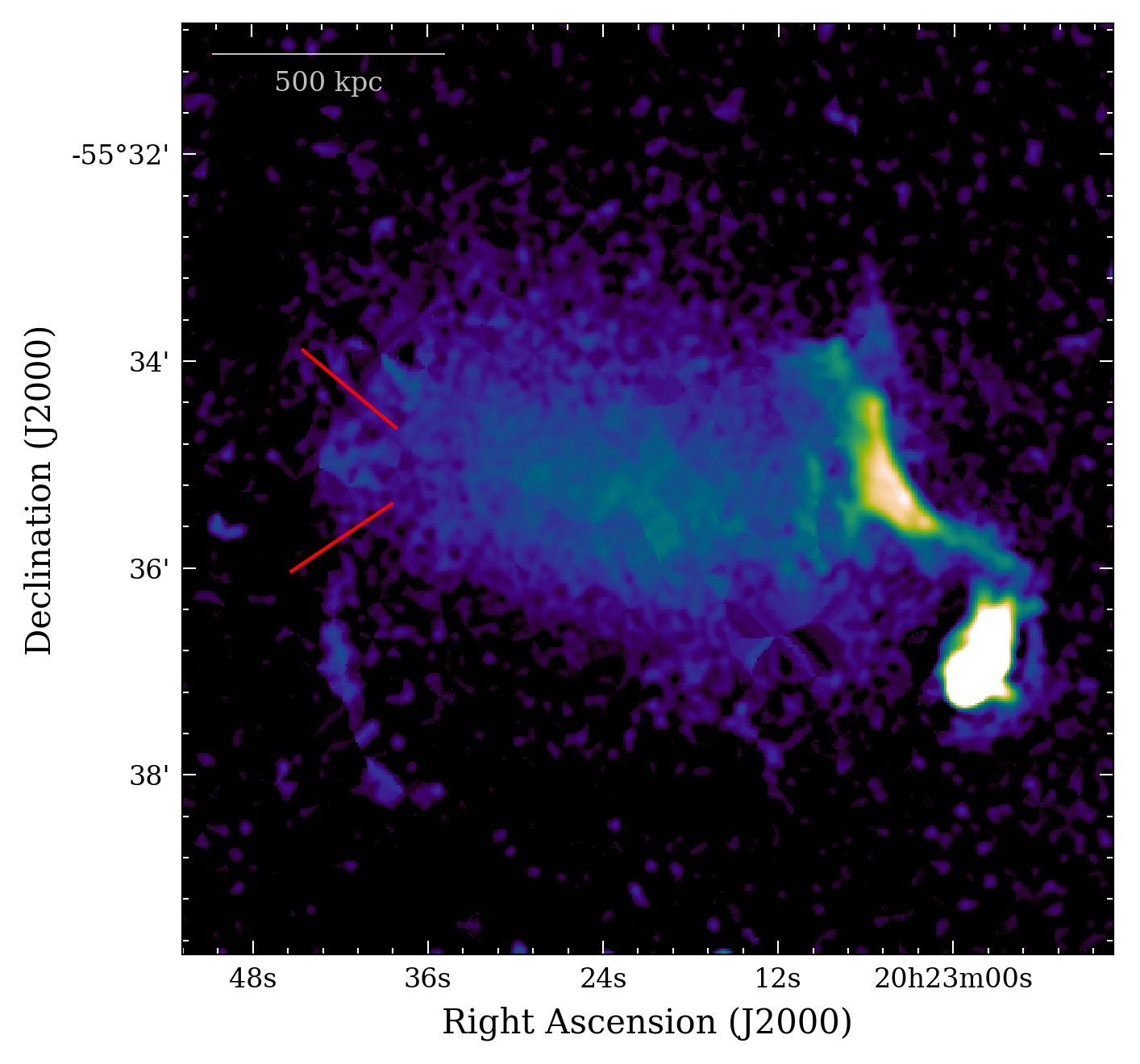}
  \includegraphics[width=.32\hsize,trim={0cm 0cm 0cm 0cm},clip,valign=c]{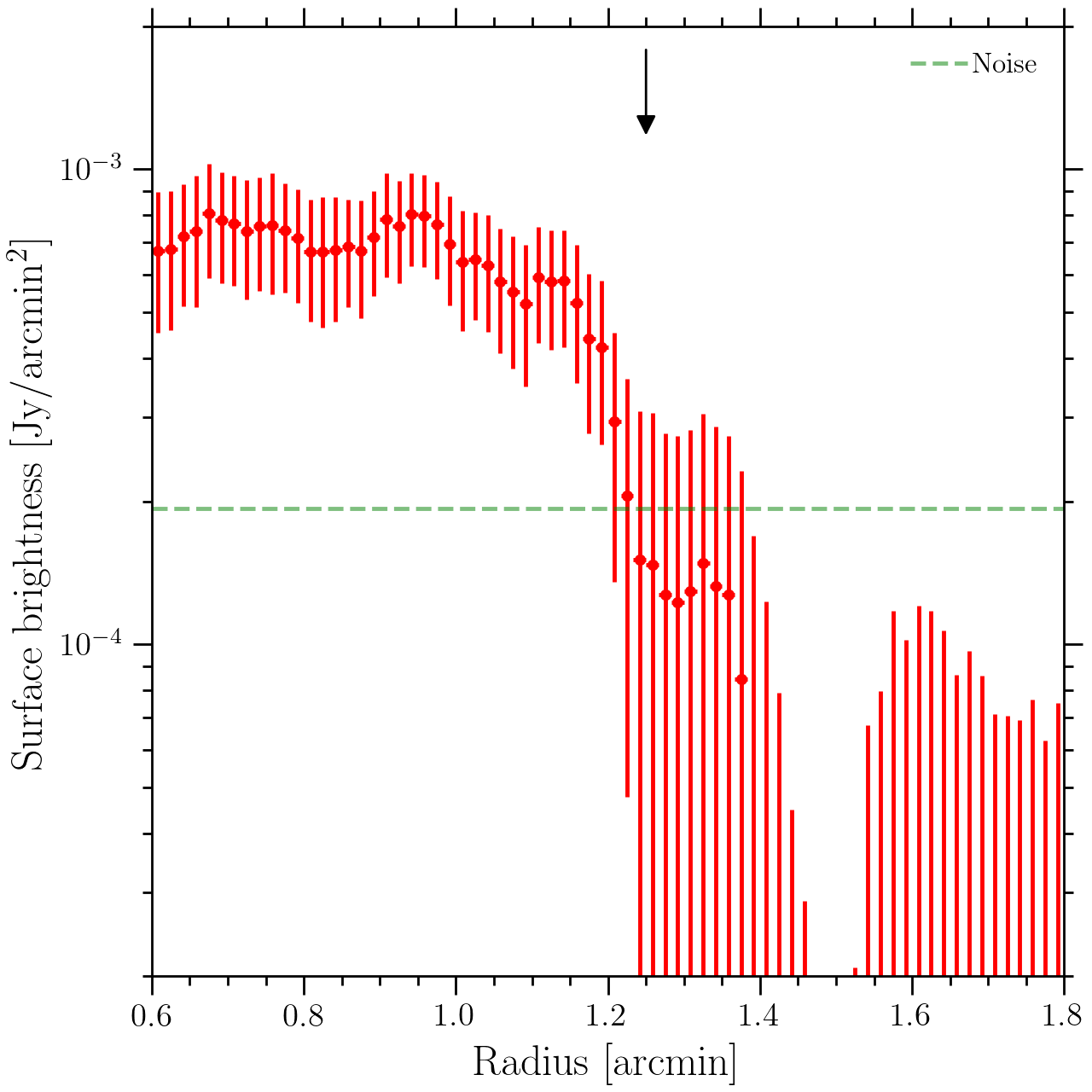}
  \caption{Same as Fig.~\ref{fig:bullet_sb} but for J2023.4-5535.}
  \label{fig:J2023.4-5535_sb}
\end{figure}
\FloatBarrier

\section{Radio halos in MGCLS without radio surface brightness gradients}\label{app:no_edges}

The radio halos where we did not identify any radio surface brightness gradient are generally characterized by low surface brightness. In a few cases, residual calibration errors (e.g.,\ Abell 545, Abell 3558, Abell S295) or artifacts introduced by the procedure of discrete source subtraction (e.g.,\ Abell 2811, J0303.7-7752) are evident in the \meerkat\ images. 

\begin{figure}[h]
  \centering
  \includegraphics[width=\hsize,trim={0cm 0cm 0cm 0cm},clip,valign=c]{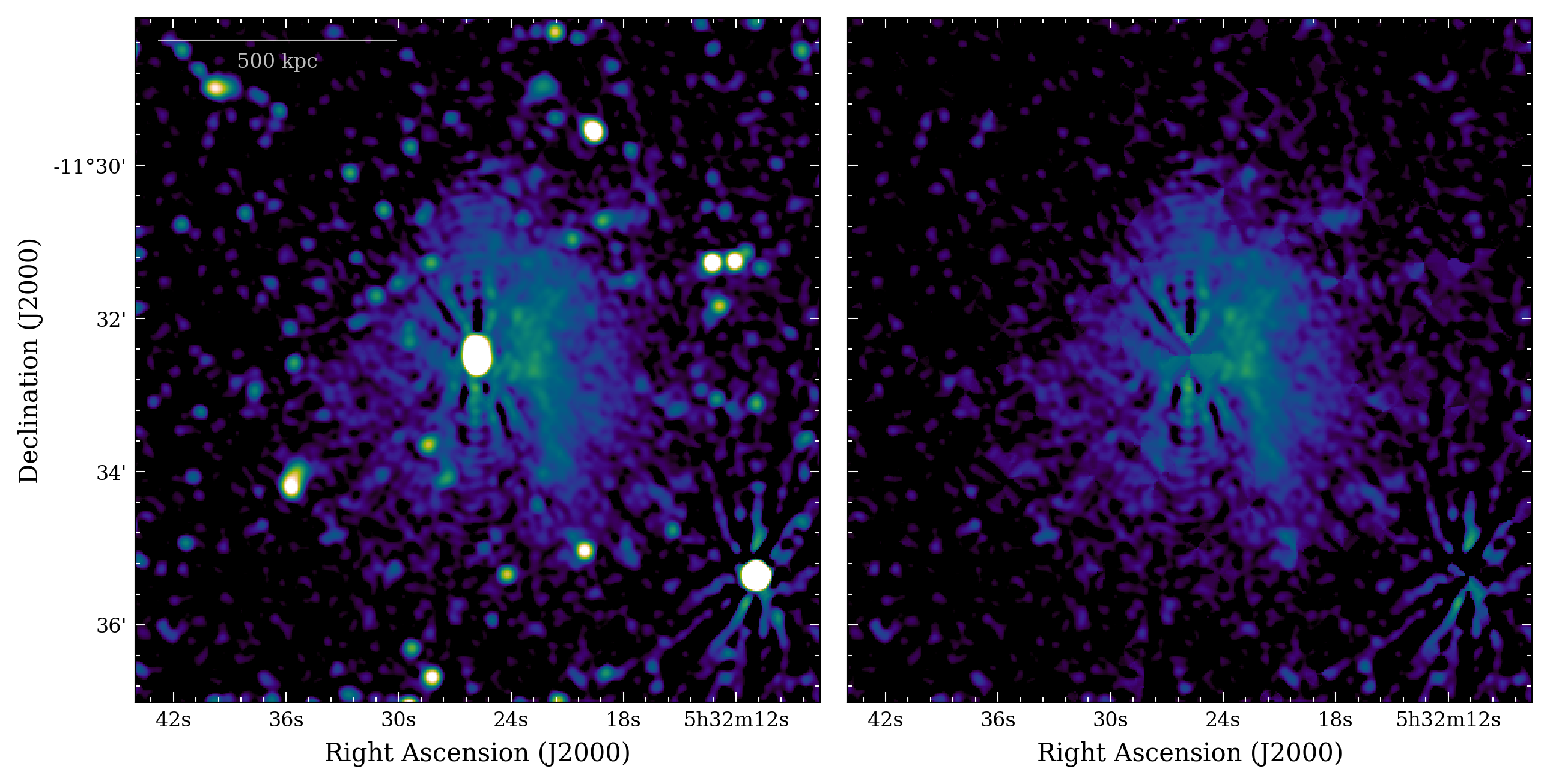}
  \includegraphics[width=\hsize,trim={0cm 0cm 0cm 0cm},clip,valign=c]{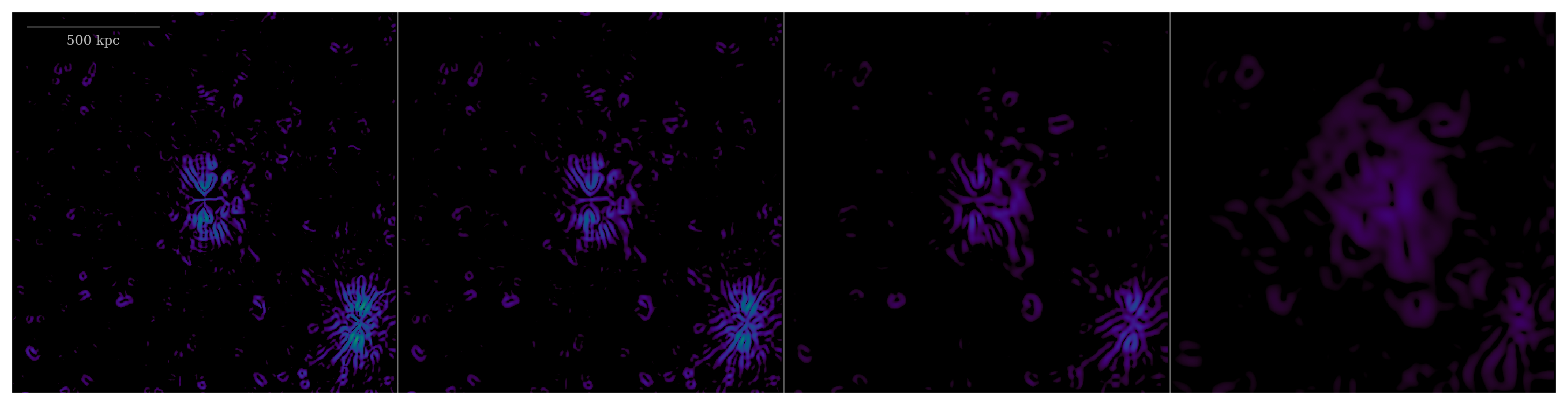}
  \caption{Same as Fig.~\ref{fig:bullet_meerkat} but for Abell 545.}
  \label{fig:Abell_545_meerkat}
\end{figure}

\begin{figure}
  \centering
  \includegraphics[width=\hsize,trim={0cm 0cm 0cm 0cm},clip,valign=c]{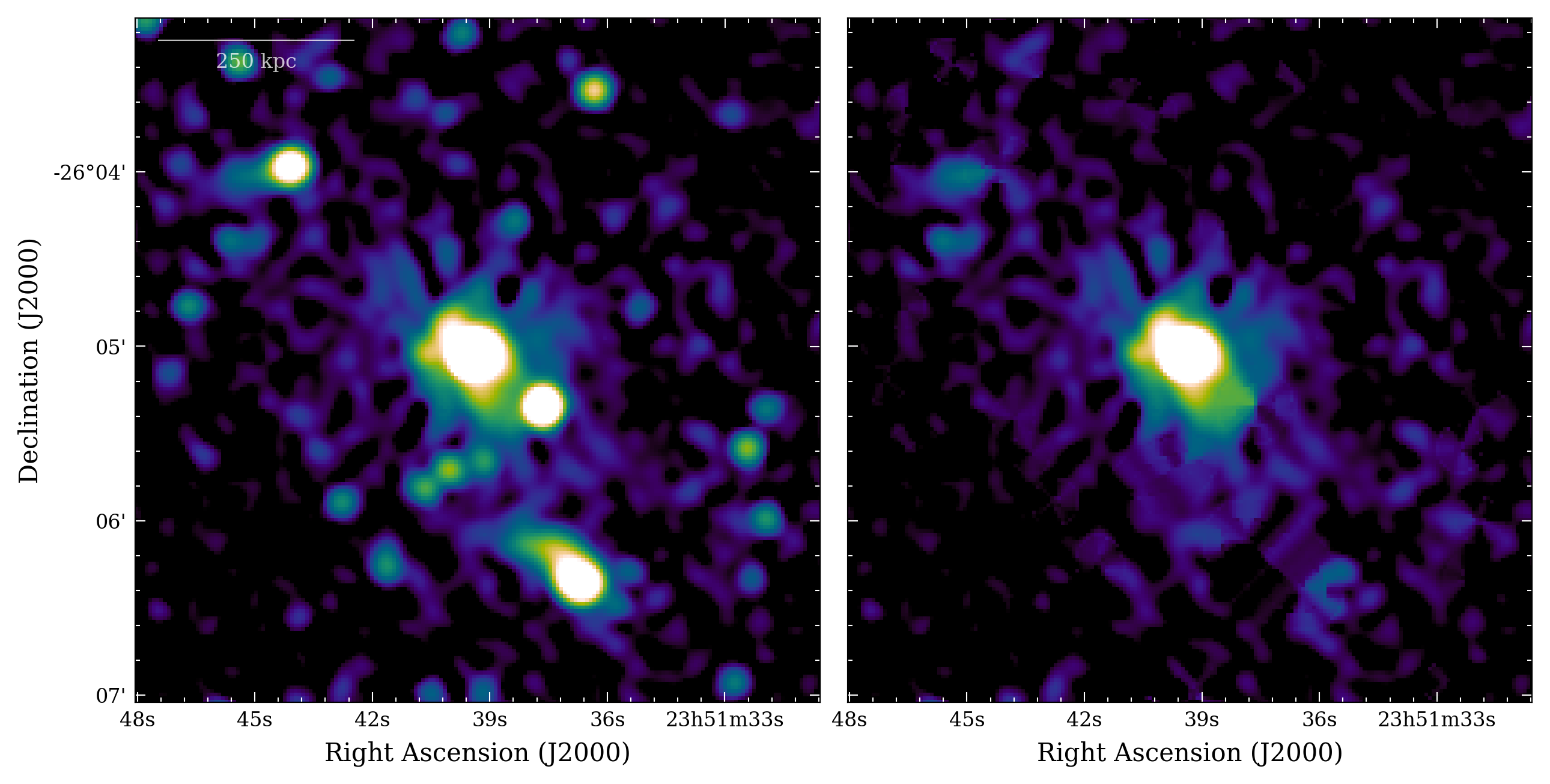}
  \includegraphics[width=\hsize,trim={0cm 0cm 0cm 0cm},clip,valign=c]{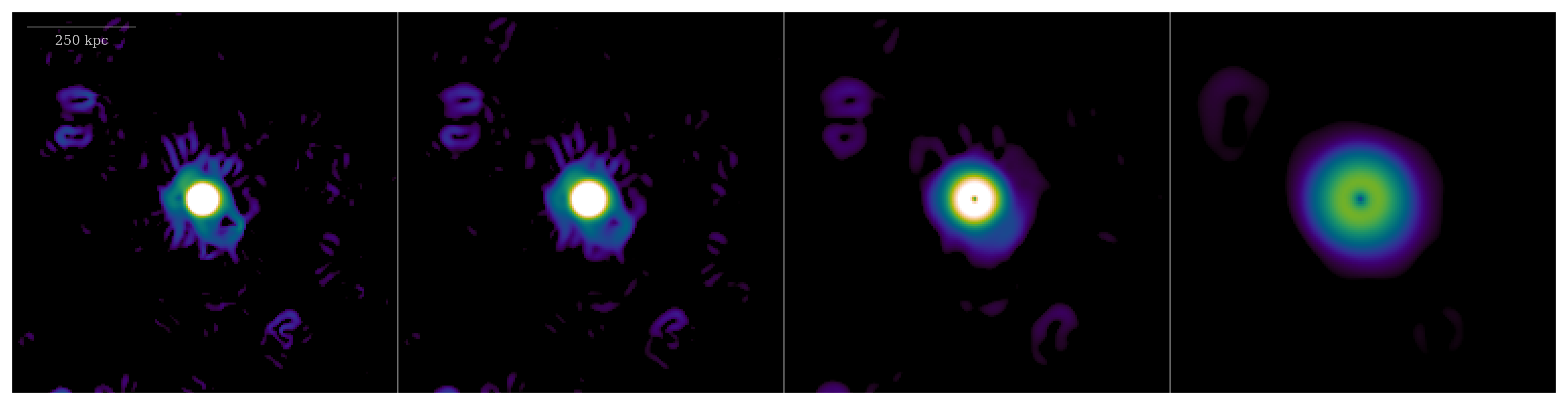}
  \caption{Same as Fig.~\ref{fig:bullet_meerkat} but for Abell 2667.}
  \label{fig:Abell_2667_meerkat}
\end{figure}

\begin{figure}
  \centering
  \includegraphics[width=\hsize,trim={0cm 0cm 0cm 0cm},clip,valign=c]{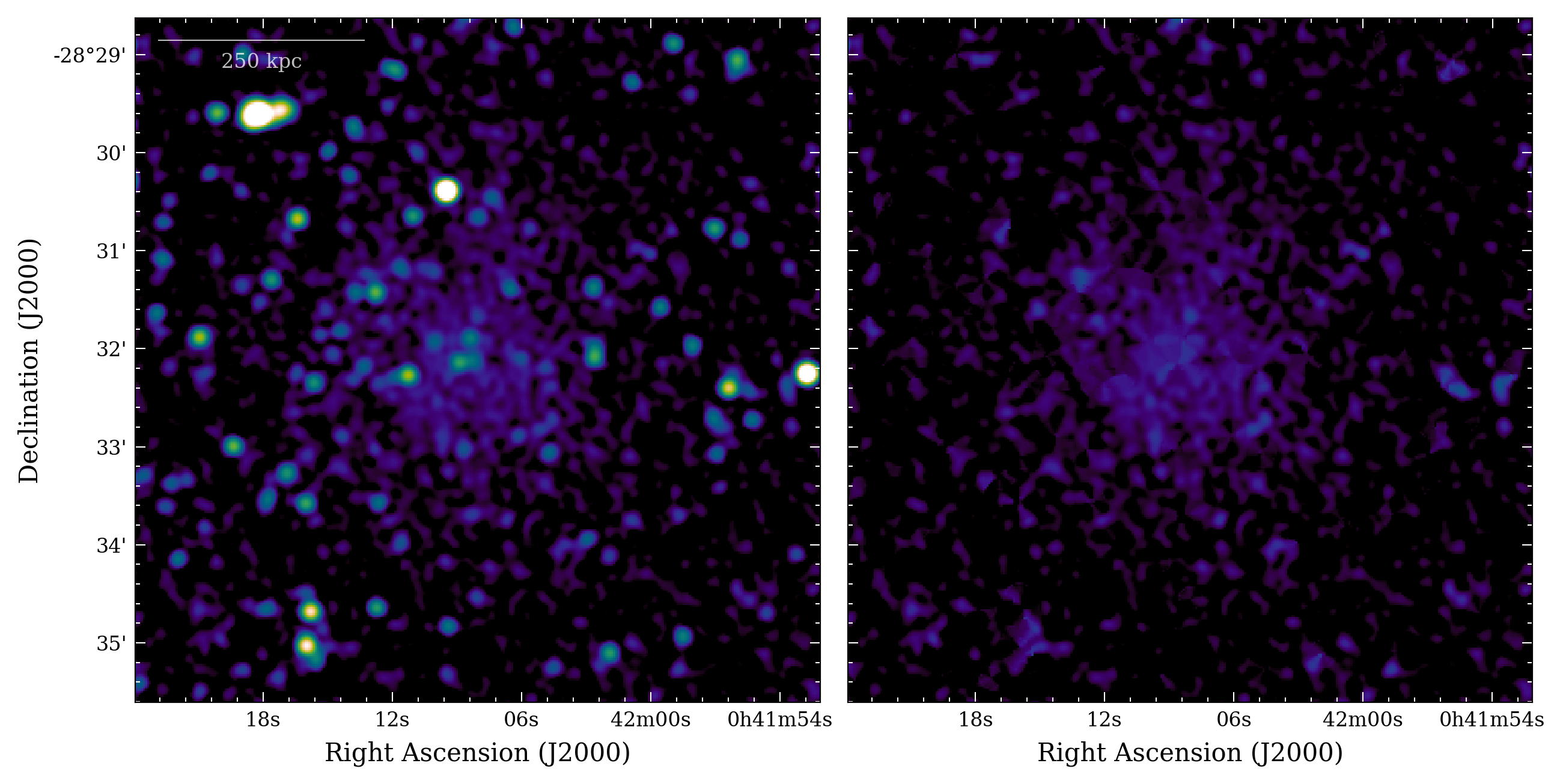}
  \includegraphics[width=\hsize,trim={0cm 0cm 0cm 0cm},clip,valign=c]{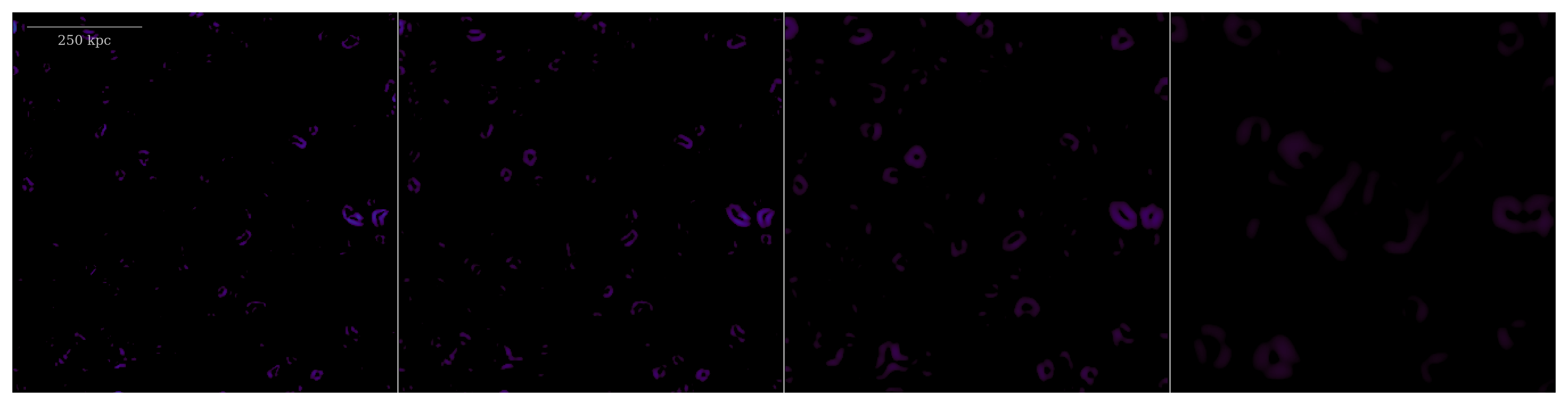}
  \caption{Same as Fig.~\ref{fig:bullet_meerkat} but for Abell 2811.}
  \label{fig:Abell_2811_meerkat}
\end{figure}

\begin{figure}
  \centering
  \includegraphics[width=\hsize,trim={0cm 0cm 0cm 0cm},clip,valign=c]{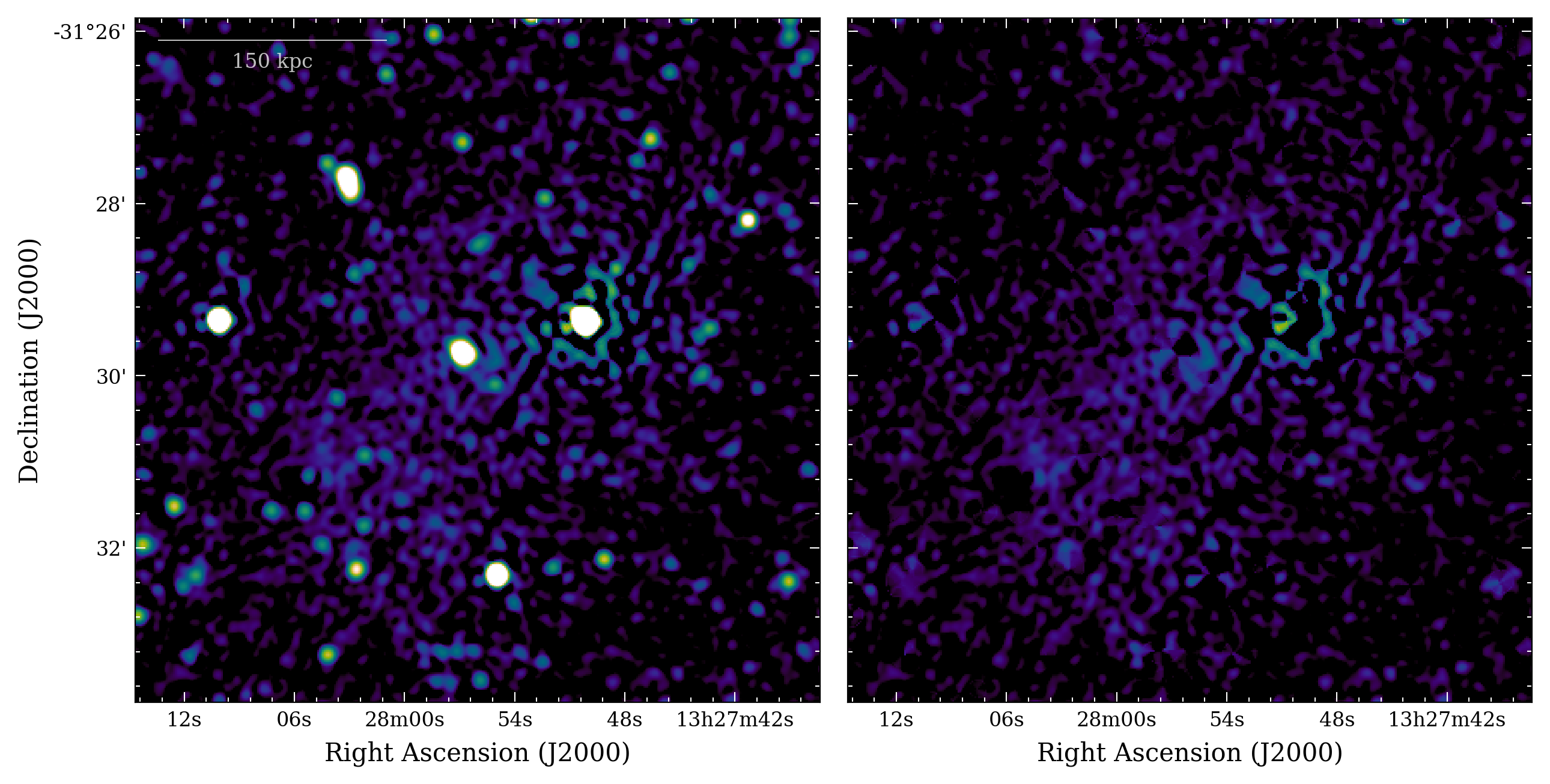}
  \includegraphics[width=\hsize,trim={0cm 0cm 0cm 0cm},clip,valign=c]{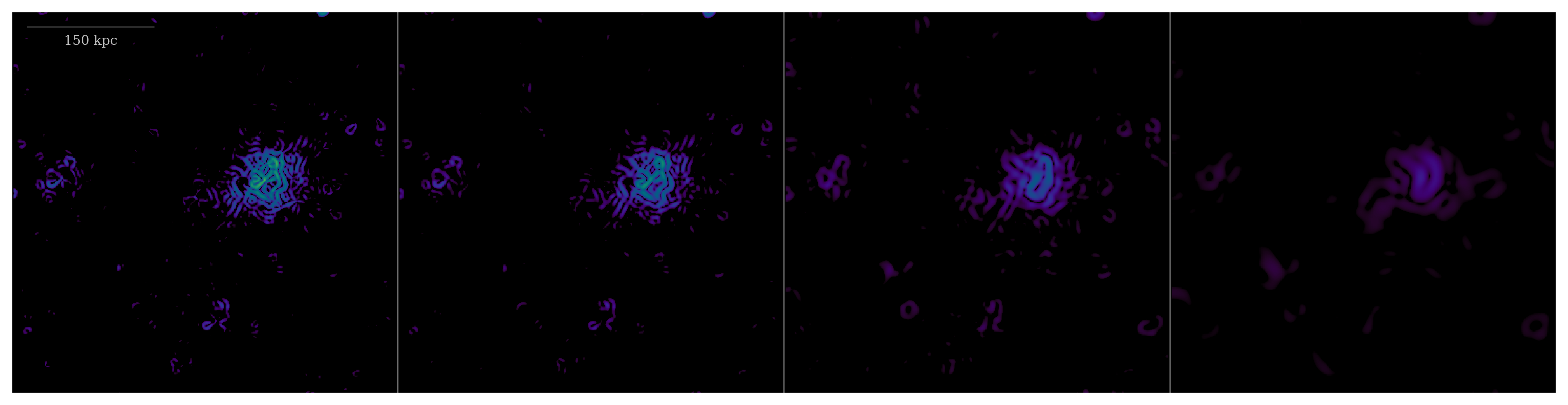}
  \caption{Same as Fig.~\ref{fig:bullet_meerkat} but for Abell 3558.}
  \label{fig:Abell_3558_meerkat}
\end{figure}

\begin{figure}
  \centering
  \includegraphics[width=\hsize,trim={0cm 0cm 0cm 0cm},clip,valign=c]{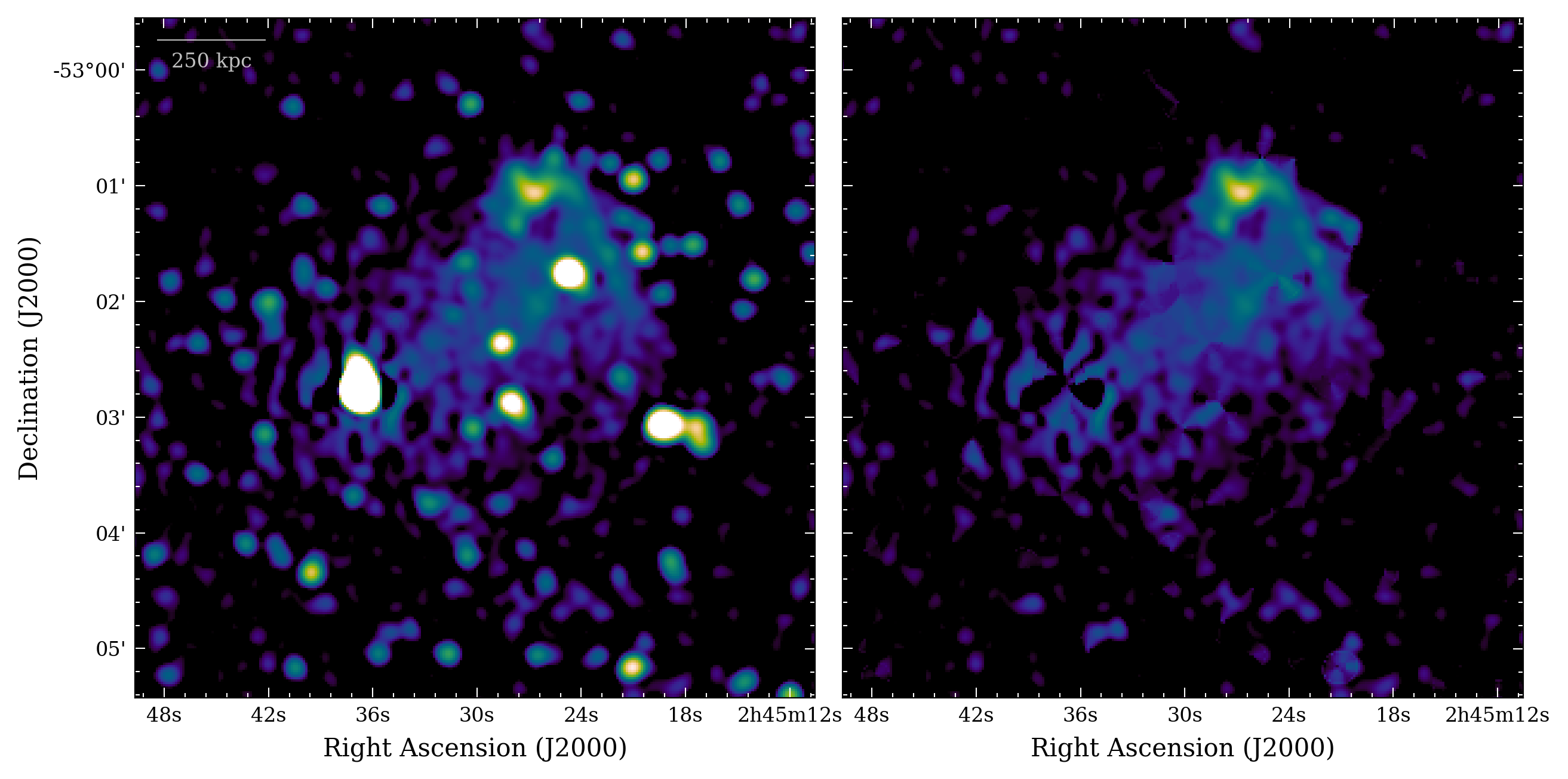}
  \includegraphics[width=\hsize,trim={0cm 0cm 0cm 0cm},clip,valign=c]{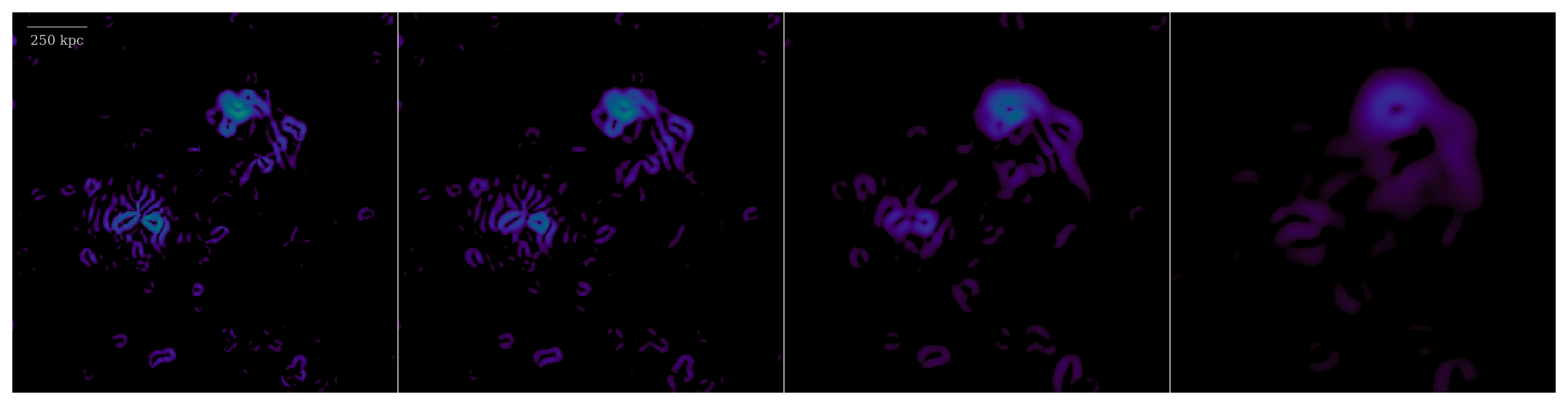}
  \caption{Same as Fig.~\ref{fig:bullet_meerkat} but for Abell S295.}
  \label{fig:Abell_S295_meerkat}
\end{figure}

\begin{figure}
  \centering
  \includegraphics[width=\hsize,trim={0cm 0cm 0cm 0cm},clip,valign=c]{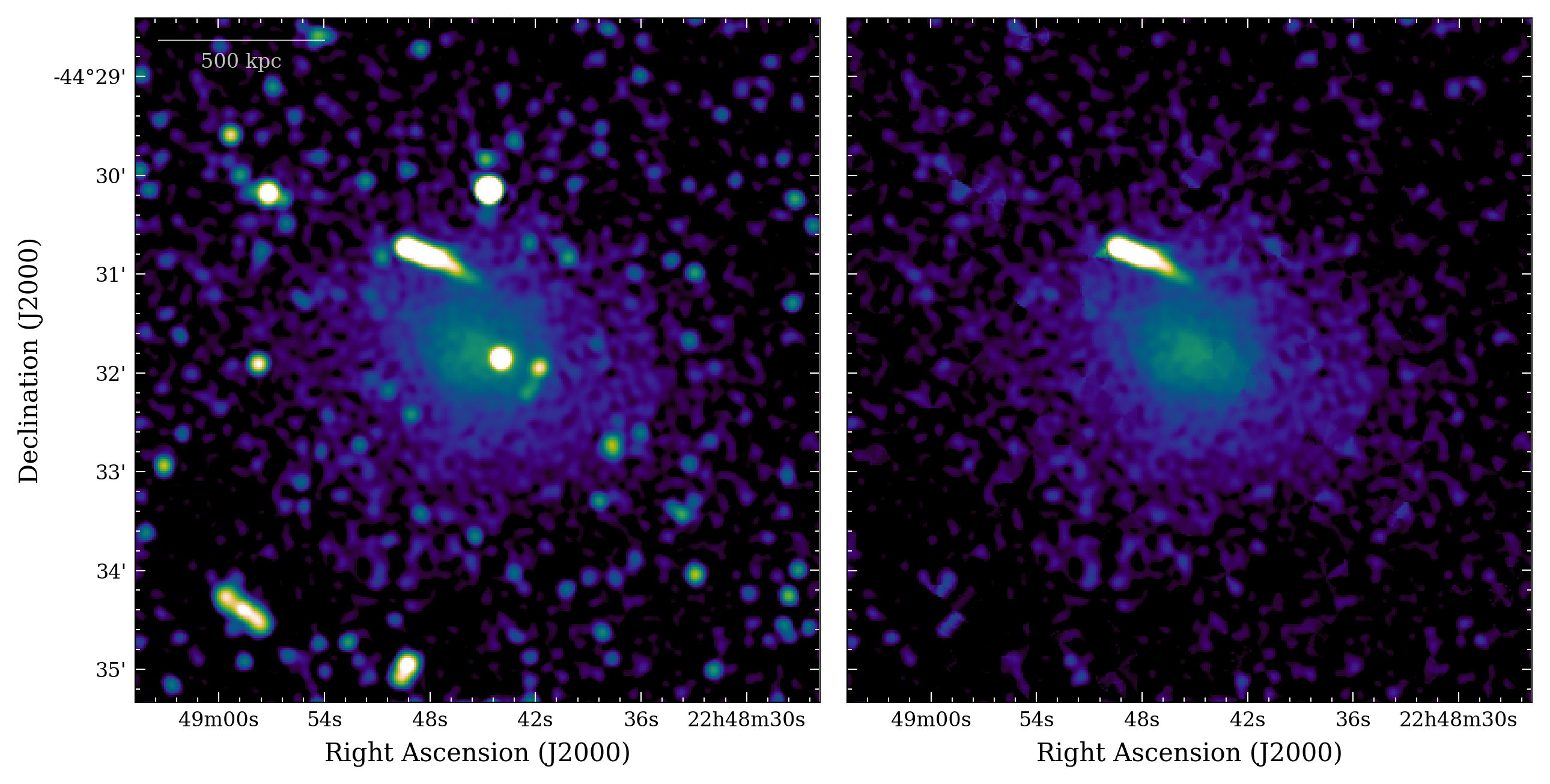}
  \includegraphics[width=\hsize,trim={0cm 0cm 0cm 0cm},clip,valign=c]{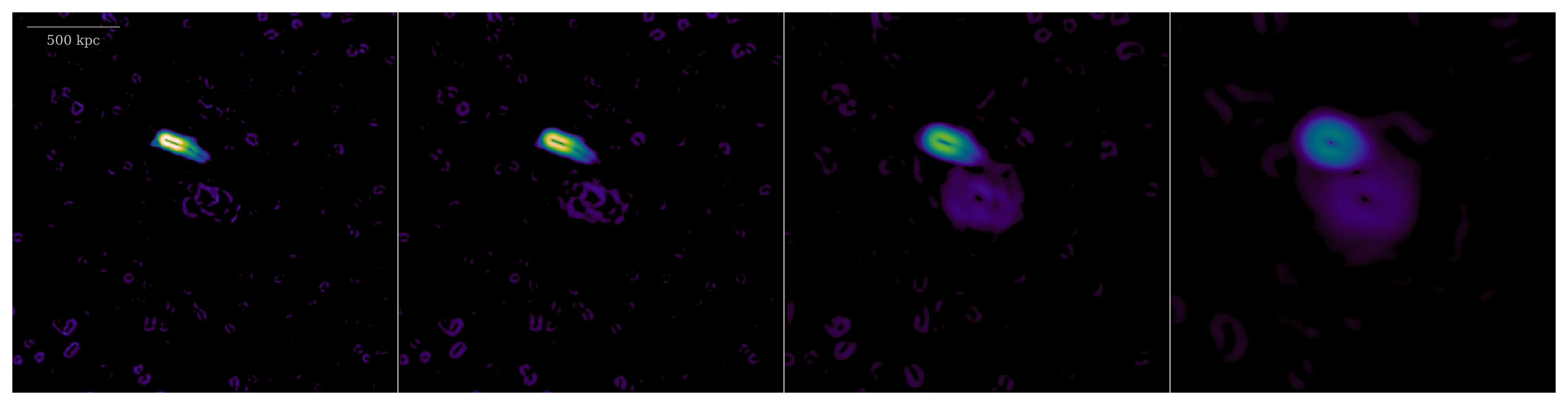}
  \caption{Same as Fig.~\ref{fig:bullet_meerkat} but for Abell S1063.}
  \label{fig:Abell_S1063_meerkat}
\end{figure}

\begin{figure}
  \centering
  \includegraphics[width=\hsize,trim={0cm 0cm 0cm 0cm},clip,valign=c]{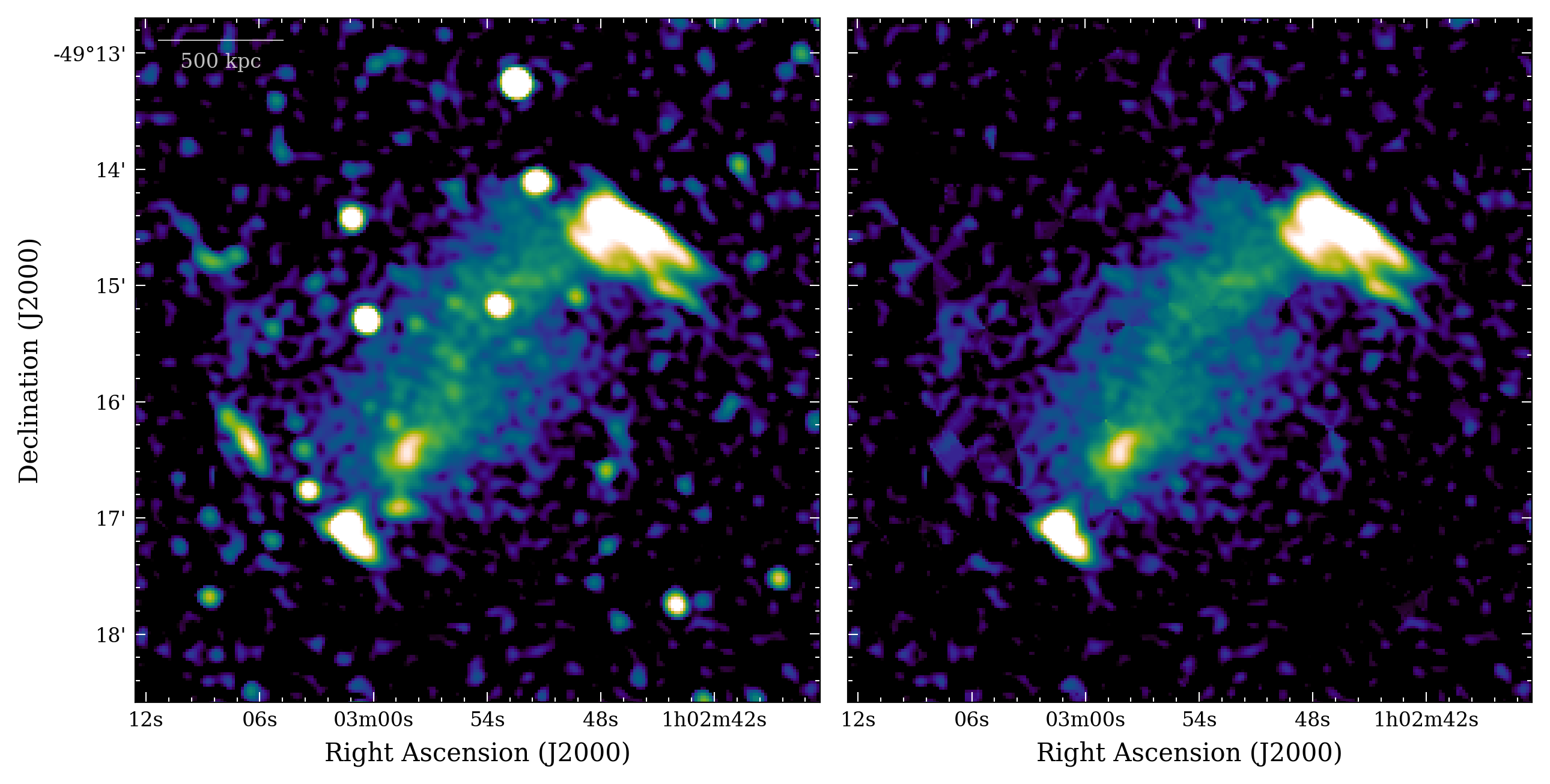}
  \includegraphics[width=\hsize,trim={0cm 0cm 0cm 0cm},clip,valign=c]{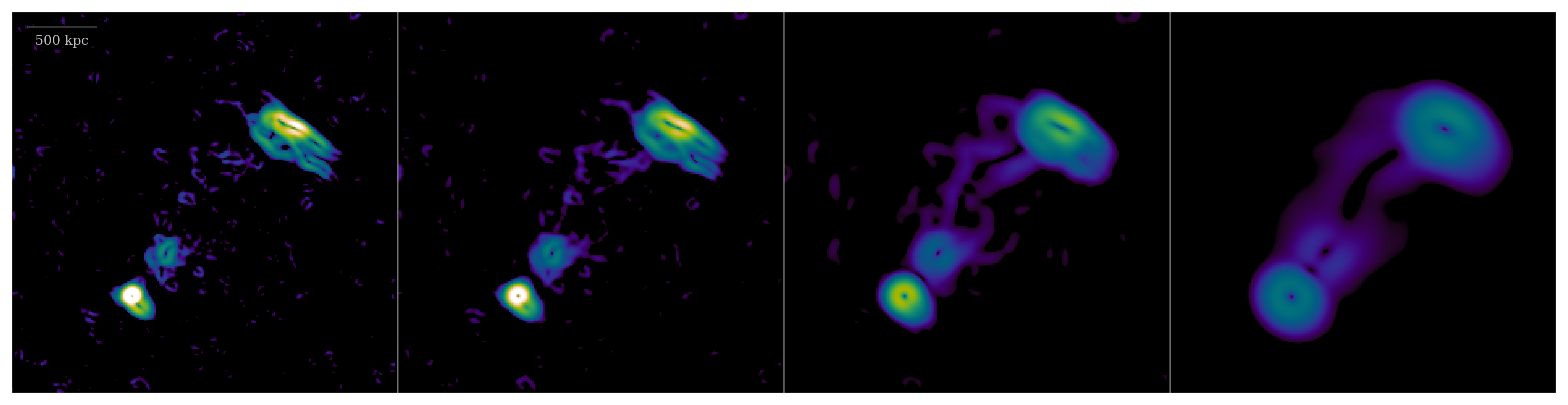}
  \caption{Same as Fig.~\ref{fig:bullet_meerkat} but for El Gordo.}
  \label{fig:ElGordo_meerkat}
\end{figure}

\begin{figure}
  \centering
  \includegraphics[width=\hsize,trim={0cm 0cm 0cm 0cm},clip,valign=c]{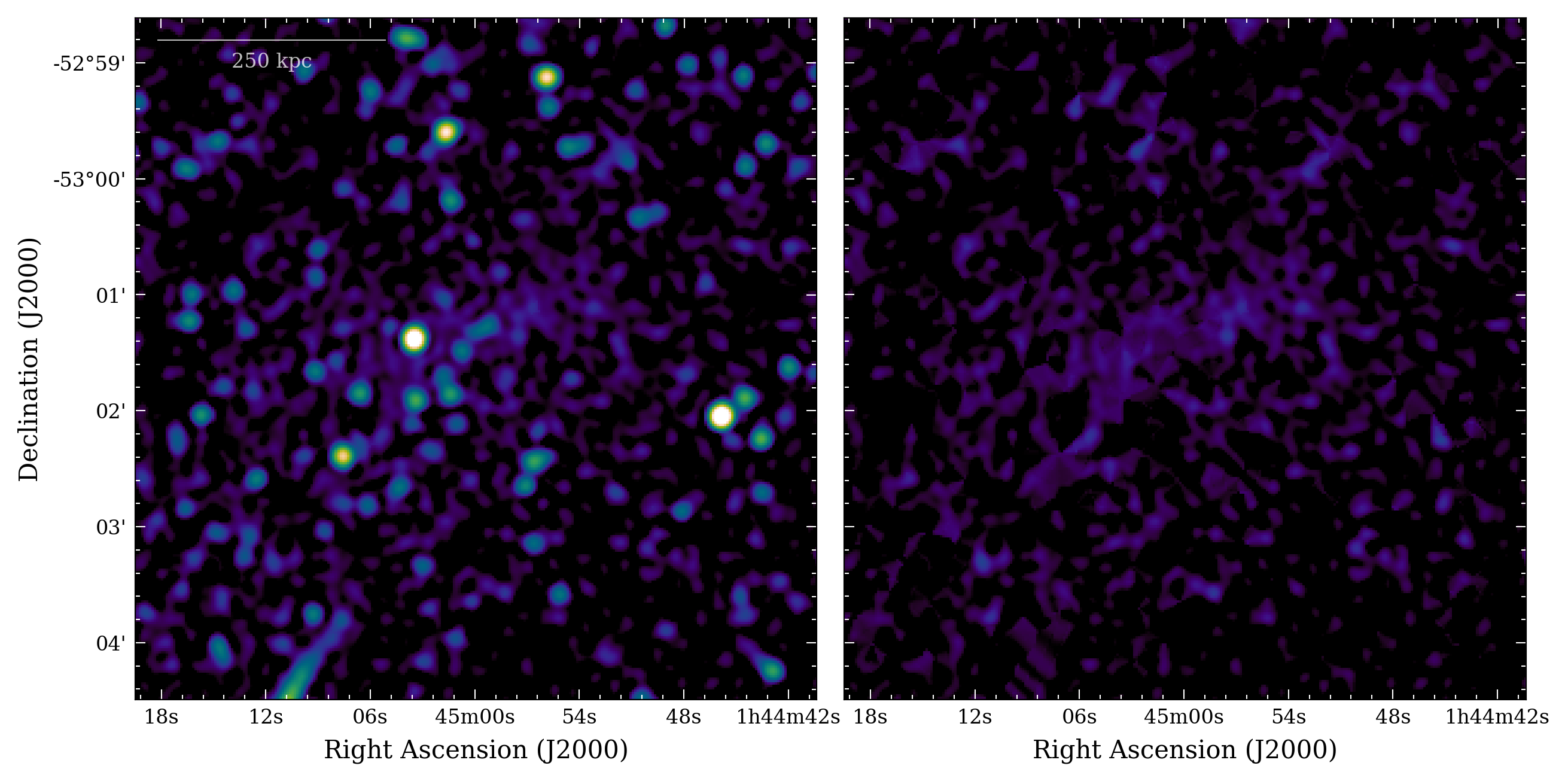}
  \includegraphics[width=\hsize,trim={0cm 0cm 0cm 0cm},clip,valign=c]{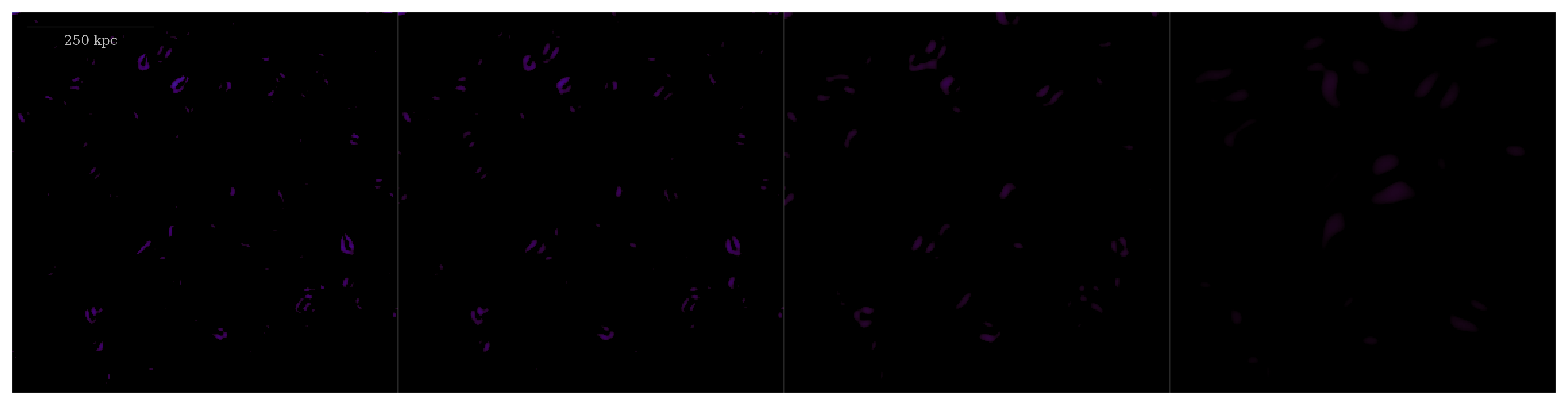}
  \caption{Same as Fig.~\ref{fig:bullet_meerkat} but for J0145.0-5300.}
  \label{fig:J0145.0-5300_meerkat}
\end{figure}

\begin{figure}
  \centering
  \includegraphics[width=\hsize,trim={0cm 0cm 0cm 0cm},clip,valign=c]{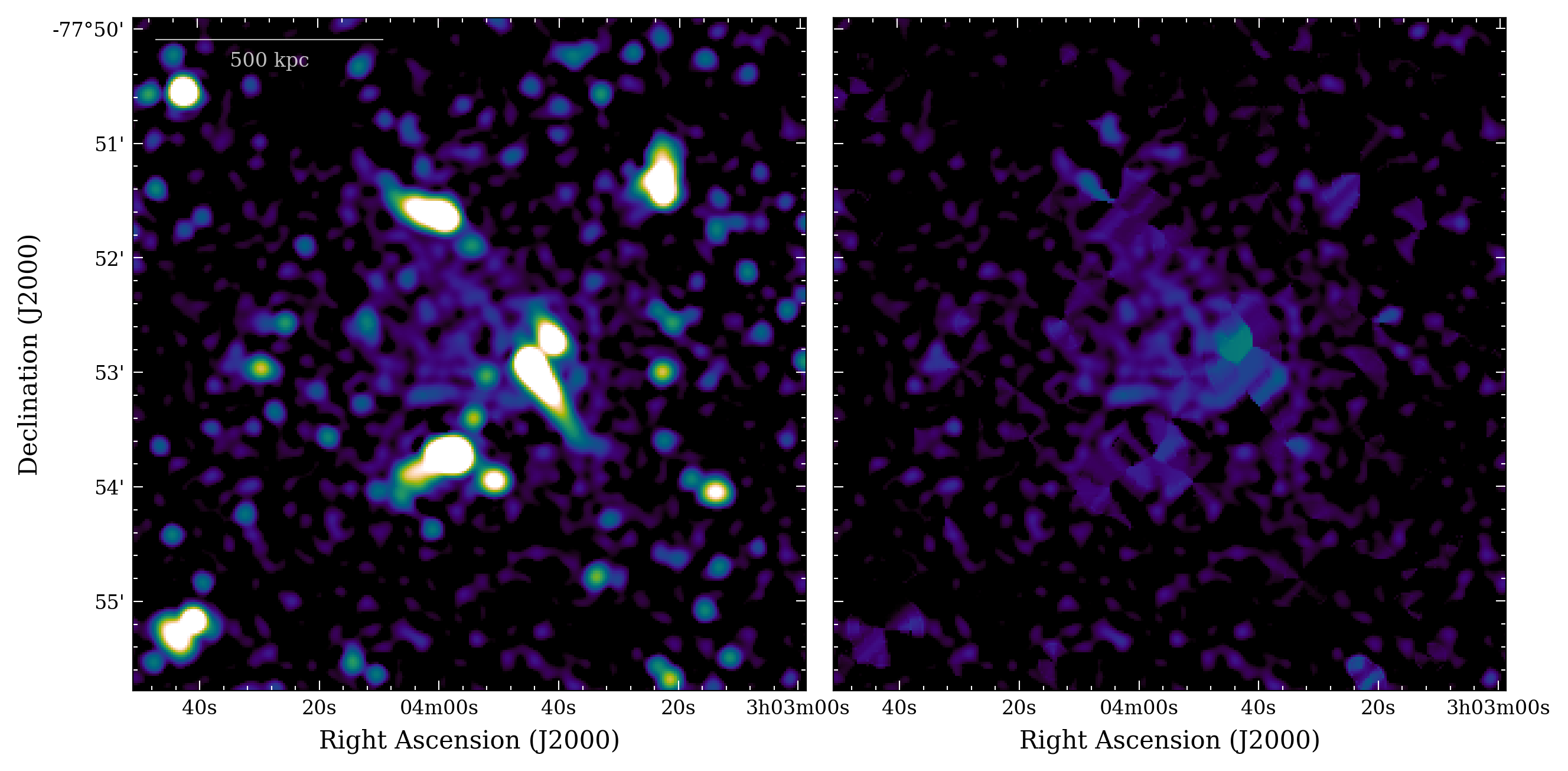}
  \includegraphics[width=\hsize,trim={0cm 0cm 0cm 0cm},clip,valign=c]{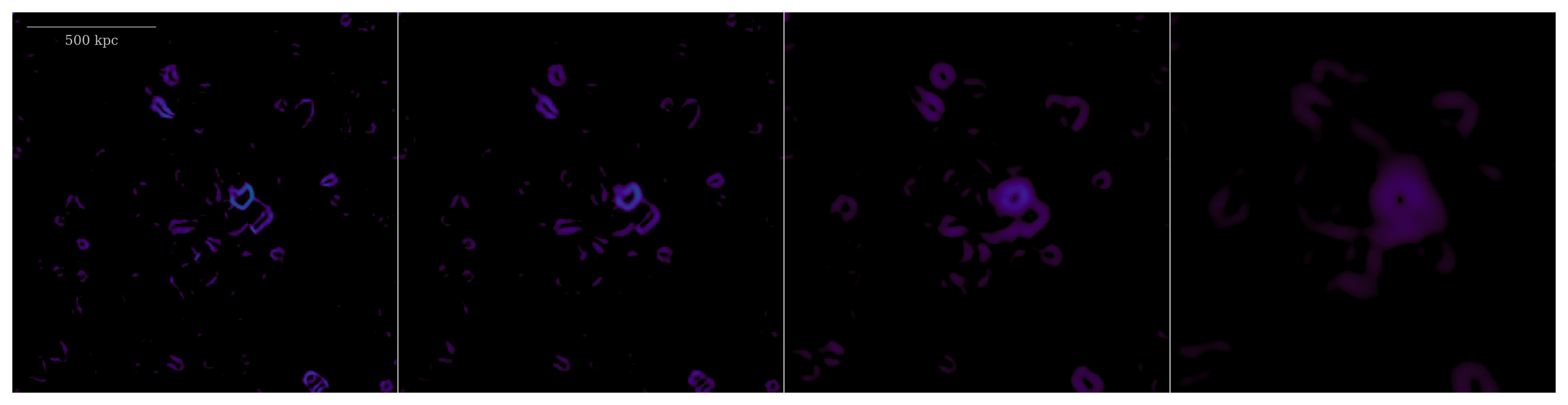}
  \caption{Same as Fig.~\ref{fig:bullet_meerkat} but for J0303.7-7752.}
  \label{fig:J0303.7-7752_meerkat}
\end{figure}

\begin{figure}
  \centering
  \includegraphics[width=\hsize,trim={0cm 0cm 0cm 0cm},clip,valign=c]{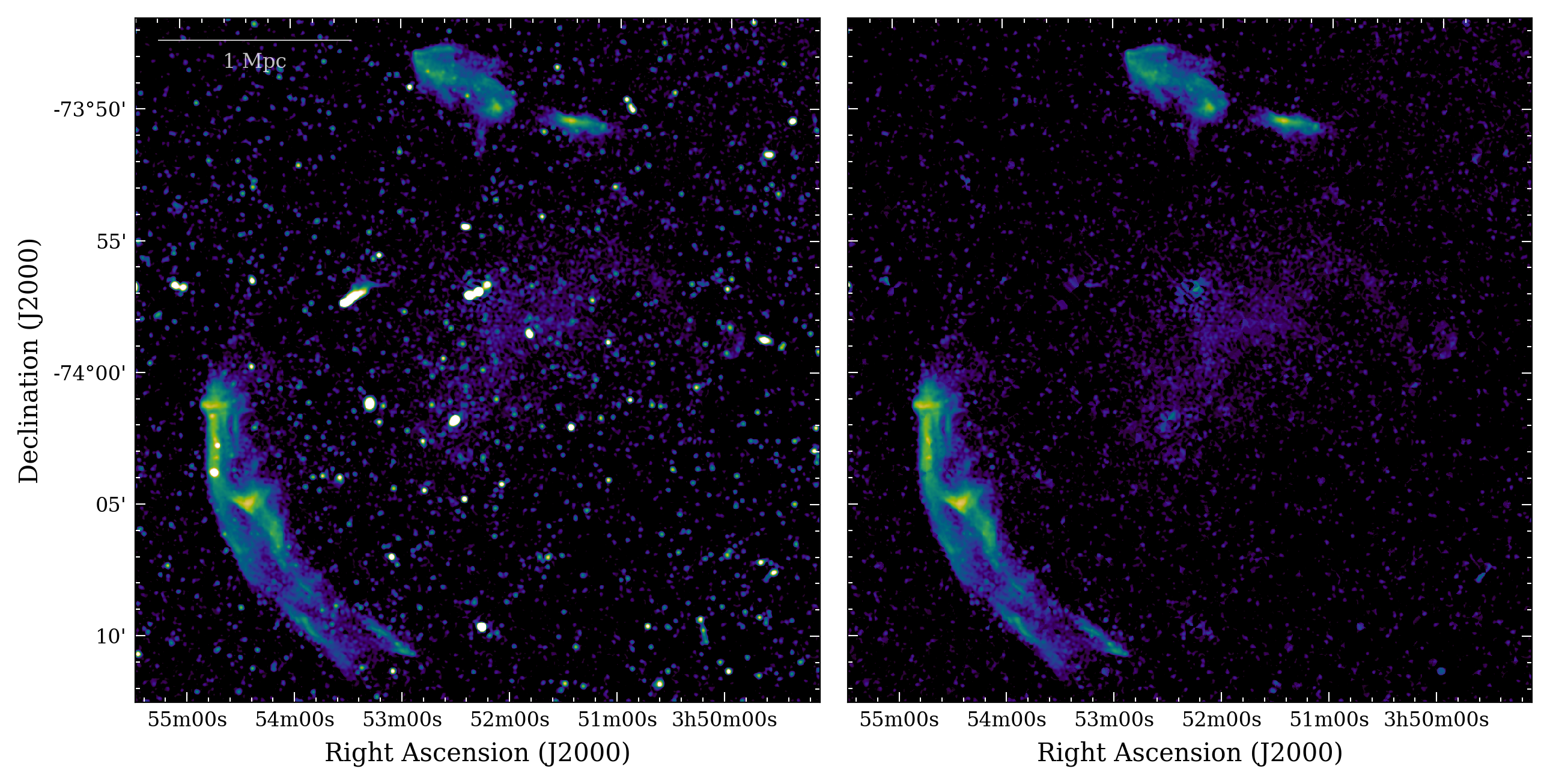}
  \includegraphics[width=\hsize,trim={0cm 0cm 0cm 0cm},clip,valign=c]{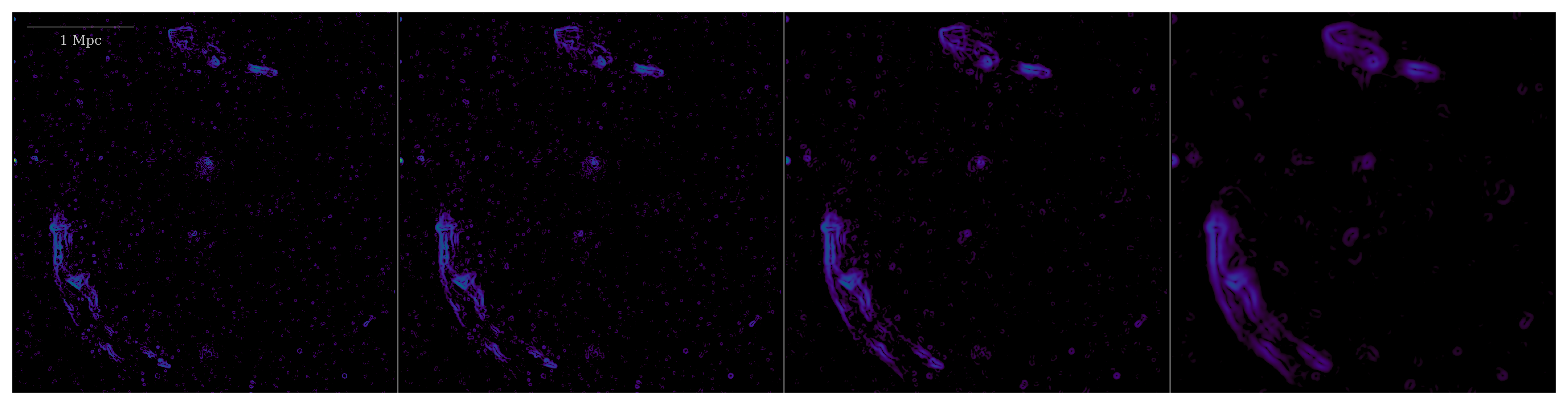}
  \caption{Same as Fig.~\ref{fig:bullet_meerkat} but for J0352.4-7401.}
  \label{fig:J0352.4-7401_meerkat}
\end{figure}

\begin{figure}
  \centering
  \includegraphics[width=\hsize,trim={0cm 0cm 0cm 0cm},clip,valign=c]{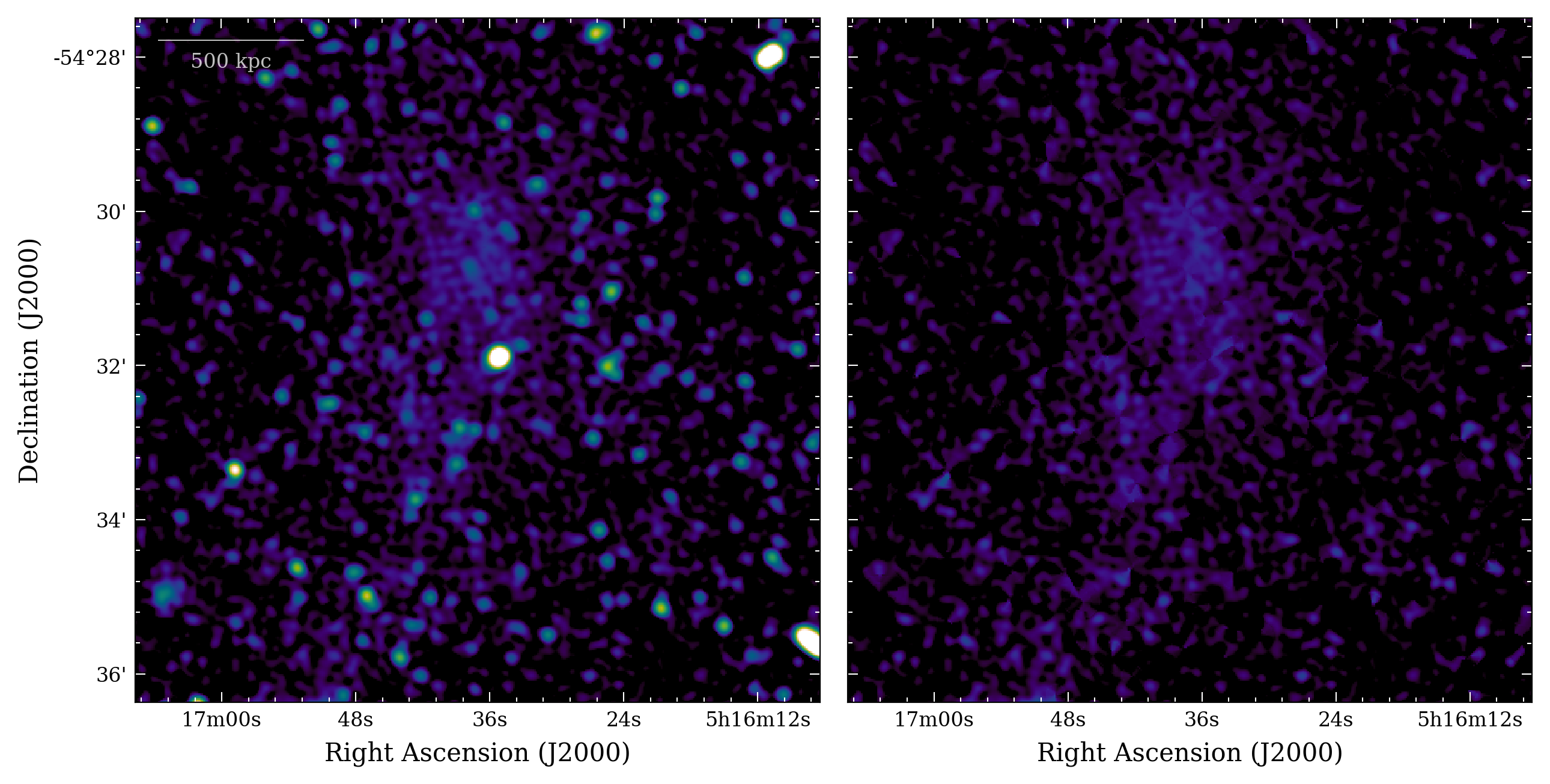}
  \includegraphics[width=\hsize,trim={0cm 0cm 0cm 0cm},clip,valign=c]{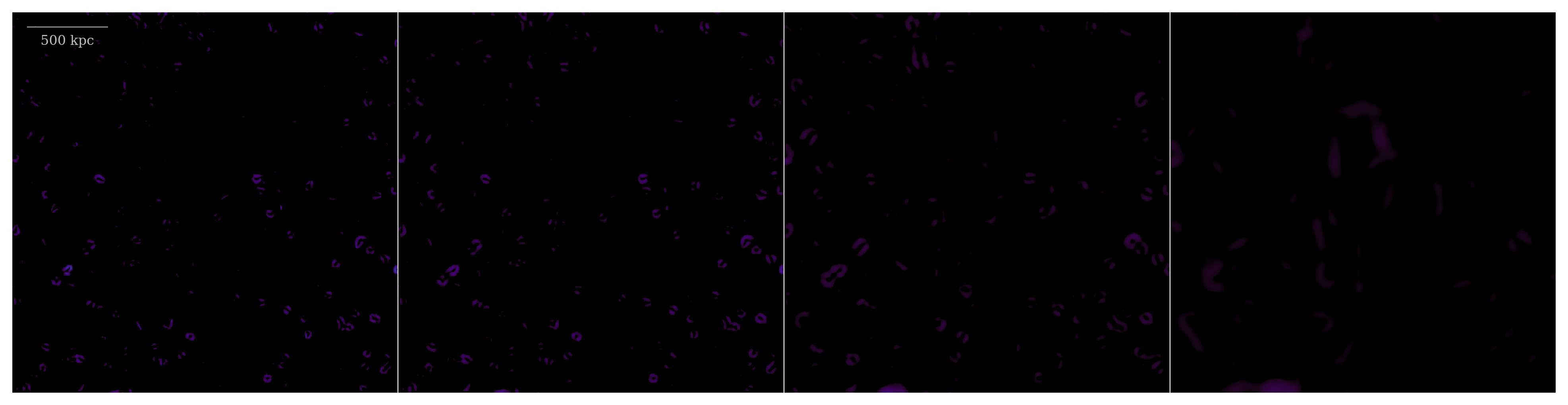}
  \caption{Same as Fig.~\ref{fig:bullet_meerkat} but for J0516.6-5430.}
  \label{fig:J0516.6-5430_meerkat}
\end{figure}

\begin{figure}
  \centering
  \includegraphics[width=\hsize,trim={0cm 0cm 0cm 0cm},clip,valign=c]{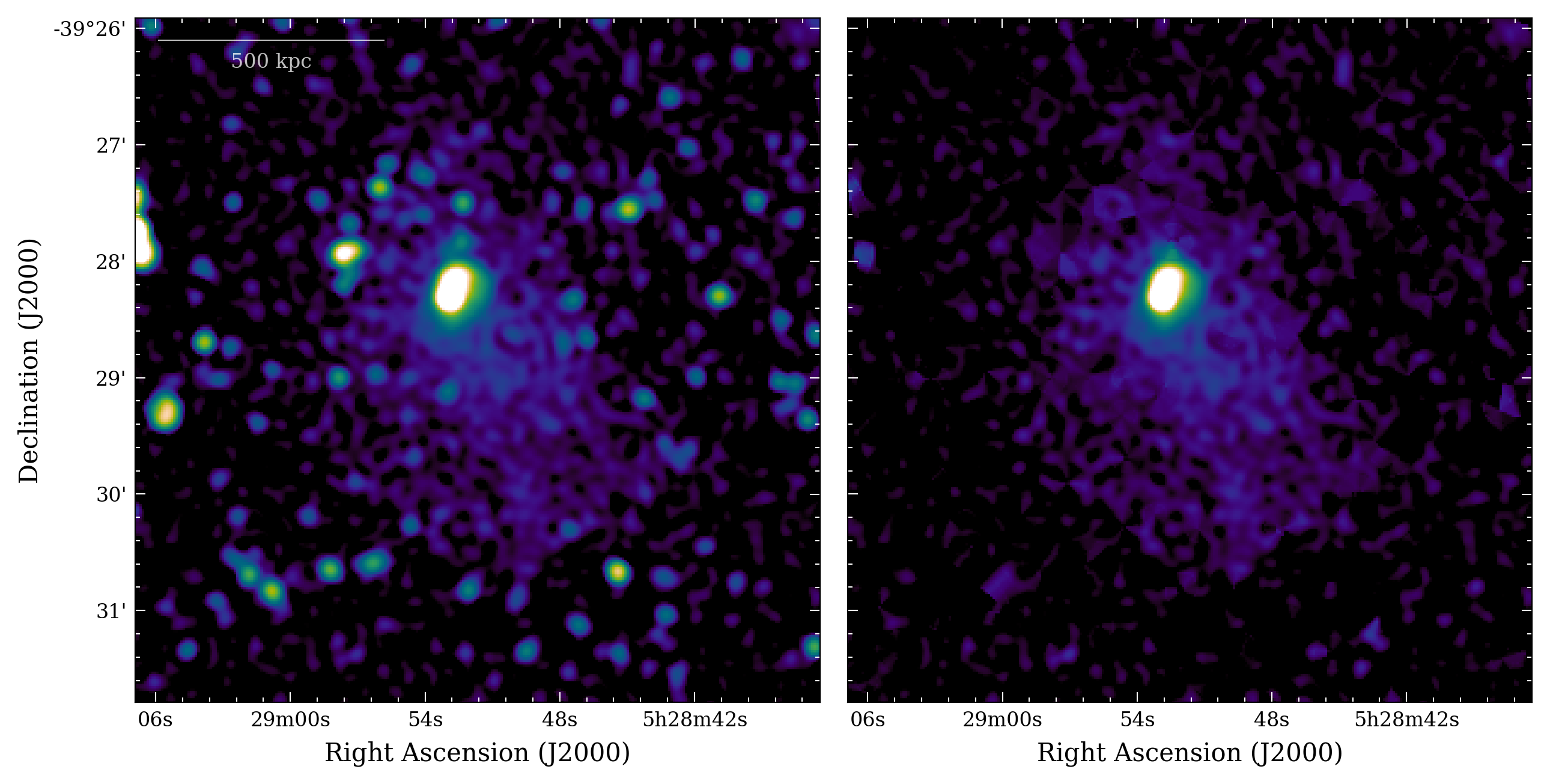}
  \includegraphics[width=\hsize,trim={0cm 0cm 0cm 0cm},clip,valign=c]{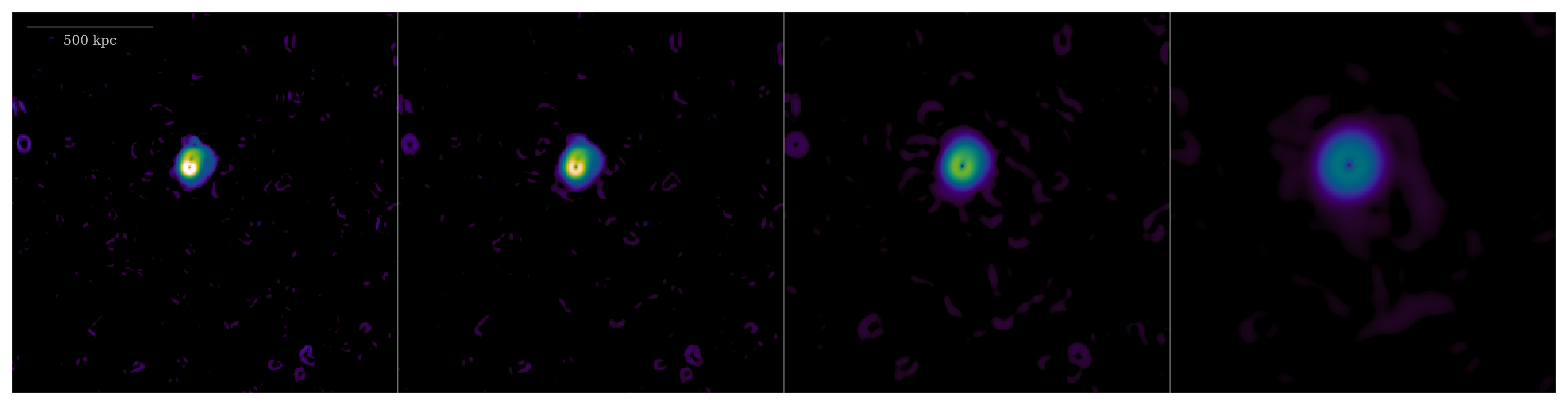}
  \caption{Same as Fig.~\ref{fig:bullet_meerkat} but for J0528.9-3927.}
  \label{fig:J0528.9-3927_meerkat}
\end{figure}

\begin{figure}
  \centering
  \includegraphics[width=\hsize,trim={0cm 0cm 0cm 0cm},clip,valign=c]{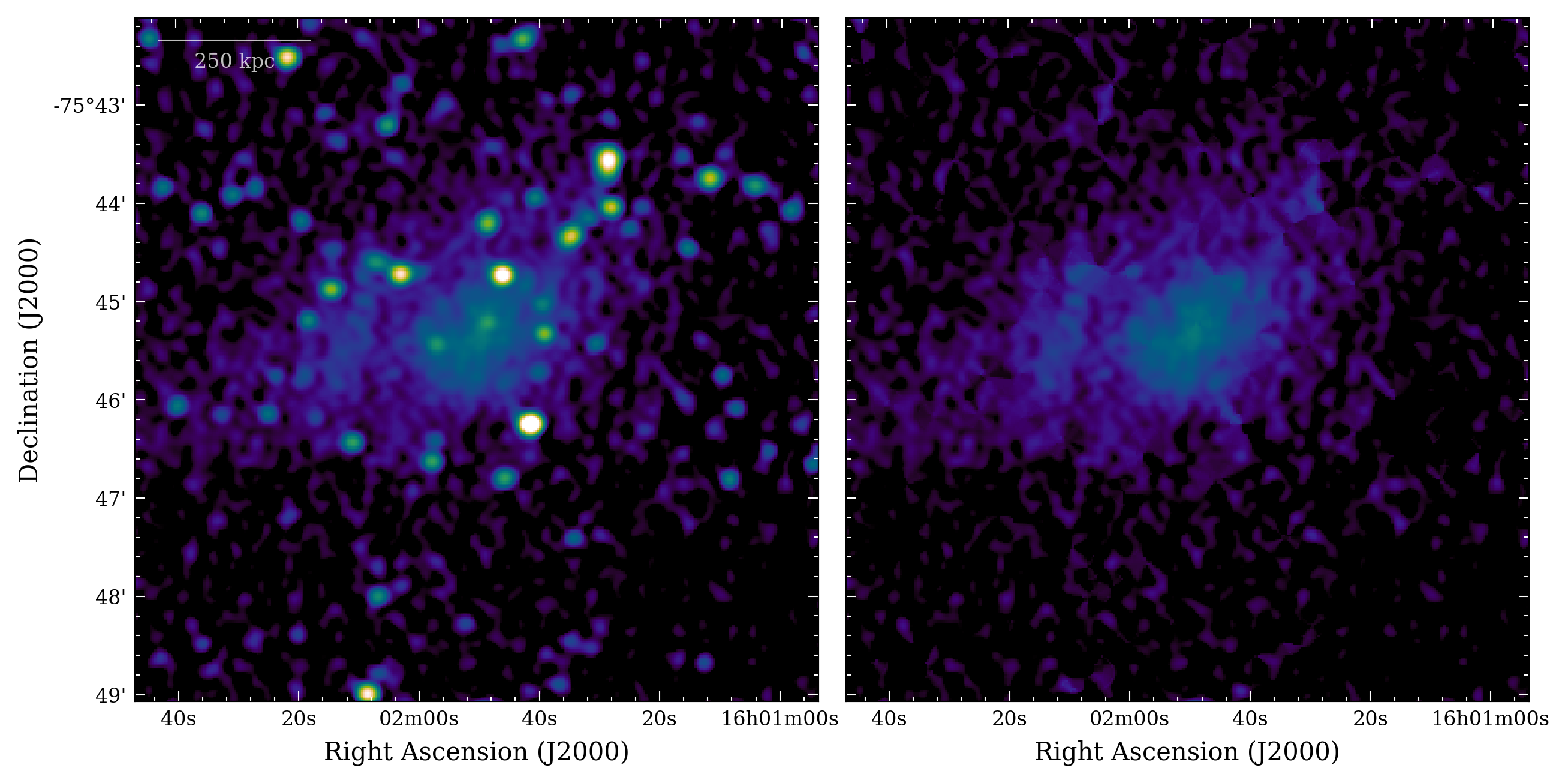}
  \includegraphics[width=\hsize,trim={0cm 0cm 0cm 0cm},clip,valign=c]{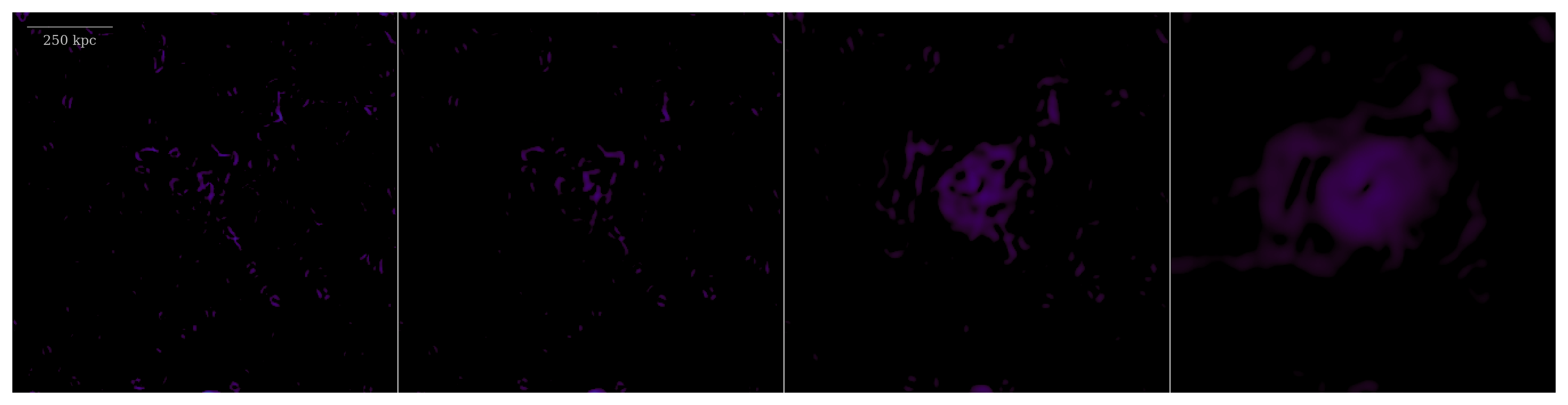}
  \caption{Same as Fig.~\ref{fig:bullet_meerkat} but for J1601.7-7544.}
  \label{fig:J1601.7-7544_meerkat}
\end{figure}

\end{appendix}

\end{document}